%% file: main.tex
\documentclass[12pt]{iopart}
\usepackage{iopams}
\usepackage{graphicx,amssymb}

\newcommand{\bilderscale}{0.12}

\newcommand{\fig}[2]{\includegraphics[width=#1\columnwidth]{./figures/#2}}

\newlength{\bilderlength}
\newcommand{\usebilderscale}{\bilderscale}
\newcommand{\bilderskip}{\hspace*{0.8ex}}

\newcommand{\diagram}[1]{%
\settowidth{\bilderlength}{\bilderskip%
\includegraphics[scale=\usebilderscale]{#1}\bilderskip}%
\parbox{\bilderlength}{\bilderskip%
\includegraphics[scale=\usebilderscale]{#1}\bilderskip}}

\graphicspath{{.}{./figures/}{./figures/icons/}}

\begin{document}
\title{A growth model for RNA secondary structures}

\author{Francois David$^1$, Christian Hagendorf$^2$, and Kay J\"org Wiese$^2$}
\address{$^1$Service de Physique Th\'eorique, CEA Saclay, 91191
Gif-sur-Yvette Cedex, FRANCE}

\address{$^2$CNRS-Laboratoire de Physique Th\'eorique, Ecole Normale
Sup\'erieure, 75231 Paris cedex 05, FRANCE}
\ead{francois.david@cea.fr, hagendor@lpt.ens.fr, wiese@lpt.ens.fr}

\date{\today}

\begin{abstract}
A hierarchical model for the growth of planar arch structures for RNA secondary structures is presented, and shown to be equivalent to a tree-growth model. Both models can be solved analytically, giving access to scaling functions for large molecules, and corrections to scaling, checked by numerical simulations of up to 6500 bases. The equivalence of both models should be helpful in understanding more general tree-growth processes. 
\end{abstract}

\pacs{87.14.gn, 87.15.bd, 02.10.Ox, 02.50.Ey}
\maketitle

\input{intro}

\input{chap2}
\input{chap3}
\input{chap4}

\input{chap5}

\input{chap6}
\input{chap7}
\input{conclusion}
\input{appendix}

\end{document}

%% file: intro.tex
\section{Introduction}

\begin{figure}[b]
  \centering
  \includegraphics[width=0.9\textwidth]{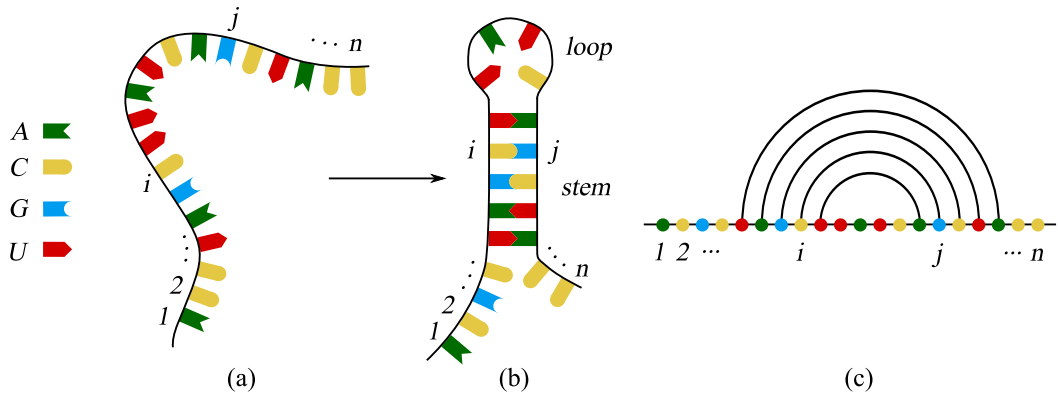}
  \caption{RNA molecules, like DNA, are long chain heteropolymers built from
four types of nucleotides: adenine (A), uracil (U), guanine (G) and
cytosine (C). In solution, a single RNA molecule bents back onto
itself and folds into a configuration of loops, stems and
terminating bonds, due to pair formation from nucleotides located on
different parts of the polymer strand. The set of base pairs,
Watson-Crick pairs A--U, G--C and the less favorable wobble pair G--U,
defines the \textit{secondary structure}. Illustration of RNA secondary structures: (a) an RNA molecule with given base sequence folds into a base pair configuration (b). In the absence of pseudo-knots the secondary structure may be represented as a diagram of non-intersecting arches (c).}
  \label{fig:secondary}
\end{figure}

RNA molecules play an important role in all living organisms \cite{Gesteland2005}. They
are usually found in a at least partially folded state, due to the
pairing of a  base with at most one other base. A given
configuration is thus characterised by the set of base pairings, see figure \ref{fig:secondary}. These pairings are mostly planar \cite{BundschuhHwa1999,Higgs2000,PagnaniParisiRicciTersenghi2000} (see \cite{OrlandZee2002} for non-planar corrections), which is what we will suppose from now on. At
high temperatures, in the so-called ``molten phase'', 
energetic considerations only play a minor role, and the probability $P_{ij}$ of two RNA-bases to pair, is  \cite{deGennes1968}
\begin{equation}
 P_{ij} \sim |i-j|^{-\rho}\ , \qquad \rho=\frac32\ ,
\end{equation}
where $i$ and $j$ are the labels of the bases counted along the
backbone/strand, and $n$ is the overall size of the RNA-molecule, i.e.\ its
total number of bases.

At low temperature, the RNA-molecule will settle
into the optimally paired (or folded) configuration, i.e.\ the
minimal energy state, as long as this state is reachable in the
available time-scales. The optimal fold for a
given molecule is a question to be answered by biology.
Since biological sequences are
 rather specific, much effort has been invested to
understand the properties of  a random sequence, termed ``random RNA''. The idea is that either
the folding properties of random RNA are close to those of
biological sequences, or if not, that they must be characterised in
order to  understand the deviations present for biological RNA,
giving eventually a hint why nature is organising in a certain way.

Characterising random RNA has proven a challenge so far: Numerical
work
\cite{BundschuhHwa1999,BundschuhHwa2002a,KrzakalaMezardMueller2002,Monthus2007}
is restricted to relatively small molecules, with up to maximally
2000 bases \cite{HuiTang2006}, despite the fact that rather efficient polynomial
algorithms exist ($\sim n^{3}$). Analytical work was pioneered by
Bundschuh and Hwa \cite{BundschuhHwa1999,BundschuhHwa2002a}. From
their numerical work, they claim that for large molecules, a
random-base model is indistinguishable from a random pairing-energy
model, where the pairing energy $\epsilon_{ij}$ between base $i$ and
$j$ is a random Gaussian variable, confirmed in \cite{KrzakalaMezardMueller2002}. Bundschuh and Hwa then conjectured
that a phase transition separates the high-temperature molten phase
from a low-temperature frozen phase. Using an RG treatment, L\"assig
and Wiese \cite{LaessigWiese2005} showed  analytically, that this
phase transition exists, and is of second order. They also
calculated the exponents characterising the transition, and using a
locking argument extended their findings to the low-temperature
(glass) phase. 
David and Wiese \cite{DavidWiese2006} substantiated these findings,
by constructing the field theory, showing its renormalisability to
all orders, and performing an explicit 2-loop calculation, yielding
\begin{equation}
\rho_{\mathrm{transition}} = \rho_{\mathrm{frozen}} \approx 1.36\ .
\end{equation}
The field theory makes some definite predictions about the
transition, which are hard to verify numerically. A major problem is that the systems are not large enough to analyze the asymptotic behavior. Under these circumstances, the  
 knowledge of a scaling function would be very helpful, as would
be the knowledge of the form of corrections to scaling.

We therefore propose a simple
{\em hierarchical} model, where all this can be calculated analytically\footnote{While working on this project, we learned
from Markus M\"uller that he had considered this model in his
PhD-thesis \cite{MarkusMuellerPhD}, but not published elsewhere. He also found the scaling exponent $\zeta$ to be discussed
below, but did not consider the scaling-functions and corrections-to-scaling which are the main purpose of this article.}. This is based on the observation, that if the
$n(n-1)/2$ possible pairing energies $\epsilon_{ij}$ are ordered
hierarchically
\begin{equation}
\epsilon_{i_{1}j_{1}} \gg \epsilon_{i_{2}j_{2}}\gg \dots \gg
\epsilon_{i_{n(n-1)/2}j_{n(n-1)/2}}\ ,
\end{equation}
then the construction of the minimal energy configuration is much
simplified: First take the largest pairing energy
$\epsilon_{i_{1}j_{1}}$, and pair bases $i_{1}$ and $j_{1}$. Among
the remaining pairings, consider only those allowed by planarity.
Among those, choose the one with the largest pairing energy, and
pair the corresponding bases. Repeat this procedure until no more
bases can be found. The same idea is at the base of the dynamics for
\textit{greedy algorithms} of RNA folding: At each time-step, choose the most favorable base pairing and fold it. 

In this article, we systematically analyse the statistical properties
of the structures built in the hierarchical
model. In particular, we compute exactly its properties as
$n\to \infty$, for example we prove that in this limit
the pairing probability reads
\begin{equation}
  P(i,j) \sim |i-j|^{-\rho},\qquad \rho = \frac{7-\sqrt{17}}{2}
  \approx 1.44\ .
\end{equation}
We then calculate scaling functions for higher moments of the ``height function'' (which encodes the pairings), and their finite-size corrections. This is achieved with two complimentary approaches: Generating functions for the arch-deposition model introduced above, and a dual tree-growth process. The advantage is that quantities which can easily be calculated in one model, are difficult to obtain in the other, and vice versa. 
This  idea may  be interesting for more general tree-growth processes, since if the dual model can be constructed there, it would allow to calculate otherwise inaccessible quantities.  For examples of tree growth processes, we refer the reader to   \cite{Pittel1985,Breiman2001,BollobasRiordanSpencerTusnady2001,MauldinSudderthWilliams1992} among the vast existing literature. 

\bigskip

The presentation is organised as follows. In section
\ref{sec:archmodel}, we provide a general framework for the
hierarchical model in terms of recursion relations for finite $n$.
The recursion relations are analysed by means of generating
functions. In the limit $n\to \infty$ we extract the scaling
behaviour of various quantities and compute sub-leading finite-size
corrections in sections \ref{sec:exponent}, \ref{s-scaling} and \ref{sec:pprobs}. We compare our results  to
numerical simulations in section \ref{sec:numerics}. In section \ref{sec:growth} we present an 
alternative tree-growth model which we show to yield
equivalent structures even though the dynamics of their construction is 
quite different.
Several technical points and extensions are relegated to three appendices.

%% file: chap2.tex
\section{Arch deposition model}
\label{sec:archmodel}
\subsection{Arch systems and height functions}
\label{archmodel.def} 
\begin{figure}[h]
\begin{center}
  \includegraphics[width=0.9\textwidth]{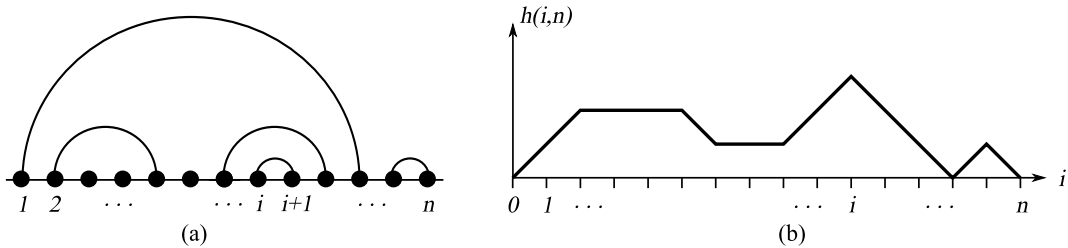}
\end{center}
\caption{(a) Arch diagram for a planar structure. (b) Corresponding
height relief, defined in (\ref{eqn:height}).}
\label{fig:structure}
\end{figure}

We consider a strand with $n$ bases labeled by
indices $i=1,\dots,n$. Similarly, we use the same index $i$ to label
the segments between consecutive bases $i$ and $i+1$. A secondary
structure $\mathcal{C}$ is a set of base pairs $(i,j)$ with $1\leq
i<j\leq n$. $\mathcal{C}$ is called \textit{planar} if any two
$(i,j), (k,l) \in \mathcal{C}$ are  either independent $i<j<k<l$ or
nested $i<k<l<j$. In what follows, the structures are supposed to be
planar. Thus, we may represent a given structure by a diagram of
non-intersecting arches (see figure \ref{fig:structure}a).

\noindent Given some structure $\mathcal{C}$, it is natural to ask
whether it contains an arch $a=(i,j)$. This is answered by
 the {\em contact operator} $\Phi_\mathcal{C}$ defined by
\begin{equation}
\Phi_\mathcal{C}(i,j):= \left\{
\begin{array}{ll}
  $1$ & \mbox{if $a\in \mathcal{C}$} \\
  $0$ & \mbox{otherwise}
\end{array}
\right.\ .
\end{equation}
For our investigations  the so-called {\em height function} for the segment $i$ will play a central role. It is defined as
\begin{equation}
  \label{eqn:height}
  h_{\mathcal{C}}(i,n) := \sum_{j=1}^{i}\sum_{k=i+1}^{n}
  \Phi_{\mathcal{C}}(j,k)
\ .
\end{equation}
It counts the number of arches above a given segment $[i,i+1]$, and thus has boundary conditions
$h_{\mathcal{C}}(0,n)=h_{\mathcal{C}}(n,n)=0$. Therefore, the height
function $h_{\mathcal{C}}(i,n)$ provides a one-to-one correspondence
between $\mathcal{C}$ and mountain reliefs (Dyck-like paths) 
on the
interval $[0,n]$ subject to vanishing boundary conditions and
$|h_{\mathcal{C}}(i+1,n)-h_{\mathcal{C}}(i,n)| = 1$ or $0$ (see figure\ \ref{fig:structure}b). We
define the average height by
\begin{equation}
  h_{\mathcal{C}}(n) := \frac{1}{n}\sum_{i=1}^{n-1}h_{\mathcal{C}}(i,n)
\ .
\end{equation}

\subsection{The random arch deposition process}

\begin{figure}[htbp]
  \centering
  \includegraphics[width=0.9\textwidth]{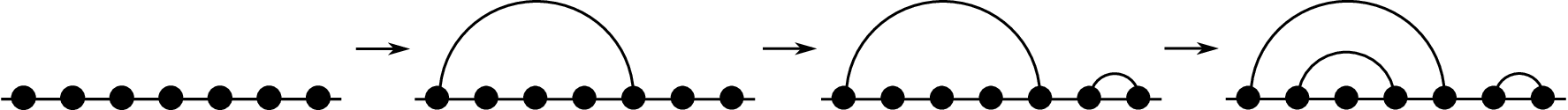}
  \caption{Building up planar structures
  via successive arch deposition.}
  \label{fig:archdep}
\end{figure}

\paragraph{Definition of the model (model A).}
The structures $\mathcal{C}$ are built up in the following way: At initial  time step
$t=0$, we start with $n$ unoccupied points on the line. At each time
step $t$, we deposit a new arch as follows. 
At time step $t-1$, we have already a planar system of $t-1$ arches linking $2(t-1)$ points.
We have  $m=n-2(t-1)$ free points left, and we may build $m(m-1)/2$ different arches. We now consider the subsets of these arches $(i,j)$ which keep the arch system planar, when added to the present structure.
We choose at random, and with equal probability, one of these arches, and add it to the system at time $t$ (as depicted on figure ~\ref{fig:archdep}).

The process is stopped as soon as no more planar deposition is possible. The stopping time ($t_{\mathrm{stop}}=\,$number of arches of the final configuration) will vary from configuration to configuration, since not all points get paired.
\medskip

We call this arch deposition
process ``model \textbf{A}". 
 
\begin{figure}[htbp]
  \centering
  \includegraphics[width=0.9\textwidth]{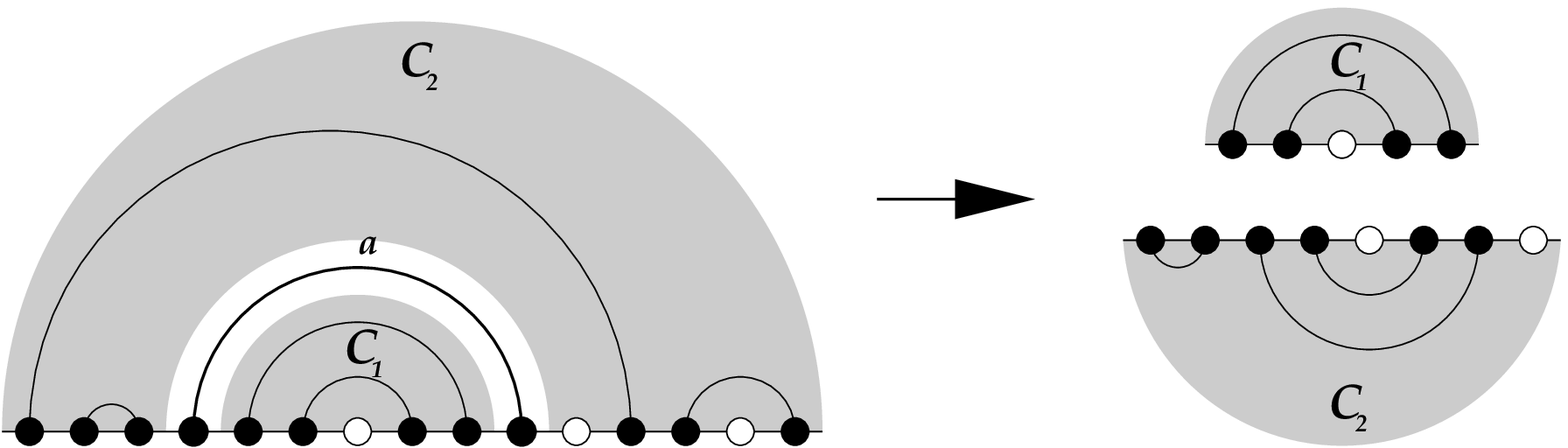}
  \caption{Decomposition of a configuration $\mathcal{C}$ in model
  \textbf{A}} 
  \label{fig:decomp}
\end{figure}

\paragraph{Hierarchy and recursion for probabilities.}
Our construction is ``hierarchical'' in the sense that each
deposition partitions the strand into two non-connected substrands.
Since this procedure is performed at random, it naturally induces a
probability measure $P_A(\mathcal{C})$ on the set of structures
$\mathcal{C}$ with a given number of points $n$. Although it turns
out to be a quite tedious exercise to compute the probabilities for
structures, even with only $n=4,5,6,\dots$ points, we can write a
formal yet powerful recursion relation for $P_A(\mathcal{C})$. Given $\mathcal{C}$, any arch
$a\in \mathcal{C}$ may have been the first arch in the construction
process (at $t=1$). Since the deposition of $a = (i,j)$, $1\leq i <
j \leq n$, is not constrained by the presence of any other arches,
its probability is uniform and simply given by $2/[n(n-1)]$. This
first step leads to a separation of the strand into an ``interior''
part with $n_1=j-i-2$ points and an ``exterior'' part with
$n_2=n-j+i$. The deposition process then grows structures
$\mathcal{C}_1$ inside and $\mathcal{C}_2$ outside the first arch (see figure \ref{fig:decomp}). The key
observation is that these structures grow independently. Therefore,
their joint probability factorises:
\begin{equation}
P_A(\mathcal{C})=\sum_{\mathrm{arch}\,a\in \mathcal{C}} {2\over n(n-1)}\,
P_A(\mathcal{C}_1)\,P_A(\mathcal{C}_2) \label{eqn:decomp}
\end{equation}
This recursion relation, together with the initial condition $P_A=1$
for the $n=0$ and the $n=1$ configurations (no point and  a single
free point), is sufficient to obtain all probabilities. In fact,
it is this relation that renders the arch-deposition model amenable to
exact analytic calculations.

With the help of $P_A(\mathcal{C})$ we can compute averages, i.e.
expectation values of an observable $\mathcal{F}_{\mathcal{C}}$ via
\begin{equation}
  {\cal F} = \langle\mathcal{F_{\cal C}}\rangle = \sum_{\mathcal{C}}P_A(\mathcal{C})
  \mathcal{F}_{\mathcal{C}}\ ,
\end{equation}
where the sum is carried out over all possible structures with a
fixed number $n$ of points. Throughout this article, we follow the
convention to note objects depending on an individual structure
$\mathcal{C}$ with a subscript.

\subsection{Summary of results}
In this article, we focus on the mean height at
a given point $h(i,n) = \langle h_{\mathcal{C}}(i,n)\rangle $ and
the probability $P(i,j)$ that two bases located at $i$ and $j$ are paired. The latter is the expectation
value of the contact operator $\Phi_{\mathcal{C}}(i,j)$, $P(i,j):=\langle \Phi_{\mathcal{C}}(i,j)\rangle$.

Before embarking into calculations, let us briefly summarise some important properties of these quantities as well as our main results.
First, note that the construction of the height function
(\ref{eqn:height}) implies that $h(1,n)$ is the probability that
point $i=1$ is paired to any other point $2\leq j \leq n$ on the
strand. Since averaging over structures will lead to translational
invariance of $\left<\Phi_{\mathcal{C}}(i,j)\right>=\left<\Phi_{\mathcal{C}}(i+m,j+m)\right> $,
for all $m$, we can interpret $h(1,n)$ as the
probability that some arbitrary point $1\le i \le n$ is involved in
a pair. We compute $h(1,n)$ for any $n$ and show that it converges to
\begin{equation}
  \lim_{n\to\infty}h(1,n) = 1-e^{-2}=0.864665\dots
\end{equation}
The full information about all possible structures is contained both in the height profiles as in the pairing probabilities. In the scaling limit $n\to \infty$, we show that the height function and the pairing probabilities take the scaling forms
\begin{eqnarray}
  h(i,n)\mathop{\sim}_{n\to\infty} n^{\zeta}\mathcal{H}_1\left(\frac{i}{n}\right)\qquad \mbox{and}\qquad
  P(i,j) \mathop{\sim}_{n\to\infty} n^{-\rho}\;\mathcal{P}\left(\frac{|i-j|}{n}\right)
  \label{eqn:scaling}
\end{eqnarray}
with scaling functions $\mathcal{H}_1$ and $\mathcal{P}$ which we
compute exactly as well as the scaling exponents $\zeta$ and $\rho$.
From eqn.\ (\ref{eqn:height}), we immediatly deduce the scaling
relation $\zeta +\rho = 2$. 

The exponent $\zeta$ is also related to the intrinsic Hausdorff dimension of the tree structure dual to the arch system by
$d_h=1/\zeta$.
Therefore it is sufficient to determine the exponent $\zeta$ which we show to be
\begin{equation}
  \zeta = \frac{\sqrt{17}-3}{2} \approx 0.561553 \ , \qquad  \rho = \frac{7-\sqrt{17}}{2} \approx 1.43845\ .
\end{equation}
This agrees with 
\cite{MarkusMuellerPhD}.
It follows that the average mean height
$h(n) = \langle h_{\mathcal{C}}(n)\rangle$ grows like $n^{\zeta}$ for large $n$. We determine its exact generating function, which allows us to compute sub-leading corrections to the scaling limit to any desired accuracy.

The analysis of higher moments $\langle h(i,n)^k\rangle$ naturally raises the question of multifractality of the arch structures/height profiles. We show that
\begin{equation}
\langle h_{\mathcal{C}}(i,n)^k\rangle\mathop{\sim}_{n\to\infty} n^{\zeta_k}\mathcal{H}_k\left(\frac{i}{n}\right)\qquad \mbox{with } \zeta_k = k\zeta
\end{equation}
with scaling functions $\mathcal{H}_k$ that we can in principle compute. Using this result we are able to prove the absence of multifractality.

\subsection{Recurrence relations and generating functions}
We now exploit the recursion relation (\ref{eqn:decomp}) to
compute the moments of the height function $\langle
h_{\mathcal{C}}(i,n)^k\rangle,\, k=0,1,2,\dots$ Our general
strategy is to extract their properties by analyzing the behavior
of their corresponding generating functions.

\subsubsection{Recurrence relation for the height function: the principle}
\begin{figure}[h]
\begin{center}
\includegraphics[width=0.9\textwidth]{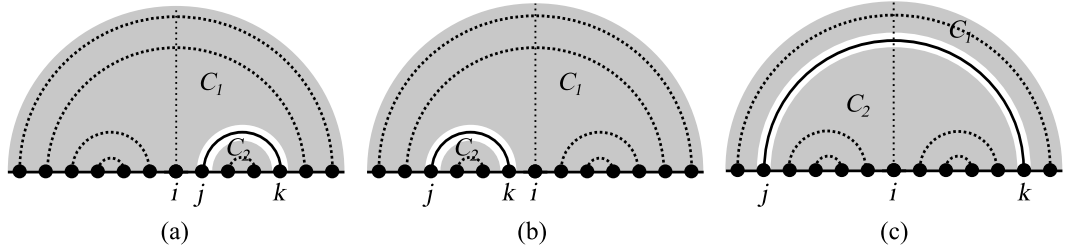}
\caption{The decomposition used in order to derive the recurrence relation for the average height function $h(i,n)$.}
\label{recurrenceH}
\end{center}
\end{figure}

We want to evaluate the height
$h_{\mathcal{C}}(i,n)$ for a given structure $\mathcal{C}$. The
first arch $a=(j,k)$ splits $\mathcal{C}$ into the two independent
substructures $\mathcal{C}_1$ and $\mathcal{C}_2$ with lengths $n_1
= n - k + j-1$ and $n_2 = k-j-1$ respectively. We now consider the height over segment $[i,i+1]$. With respect to first arch $a$,
this segment  may have three different locations, as
indicated on figure\ \ref{recurrenceH}. (a) if $i<j$, the segment is
situated on the part of the strand which belongs to $\mathcal{C}_1$
and thus the height is given by $h_{\mathcal{C}_1}(i,n-k+j-1)$. (b)
The case $i\geq k$ is similar, but
 we must shift the position $i\rightarrow i - k+j-1$, we
 thus find the height $h_{\mathcal{C}_1}(i-k+j-1,n-k+j-1)$. (c)
Finally, if $j\leq i < k$, we have to count the height for the
structure $\mathcal{C}_2$ with the readjusted position  $i \to i-j$, the
arches in $\mathcal{C}_{1}$ over $\mathcal{C}_2$ and the contribution from $a$. These
three terms together are
$h_{\mathcal{C}_1}(i-1,n-k+j-1)+h_{\mathcal{C}_2}(i-j,k-j-1)+1$.

Upon averaging and using (\ref{eqn:decomp}) we
obtain the recursion relation for the average height function
\begin{eqnarray}
\label{rec4h}
 \fl  \frac{n(n-1)}{2}h(i,n)
   =&
 \sum_{i<j<k<n}h(i,n{-}k{+}j{-}1)
 +\sum_{0<j<k\leq i}h(i{-}k{+}j{-}1,n{-}k{+}j{-}1)\nonumber\\
  &+ \sum_{0<j\leq i < k < n} \left[ h(j{-}1,n{-}k{+}j{-}1)
  +h(i{-}j,k{-}j{-}1) + 1 \right]
\end{eqnarray}
In the scaling limit $n\to \infty$, we may insert  the 
 scaling ansatz $h(i,n) \sim n^\zeta {\mathcal H}(i/n)$ from (\ref{eqn:scaling}) and replace sums by integrals, which yields after a few manipulations a Volterra-like double integral equation for the scaling function $\mathcal{H}(x)$. It is possible, though tedious, to show that the integral equation allows a solution $\mathcal{H}(x) \propto x^\zeta(1-x)^\zeta$ with the scaling exponent $\zeta = (\sqrt{17}-3)/2$. Besides the quite complicated treatment of the integral equation we have found evidence for this scaling form from numerical simulations (see section \ref{sec:numerics}).

In the sequel, we shall develop a more systematic approach to extract the scaling behaviour which is based on recursion relations like (\ref{rec4h}). Furthermore, this allows to compute sub-leading corrections to the scaling limit and therefore to exactly quantify finite-size contributions. 

\subsubsection{Generating functions for the local height moments}
Since the relations for the height $h$ are additive in $h$, it is
convenient to deal with the exponential function as a generating
function of the moments. We thus consider the generating function
for the height $h$ at site $i$ for a strand of length $n$
\begin{equation}
\label{lf49} E(i,n;z)=\langle \exp\left(z\,
h_{\mathcal{C}}(i,n)\right)\rangle=\sum_{k=0}^\infty {z^k\over k!}\
\langle h_{\mathcal{C}}(i,n)^k\rangle\ .
\end{equation}
We obtain the recurrence equation for $E$ 
\begin{eqnarray}
\label{lf50}
 \fl\frac{n(n-1)}{2}E(i,n;z)=&\sum_{i<j<k}E(i,n-k+j-1;z)\nonumber\\
 &+\sum_{j<k\leq i}E(i-k+j-1,n-k+j-1;z)\nonumber\\
 &+e^z \sum_{j\leq i < k}E(i-1,n-k+j-1;z)E(i-j,k-j-1;z)
\ .\qquad
\end{eqnarray}
Note the crucial factorisation in the last term due to the
independence of the substructures $\mathcal{C}_1$ and
$\mathcal{C}_2$ inside $\mathcal{C}$ once the first arch $a$ is chosen. It is convenient to introduce the ``grand-canonical'' generating
function
\begin{equation}
\label{lf51} G(u,v;z) = \sum_{n=0}^{\infty}\sum_{i=0}^n \left\langle
e^{zh_{\mathcal{C}}(i,n)}\right\rangle u^i
v^{n-i}=\sum_{n=0}^{\infty}\sum_{i=0}^n E(i,n;z) u^i v^{n-i} \ ,
\end{equation}
which contains the contribution of strands with arbitrary length
$n$, and which is left/right symmetric $G(u,v;z)=G(v,u;z)$.
The discrete recursion
relation for $E$ becomes the non-linear partial differential
equation
\begin{eqnarray}
\label{PDE4H}
    &
    \fl\left[ {1\over 2}\left(u^2{\partial^2\over\partial
u^2}+v^2{\partial^2\over\partial
v^2}\right)+uv{\partial^2\over\partial u\partial v}\right]G(u,v;z)\nonumber\\
    &\fl\quad
    \quad = \left[ {u^2\over
(1-u)}\left(u{\partial\over\partial u}+1\right) +{v^2\over
(1-v)}\left(v{\partial\over\partial v}+1\right)\right] G(u,v;z) +
uv\,e^z\,G(u,v;z)^2
\end{eqnarray}
We directly derive initial conditions for $G(u,v;z)$ at $u=0$ (or $v=0$)
from the series development (\ref{lf51}). Since by definition the height
function vanishes at the ends of the strand, we have
\begin{equation}
\label{Gu0} G(0,v;z)=\sum_{n=0}^\infty \left\langle
e^{zh_\mathcal{C}(0,n)}\right\rangle v^n=\sum_{n=0}^\infty v^n={1\over
1-v} \ .
\end{equation}
For $z=0$ we find $G(u,v;0) = (1-u)^{-1}(1-v)^{-1}$.

\subsubsection{Generating functions for the local height $h(i,n)$ and $h(n)$}
From $G$ we obtain the generating function for the height
$h(i,n)$ itself
\begin{equation}
\label{Fdef} F(u,v)=\sum_{n=0}^\infty\sum_{i=0}^n \langle h_\mathcal{C}(i,n)\rangle
u^iv^{n-i} =  \left.{\partial\over\partial
z}G(u,v;z)\right|_{z=0}\ .
\end{equation}
Using (\ref{PDE4H}) and $G(u,v;0)=(1-u)^{-1}(1-v)^{-1}$, we conclude that $F$ satisfies the linear partial differential equation
\begin{eqnarray}
\fl \Bigg[ {1\over 2}\Big(u^2{\partial^2\over\partial
u^2}+v^2{\partial^2\over\partial
v^2}\Big)&+uv{\partial^2\over\partial u\partial
v}\Bigg]F(u,v) =\Bigg[ {u^2\over
(1-u)}\left(u{\partial\over\partial u}+1\right)
 +{v^2\over
(1-v)}\left(v{\partial\over\partial v}+1\right)\nonumber\\\fl &+2{u\over1-u}{v\over
1-v} \Bigg] F(u,v) +{u\over (1-u)^2}{v\over (1-v)^2} \label{LPDE4F}
\end{eqnarray}
with initial conditions $F(0,v)=F(u,0)=0$. It is straightforward to obtain the generating function of the 
sum of the heights $nh(n)$ (or total area below the height curve $h(i,n)$, $0\le i\le n$) 
from $F(u,v)$ by setting $u=v$:
\begin{equation}
  K(v) := F(v,v) = \sum_{n=0}^\infty \left(\sum_{i=0}^n\langle h_\mathcal{C}(i,n)\rangle\right)v^n=\sum_{n=0}^\infty nh(n) v^n
  \label{Kdef}
\end{equation}
(\ref{LPDE4F}) implies that $K$ is solution of the ordinary differential equation
\begin{equation}
  (1-v)^2K''(v)-2v(1-v)K'(v)-4(2-v)K(v) = \frac{2}{(1-v)^2}
  \label{ODE4K}
\end{equation}
From $h(0)=h(1)=0$ we infer the intial conditions $K(0)=K'(0)=0$. 
Analysis of (\ref{LPDE4F}) and (\ref{ODE4K}) in the limit $u,v\to 1$ will give access to the scaling limits of the height function as well as its average $h(n)$.

%% file: chap3.tex
\section{Mean height and the scaling exponent $\bzeta$}
\label{sec:exponent}
\label{ss-zeta&h}

In this section, we derive the exact scaling form of the mean height $h(n)$ from the differential equation (\ref{ODE4K}) for $K(v)$. In order to get an idea of the scaling limit, let us suppose that $h(n)$ scales like
\begin{equation}
\label{zetahn} h(n)\,\mathop{\sim}_{n\to\infty} \,c\, n^\zeta\,.
\end{equation}
Since $\zeta>0$, insertion of this \textit{ansatz} into (\ref{Kdef}) implies
that the generating function $K(v)$ is analytic in the vicinity of $v=0$,
with convergence radius $1$. Its closest singularity is situated at $v=1$,
with a power-like divergence
\begin{equation}
\label{Hsingz} K(v)\,\mathop{\sim}_{v\to 1^-}\, {c\,\Gamma(\zeta+2)\over(1-v)^{2+\zeta}}\, .
\end{equation}
Inserting this ansatz into (\ref{ODE4K}), the most singular term  is $p(\zeta) (1-v)^{-2-\zeta} \stackrel!= 0$, with   $p(\zeta)=\zeta^2 +3\zeta-2$. The roughness exponent $\zeta$ is thus solution of  $p(\zeta)=0$, i.e.\
\begin{equation}
\label{zetapm} \zeta_\pm={-3\pm\sqrt{17}\over 2}
\end{equation}
We thus identify the roughness exponent with the larger solution
\begin{equation}
\label{thezeta} \zeta=\zeta_+={\sqrt{17}-3\over 2}=0.561552\dots \ .
\end{equation}
This is the value obtained (using a different argument) by Markus
M\"uller in  \cite{MarkusMuellerPhD}. Let us recall that the roughness exponent $\zeta$ is related to the
pairing-probability exponent $\rho$ by $\zeta+\rho=2$. Moreover, its
inverse is equivalent to the intrinsic fractal dimension (or intrinsic Hausdorff
dimension) $d^\mathrm{f}={1/\zeta}$. Thus
for our model
\begin{equation}
\label{rho&df} \rho={7-\sqrt{17}\over 2}=1.438447\dots \quad,\qquad
d^\mathrm{f}={\sqrt{17}+3\over 4}=1.78077\dots \ .
\end{equation}
This exact value for the roughness exponent $\zeta$ is larger than
the one for generic arch systems (with weight factors given by the
Catalan statistics, i.e.\ generic trees or branched polymers in the
dual picture), which corresponds to RNA in the homopolymer phase (no
disorder), where
\begin{equation}
\label{zeta0} \zeta_{{0}}={1\over 2}\ , \qquad \rho_{{0}}=\frac32\ ,
\qquad d^\mathrm{f}_{{{0}}} = 2\ .
\end{equation}
However, it is smaller than the value observed in numerical simulations for
random RNA \cite{KrzakalaMezardMueller2002,BundschuhHwa2002a}
\begin{equation}
\label{zetaRNA}
\zeta_{\scriptscriptstyle{\mathrm{random\,RNA}}}\approx 0.66 \ ,
\end{equation}
and that of 2-loop RG \cite{DavidWiese2006} $\zeta=0.64$ for random RNA.

We now solve equation (\ref{ODE4K}) exactly.
First, note that a particular solution of the full equation is given by
\begin{equation}
\label{K0expl} K_0(v)=-{1\over (1-v)^2} \ .
\end{equation}
Consequently, we need an appropriate solution $K_1(v)$ of the homogeneous version of (\ref{ODE4K}). 
Performing the transformations
\begin{equation}
\label{K1&u}
K_1(v)=
e^{-2 v} {(v-1)}^{\zeta +1}\, u(z) \ ,\quad z=2(1-v) \ ,\quad
\zeta = \frac{\sqrt{17}-3}{2}
\end{equation}
the equation for $K_1$ is changed to a confluent hypergeometric equation for $u(z)$
\begin{equation}
\label{eq4u}
 z u''(z)+ \left[ 2(\zeta+2)-z\right]u'(z)-(\zeta+1)u(z)=0
\ .
\end{equation}
After a few manipulations, the (appropriate) general solution of this
differential equation for $K(v)=K_0(v)+K_1(v)$ is of the form
\begin{eqnarray}
   K(v) =& \frac{\mathbf{C}_+}{(1-v)^{\zeta
   +2}}\,M(-\zeta,-2-2\zeta;2-2v) \nonumber\\&+ \mathbf{C}_-
   (1-v)^{1+\zeta}M(\zeta+3,2\zeta + 4;2-2v)-\frac{1}{(1-v)^2}
\label{Kexpl}
\end{eqnarray}
where $M(a,b,z)$ is the confluent hypergeometric function
\cite{AbramowitzStegun}. The coefficients $\mathbf{C}_+$ and
$\mathbf{C}_-$ are fixed by the constraint that $K(v)$ be analytic
at $v=0$ and that its Taylor expansion start at order $v^2$. Hence
$\mathbf{C}_+$ and $\mathbf{C}_-$ are given by complicated and not
especially enlightening  combinations of confluent hypergeometric
functions at $z=2$. Numerically we find
\begin{equation}
\label{C1&2expl} \mathbf{C}_+=0.713263\dots\qquad\mathbf{C}_-=0.519299\dots\ .
\end{equation}
The first terms of the Taylor expansion of $K$ are rationals
\begin{equation}
   K(v)=  v^2 + \frac{4}{3}\, v^3 +\frac{8}{3} v^4 +
   \frac{56}{15}\,v^5+\dots
\ .
\end{equation}
The asymptotic limit $n\to \infty$ is equivalent to $v\to 1^-$. In this case, $K(v)$ has a power-law divergence and its most singular terms contribute to the leading orders of $h(n)$. A Taylor expansion of (\ref{Kexpl}) yields
\begin{equation}
  \fl K(v) = \frac{\mathbf{C}_+}{(1-v)^{2+\zeta}}\left(1+\frac{\zeta}{1+\zeta}\;(1-v)
  -\frac{\zeta(1-\zeta)}{(1+\zeta)(1+2\zeta)}\;(1-v)^2\right)-\frac{1}{(1-v)^2}+\dots
\end{equation}
where we have omitted the terms which remains finite $v\to 1^-$. This expression allows to compute the scaling behaviour for the average height $h(n)$ by inversion of the transformation. After some algebra, we find for $n\gg 1$
\begin{eqnarray}
   \fl h(n) &=&\frac{\mathbf{C}_+}{\Gamma(\zeta+2)\,n\,\Gamma(n+1)}\Bigg\{\Gamma(2+\zeta+n)+\zeta\;\Gamma(1+\zeta+n)\nonumber\\
  \fl&&\qquad\qquad\qquad-\frac{\zeta(1-\zeta)(1+\zeta)}{(1+2\zeta)}\;\Gamma(\zeta+n)\Bigg\}-1-n^{-1}+\dots\nonumber\\
 \fl &=& 0.51334\, n^\zeta -1 +1.31498\, n^{\zeta-1}-n^{-1}+0.41413\, n^{\zeta-2} +O(n^{\zeta-3})\qquad
\label{eqn:hasymp}
\end{eqnarray}
Therefore at leading order we indeed find the scaling law $h(n)\sim c\,n^\zeta$ with $c =\mathbf{C}_+/\Gamma(\zeta+2)= 0.51334\dots$. Note that in principle, all amplitudes in (\ref{eqn:hasymp}) as well as subsequent corrections may be computed exactly in terms of hypergeometric functions and the gamma function. However, we shall omit these rather lengthy expressions and content ourselves with numerical values. This explicit solution will be useful to test numerical simulations and the domain of validity for the scaling ansatz (see section \ref{sec:numerics}).

%% file: chap4.tex
\section{Scaling behaviour and scaling functions}
\label{s-scaling}
\subsection{Scaling form for the $F$ function}
\label{ss-sfFf}
In this section we show that 
the average-height function $h(i,n)=\langle h_{\mathcal C}(i,n)\rangle$ takes the following scaling
form in the limit of long strands
\begin{equation}
\label{scalingH1} \langle h_{\mathcal C}(i,n)\rangle\mathop{=}_{n\to\infty}
n^\zeta\, \mathcal{H}_1(x)\quad,\qquad x={i\over n} \ .
\end{equation}
This is in fact a particular case of the general scaling form for
the moments of $h$
\begin{equation}
\label{scalingHk} \langle h_{\mathcal C}(i,n)^k\rangle\mathop{=}_{n\to\infty}
n^{k\zeta}\, \mathcal{H}_k(x)\quad,\qquad x={i\over n} \ .
\end{equation}
The partial differential equation (\ref{LPDE4F}) indicates that the generating function $F(u,v)$ has in
$\mathbb{R}^2$
singular lines at $u=1$ and at $v=1$. These singularities govern the long-strand limit
$n\to\infty$ with respectively $n-i=\mathcal{O}(1)$ and
$i=\mathcal{O}(1)$. The scaling limit $n\to\infty$,
$i/n=\mathcal{O}(1)$ is governed by the singularity at $u=v=1$.

\noindent To prove validity of the scaling (\ref{scalingH1}), it is sufficient to
show that in the limit $u,v\to 1$ the generating function $F(u,v)$
scales as
\begin{equation}
\label{scalingF} F(u,v)\mathop{=}_{u,v\to
1}\tau^{-2-\zeta}\mathcal{F}_1(\omega)
\end{equation}
with
\begin{equation}
\tau=1-{u+v\over 2}\
,\quad \sigma={v-u\over 2}\ ,\quad \omega= \sigma^2/\tau^2\ .
\end{equation}
In terms of the new variables $\sigma$ and $\tau$, the scaling limit is $\tau$ and $\sigma\to 0$,
$\omega=\sigma^2/\tau^2=\mathcal{O}(1)$ fixed. 
In this scaling limit the transformation (\ref{Fdef}) 
$\langle h_{\mathcal C}\rangle\to F$ 
becomes a double Laplace  transform. The corresponding   transformation $\mathcal{H}_1(x)\to\mathcal{F}_1(\omega)$ is
\begin{eqnarray}
\label{FfromH}
 \mathcal{F}_1(\omega)&\approx \int_{0}^{\infty} \mbox{d}n\, \int_{0}^{1}\mbox{d}x\, 
n^{\zeta +1} (1-\sqrt\omega  \tau -\tau)^{n x} (1+\sqrt\omega  \tau -\tau)^{n-n x} {\cal H}_{1}(x) \nonumber\\ 
&\approx\int_{0}^{\infty} \mbox{d}n\, \int_{0}^{1}\mbox{d}x\, 
n^{\zeta +1} e^{-(\sqrt\omega +1) \tau
   n x}\, e^{- (1-\sqrt\omega ) \tau 
   n(1- x)} {\cal H}_{1}(x) \nonumber\\ 
&=\Gamma(2+\zeta)\int_0^1
\mbox{d}x\,\mathcal{H}_1(x)\,\left[
1-\sqrt{\omega}+2x\sqrt{\omega}
\right]^{-(2+\zeta)}.
\end{eqnarray}
To obtain the equation for $\mathcal{F}_1(\omega)$, we keep the most singular terms in  (\ref{LPDE4F}) when $u,v\to 1$. This gives
\begin{eqnarray}
\label{LPDE4Fs} \fl\left[{1\over 2}\left({\partial^2\over\partial
u^2}+{\partial^2\over\partial v^2}\right) +{\partial^2\over\partial
u\partial v}-{1\over 1-u}{\partial\over\partial u}-{1\over
1-v}{\partial\over\partial v}-{2\over (1-u)(1-v)}\right]F(u,v)\simeq
0 \ .
\end{eqnarray}
Using ansatz (\ref{scalingF}), we obtain a hypergeometric
differential equation for $\mathcal{F}_1(\omega)$
\begin{equation}
\label{hgeq4F}
\omega(1-\omega)\mathcal{F}''_1(\omega)+\left[\frac{3+2 \zeta}2 -\frac{7+2 \zeta}2\omega\right]\mathcal{F}'_1(\omega)-\frac{(\zeta +4)}{2}  \mathcal{F}_1(\omega)=0\ .
\end{equation}
Thus the scaling function $\mathcal{F}_1(\omega)$ is a hypergeometric
function
\begin{equation}
\label{2F1-a} \mathcal{F}_1(\omega)\
=\ \mathbf{D} \  {_2F_1}(1+\zeta/2,(3+\zeta)/2,\zeta+3/2,\omega) \ .
\end{equation}
where $\mathbf{D}$ denotes some constant. We explicitly compute $\mathbf{D}$ 
 from the constants $\mathbf{C}_+$ and $\mathbf{C}_-$ obtained in (\ref{Kexpl}) and ({\ref{C1&2expl}) since
$K(v)=F(v,v)\simeq(1-v)^{-(2+\zeta)}\mathcal{F}_1(0)=\tau^{-(2+\zeta)}\mathcal{F}_1(0)$ as $v\to 1$.
One obtains
\begin{equation}
\label{D1expl} \mathbf{D}=\mathbf{C}_+= 0.713263 \cdots\quad\ .
\end{equation}
As we shall see in the next sub-section, the form (\ref{2F1-a}) for $\mathcal{F}_1(z)$ implies a very simple scaling function for the average height.

\subsection{Scaling form for the average height function $h(i,n)$}
\label{ss-sahf}

\noindent\textbf{Proposition:} The scaling limit $\mathcal{H}_1$ of the
average height distribution $h(i,n)$, as defined in
(\ref{scalingH1}), is given by a simple ``beta-law" with exponent $\zeta$
\begin{equation}
\label{H1expl} \mathcal{H}_1(x)=\mathbf{E}\,
x^\zeta(1-x)^\zeta 
\ , \qquad  
\zeta={\sqrt{17}-3\over 2}
\end{equation} 
and the amplitude
\begin{equation}
\label{E1D1expl}
\mathbf{E}
= {\Gamma(2+2\zeta)\over (1+\zeta)\Gamma(1+\zeta)^3}\,\mathbf{C_+} = 1.45717
\ldots
 \end{equation} where
$\mathbf{C_+}$ is given in (\ref{D1expl}). 

\noindent\textbf{Discussion and proof:} The fact that the average heigth $\mathcal{H}_1(x)$ scales  as $x^\zeta$ for small $x$ was already known by
M\"uller   \cite{MarkusMuellerPhD}. The simple exact form for $\mathcal{H}_1(x)$ is quite remarkable and unexpected. Our first hints for (\ref{H1expl}) came from the numerical
simulations that we describe in section \ref{sec:numerics}.

To prove  (\ref{H1expl})   it is
simpler to start from  (\ref{H1expl}) and to show that it implies
the form (\ref{2F1-a}) for $\mathcal{F}_1(z)$ (the transformation
$\mathcal{H}_1\to\mathcal{F}_1$ is linear and one-to-one). Inserting
(\ref{H1expl}) into the definition for $F(u,v)$ when $u,v\to 1$ we
have (this is equivalent to use (\ref{FfromH}))
\begin{eqnarray}
\label{lf60}
   \fl F(u,v)&\simeq \  \mathbf{E} \sum_{i,j} {i^\zeta\,j^\zeta\over (i+j)^\zeta}\,u^i\,v^j\
    \simeq
    \  \mathbf{E}\int_0^\infty \mbox{d}i\int_0^\infty \mbox{d}j\, {i^\zeta\ j^\zeta\over (i+j)^\zeta}\ e^{-i(1-u)}\,e^{-j(1-v)}\nonumber\\
    \fl&=\ {\sqrt{\pi}\mathbf{E}\over 2^{1+2\zeta}}{\Gamma(1+\zeta)\Gamma(2+\zeta)\over\Gamma(3/2+\zeta)}\,(1-v)^{-(2+\zeta)}\,{_2F_1}\left(1+\zeta,2+\zeta,2+2\zeta;{u-v\over 1-v}\right)
\end{eqnarray}
We now use the quadratic identity for hypergeometric functions \cite{AbramowitzStegun}
\begin{equation}
\label{quadrid}
   _2F_1(a,b,2b,z) = \left(1-\frac{z}{2}\right)^{-a}{_2}
   F_1\left(\frac{a}{2},\frac{1+a}{2},\frac{2b+1}{2},
   \left(\frac{z}{2-z}\right)^2\right)
\end{equation}
in the special case $a=2+\zeta$, $b=1+\zeta$. 
We obtain
\begin{eqnarray}
\label{lf61}
\fl F(u,v)= {\sqrt{\pi}\mathbf{E}\over
2^{1+2\zeta}}{\Gamma(1+\zeta)\Gamma(2+\zeta)\over\Gamma(3/2+\zeta)}\,&\left[1-{u+v\over
2}\right]^{-2-\zeta}\,\\ &\times{_2}F_1\left({2+\zeta\over 2},{3+\zeta\over
2},{3\over 2}+\zeta;\left[{v-u\over 2-u-v}\right]^2\right) \ .
\end{eqnarray}
Upon identification with (\ref{scalingF}) (and using the duplication formula for the $\Gamma$-function, \cite{AbramowitzStegun}) we recover the scaling
solution (\ref{2F1-a}) for $\mathcal{F}_1$. Q.E.D.

\subsection{Scaling for higher moments of the height function}
\label{ss-scalhk}
In this section, we study the higher moments of the local height $h_{\mathcal C}(i,n)$ at site $i$ for a strand of length $n$, $\langle h_{\mathcal C}(i,n)^k\rangle$. Once again, the starting point is a generating function
\begin{equation}
G_k(u,v)=\sum_{n=0}^\infty\sum_{i=0}^n u^i v^{n-i}\,\langle h_{\mathcal C}(i,n)^k\rangle\ =\ \left.{\partial^k\over\partial z^k}G(u,v;z)\right|_{z=0}.
\end{equation}
$G(u,v;z)$ denotes the solution of (\ref{PDE4H}). From this equation, we are able to recursively determine a generating function $G_k$ by a partial differential equation involving functions $G_{k'}$, $k'<k$.
For example, $G_2$ is a solution of the linear equations
\begin{eqnarray}
\label{PDE4H2}
    \fl
    \left[ {1\over 2}\left(u^2{\partial^2\over\partial
u^2}{+}v^2{\partial^2\over\partial
v^2}\right){+}uv{\partial^2\over\partial u\partial v}
-
{u^2\over
(1{-}u)}\left(u{\partial\over\partial u}{+}1\right) 
{-}{v^2\over
(1{-}v)}\left(v{\partial\over\partial v}+1\right)\right] G_2(u,v)\nonumber\\
\fl =
uv\left({1\over (1{-}u)^2(1{-}v)^2}{+}{4\over (1{-}u)(1{-}v)}F(u,v){+}2F(u,v)^2{+}{2\over (1{-}u)(1{-}v)}G_2(u,v)\right)
\end{eqnarray}
where the right-hand side involves the $k=0$ and $k=1$ moments.
$F(u,v)=G_1(u,v)$ is the generating function for the average height $\langle h_{\mathcal C}(i,n)\rangle$ studied previously.

\subsubsection{Averaged k-moments}
\label{sss-akm}
Let us first consider the generating function for the ``integral'' of the averaged $k$-moment
\begin{equation}
K(t;z)=G(t,t;z)=\sum_{k=0}^\infty {z^k\over k!} K_k(t)
\end{equation}
with
\begin{equation}
K_k(t)=\sum_{n=0}^\infty t^n \sum_{i =0}^n\langle {h_{\mathcal{C}}(i,n)^k}\rangle
\end{equation} 
We already know $K_0(t)=(1-t)^{-2}$ and $K_1(t)$ from (\ref{Kexpl}).
$K(t;z)$ satisfies the non-linear differential equation
\begin{equation}
\left[{1\over 2}{\partial^2\over\partial t^2}-{t\over 1-t}{\partial\over\partial t}-{2\over 1-t}\right]K(t,z)=\mathrm{e}^z K^2(t;z)
\label{PDE4K}
\end{equation}
The scaling limit corresponds to $t\to 1^-$. In this limit, we expect the functions $K_k(t)$ to scale as
\begin{equation}
\label{Kktscal}
K_k(t)\ \simeq\  b_k (1-t)^{-2-\zeta_k}\ ,\quad \zeta_k=k\,\zeta
\end{equation}
This implies that the average of the $k$-moment of the local height scales with the length of the strand $n$ as
\begin{equation}
{1\over n}\sum_{i=0}^n \langle h_{\mathcal C}(i,n)^k\rangle
\ \simeq\ {a_k}\,n^{\zeta k}\ ,\quad a_k={b_k\over\Gamma(2+k\zeta)}
\end{equation}
The coefficients $a_k$ can be computed recursively from the first non-trivial one $a_1\simeq 0.51334$, that we computed previously, see (\ref{eqn:hasymp}).
Indeed from (\ref{Kktscal}) it follows that $K(t,z)$ takes the scaling form
\begin{equation}
K(t,z)\ \mathop{=}_{t\to 1^-}\ {1\over (1-t)^2}\mathcal{K}(u) ={1\over (1-t)^2}
\sum_{k=0}^\infty {b_k\over k!}\,u^k\ ,
\end{equation}
with the scaling variable $u=z(1-t)^{-\zeta}$.
Corrections are of order $(1-t)^{1-\zeta}$; considering $\rme^{z} K(t,z)$, they would be of order $(1-t)$.

Using (\ref{PDE4K}) and inserting the scaling function $\mathcal{K}(u)$, we obtain up to terms of order $(1-t)$ the equation
\begin{equation}
\mathcal{K}(u)+u \mathcal{K}'(u)+{1\over 2}\zeta^2u^2 \mathcal{K}''(u)= \mathcal{K}(u)^2\ .
\end{equation}
We obtain
\begin{equation}
b_2=b_1^2\,{5+\sqrt{17}\over 6}\ ,\quad b_3= b_1^3\,\frac{92 + 22\,{\sqrt{17}}}{59}
\ ,\qquad\cdots
\end{equation}
Numerically, we find 
\begin{eqnarray}
\fl{\cal K}(u) &= 1 + 0.713243 u + 0.386756 u^2 + 0.18727 u^3 
+ 0.0851827 u^4 \nonumber\\
&+ 
 0.0372364 u^5 + 0.015835 u^6
 + 0.00659914 u^7 + 0.00270789 u^8 + \ldots
\end{eqnarray}
As an application, let us evaluate the average height fluctuations by considering the quantity
\begin{eqnarray}
\fl  \Delta_2 &= \frac{1}{n}\sum_{k=1}^n\left(\langle h_{\mathcal C}(k,n)^2\rangle -\langle h_{\mathcal C}(k,n)\rangle^2 \right)
  \approx \left(a_2-\frac{\mathbf{E}^2\Gamma(2\zeta+1)}{\Gamma(4\zeta+2)}\right)n^{2\zeta}\approx 0.055658\,n^{2\zeta}
  \label{eqn:delta2}
\end{eqnarray}
We thus conclude that the fluctuations of the height function remain large in the scaling limit (see section \ref{sec:numerics}).

\subsubsection{General scaling function}
\label{sss-gsf}
We now consider the general scaling limit of the generating function $G(u,v;z)$ for the moments of $h_{\mathcal C}(i,n)$. The correct ansatz is
\begin{equation}
\label{scaling4G}
G(u,v;z)\mathop{=}_{u,v\to 1} \tau^{-2}\mathcal{G}(\tilde z,\omega)
\end{equation}
with 
\begin{equation}
\ \tilde z= z\tau^{-\zeta},\quad\tau=1-{u+v\over 2},\quad \omega=\left({v-u\over 2-u-v}\right)^2
\end{equation}
In the scaling limit we obtain for $\mathcal{G}(\tilde z,\omega)$ the equation
\begin{eqnarray}
\label{scalingEqG}
 \fl   \left[2 \omega^2{\partial^2\over \partial \omega^2} {+}{\zeta^2\over 2}  \tilde z^2{\partial^2\over\partial \tilde z^2}{+}2\zeta \omega{\partial^2\over\partial \omega\partial  \tilde z}{+}{3{-}7\omega\over 1{-}\omega}\omega{\partial\over\partial \omega}
    {+}{1{-}(1{+}\zeta)\omega\over 1{-}\omega}  \tilde z {\partial\over\partial  \tilde z}+{1-3\omega\over 1-\omega}\right]\mathcal{G}(  \tilde z,\omega)  \nonumber\\
     =  \mathcal{G}(  \tilde z,\omega)^2
\end{eqnarray}
Expanding in $\tilde z$ we find the scaling limit for the generating functions of the moments of $h_{\mathcal C}(i,n)$ via a Taylor expansion
\begin{equation}
\label{scGexp}
\mathcal{G}(\tilde z,\omega)=\sum_{k=0}^\infty {\tilde z^k\over k!}\mathcal{G}_k(\omega)
\end{equation}
At order $k=0$ and $k=1$ we recover our previous results (\ref{2F1-a})
\begin{equation}
\label{scG0G1}
\mathcal{G}_0(\omega)={1\over 1-\omega}
\ ,\quad
\mathcal{G}_1(\omega)=\mathcal{F}_1(\omega)
\end{equation}
and for $k\ge 2$ recursive second-order linear differential equations for the $\mathcal{G}_k(\omega)$ with coefficients and second members depending on the previous $\mathcal{G}_{k'}(\omega)$ ($k'<k$). From $\mathcal{G}_{k}(\omega)$ we obtain the scaling form for the moments
\begin{equation}
\label{scalHk2}
\langle h_{\mathcal C}(i,n)^k\rangle\mathop{\sim}_{n\to\infty} n^{k\zeta}\,\mathcal{H}_k(x)\ ,\quad x={i\over n}
\end{equation}
The scaling function $\mathcal{H}_k(x)$ is related to $\mathcal{G}_k(\omega)$ by the integral transformation
\begin{equation}
\label{GkfromHk}
 \mathcal{G}_k(\omega)=\Gamma(2+k\zeta)\int_0^1
dx\,\mathcal{H}_k(x)\,\left[
1-\sqrt{\omega}+2x\sqrt{\omega}
\right]^{-(2+k\zeta)} \ .
\end{equation}
which generalises (\ref{FfromH}).

\subsection{Simple scaling or multifractality?}
\label{ss-mfrac}
Studying the 
roughness properties of the height function in the scaling limit, we naturally are led to the question of multifractality.
We shall argue that within our model the height-profile statistics is solely governed by the scaling exponent $\zeta$. This excludes strong fluctuations which might lead to multifractal scaling.
To this end, let us consider the moments of the local height variations
\begin{equation}
\Delta h_{k}=\langle \left|h_{\mathcal C}(i,n)-h_{\mathcal C}(j,n)\right|^{k}\rangle
\end{equation}
In the scaling limit $n\to \infty$, we expect a relation of type
\begin{equation}
\langle \left|h_{\mathcal C}(i,n)-h_{\mathcal C}(j,n)\right|^{k}\rangle \ \propto\ | i-j|^{\zeta_{k}}
\end{equation}
in the regime $1\ll |i-j] \ll  n$.
If $\zeta_{k}=k\,\zeta$ there is simple scaling, whereas $\zeta_{k}>k\zeta$ (at least for large enough $k$) implies multifractal behaviour.

We now argue that we are in the first case and there is no evidence for multifractality.
Indeed, it is easy to show (using the height picture, and using translation invariance to move the point $i$ to the origin of the strand) that the following general inequality holds
\begin{equation}
\langle \left|h_{\mathcal C}(i,n)-h_{\mathcal C}(j,n)\right|^{k}\rangle\ \le \ \langle \left|h_{\mathcal C}(\ell,n)\right|^{k}\rangle
\ ,\quad \ell =|j-i|
\end{equation}
We know from (\ref{scalHk2}) that for {$\ell \ll n$} this scales as
\begin{equation}
\langle \left|h_{\mathcal C}(\ell,n)\right|^{k}\rangle\ \propto\ n^{k\zeta}\;\mathcal{H}_k(\ell/n)
\end{equation}
In the limit $n\to\infty$, $\ell$ finite, $\langle \left|h_{\mathcal C}(\ell,n)\right|^{k}\rangle$ remains finite, since it is bounded by $|\ell|^k$. This implies that $\mathcal{H}_k(x)$ should behave for small $x$ as
\begin{equation}
\mathcal{H}_k(x)\ \mathop{\simeq}_{x\to 0} \ x^{k\zeta}\ .
\end{equation}
This can be shown more rigorously using (\ref{scalingEqG}) for the generating function $\mathcal{G}(\tilde z,\omega)$ of the $\mathcal{G}_k$ and the integral relation 
(\ref{GkfromHk}) between the  $\mathcal{G}_k$ and the  $\mathcal{H}_k$.
The small-$x$ behavior of $\mathcal{H}_k(x)$ is related to the $\omega\to 1$ behavior of $\mathcal{G}_k(\omega)$.
One can check from (\ref{scalingEqG}) that the function $\mathcal{G}(\tilde z,\omega)$ must behave when $\omega\to 1$ as
\begin{equation}
\mathcal{G}(\tilde z,\omega)\ \mathop{\sim}_{\omega\to 1}\ {\Omega(\tilde z)\over 1-\omega}+\mathcal{O}(\log(1-\omega))
\end{equation}
Using (\ref{GkfromHk}), this implies that
\begin{equation}
\mathcal{G}_k(\omega)\ \mathop{\sim}_{\omega\to 1}\ {\Omega_k\over 1-\omega}
\quad\Rightarrow\quad \mathcal{H}_k(x)\ \mathop{\simeq}_{x\to 0}\ x^{k\zeta}
\end{equation}
We conclude that
\begin{equation}
\langle \left|h_{\mathcal C}(i,n)-h_{\mathcal C}(j,n)\right|^{k}\rangle\ \le\ \mbox{const.}\, |i-j|^{k\zeta},\quad\mbox{for}\ 1\ll\ell\ll n.
\end{equation}
what implies that $\zeta_k\le k\zeta$. However, we know that $\zeta_k\ge k\zeta$ from general correlation inequalities. Hence it follows
\begin{equation}
\zeta_k=k\,\zeta
\end{equation}
what proves the abscence of multifractal behaviour, at least for moments of  $|h_{\mathcal C}(i,n)-h_{\mathcal C}(j,n)|$.

\subsection{Corrections to scaling}
\label{ss-corsca}
We can study the corrections to scaling for the height function $\langle h_{\mathcal C}(i,n)\rangle$.
Let us come back to equation (\ref{LPDE4F}) for the generating function $F(u,v)$ defined by (\ref{Fdef}).
A particular solution of (\ref{LPDE4F}) is
\begin{equation}
F_0(u,v)=-{1\over (1-u)(1-v)}
\end{equation}
Thus the general solution of (\ref{LPDE4F}) is of the form
\begin{equation}
\label{GenSolF}
F(u,v)=F_0(u,v)+\mathbf{C_+}F_+(u,v)+\mathbf{C_-}F_-(u,v)
\end{equation}
where $F_+(u,v)$ and $F_-(u,v)$ are two linearly independent solutions of the linear equation with no r.h.s.
\begin{eqnarray}
&\left[{u^2\over 2}{\partial^2\over\partial
u^2}+{v^2\over 2}{\partial^2\over\partial
v^2}+uv{\partial^2\over\partial u\partial
v}
- {u^2\over(1-u)}\left(u{\partial\over\partial u}+1\right)
\right.\nonumber \\&\left.
-{v^2\over(1-v)}\left(v{\partial\over\partial v}+1\right) 
-{2uv\over (1-u)(1-v)} \right] F(u,v)
=0 
\label{PDE4Fpm}
\end{eqnarray}
It is possible to go to the scaling variable $\tau$ and $\omega$ used in equations~(\ref{scalingF}) and (\ref{scaling4G})
\begin{equation}
u=1-\tau(1+y)\ ,\quad v=1-\tau(1-y)\ ,\quad \omega=y^2
\end{equation}
and to take for $F_+$ and $F_-$ the solutions which can be written respectively as
\begin{equation}
\fl
F_+(u,v)=\tau^{-2-\zeta_+}\tilde{\mathcal{F}}_+(\tau,\omega)
\ ,\quad
F_-(u,v)=\tau^{-2-\zeta_-}\tilde{\mathcal{F}}_-(\tau,\omega)
\ ,\quad
\zeta_\pm={\pm \sqrt{17}- 3\over 2}
\end{equation}
($\zeta_+=\zeta$ is the roughness exponent), such that $\tilde{\mathcal{F}}_+(\tau,\omega)$ and
$\tilde{\mathcal{F}}_-(\tau,\omega)$ have an asymptotic expansion in powers of $\tau$ in the scaling limit $\tau\to 0$, and are regular in the domain $\omega\in[0,1[$.
Indeed (\ref{PDE4Fpm}) becomes for $\tilde{\mathcal{F}}_\pm$ the linear equations
\begin{eqnarray}
\label{PDE4tF}
\fl	\left[ {-}2{\left(\tau{-}1 \right) }^2\tau\zeta_\pm {+} 
     		2\tau^3\omega^2\left( 2 {+} \zeta_\pm \right)  {-} 
    		 2 \left( \tau{-}1 \right) \omega
      		\left( 4 {+} 2\tau^2 {+} \zeta_\pm {+} 
        		2\tau\left( 1 {+} \zeta_\pm \right)  \right) 
	\right]  \tilde{\mathcal{F}}_\pm (\tau,\omega)
\nonumber \\
\fl{+} 2\left[ {-}3 {+} 2\tau^3{\left( \omega{-}1 \right) }^2 {-} 
     		2\zeta_\pm {-} 
    		 2\tau\left( \omega {-}1 \right) \zeta_\pm {+} \omega\left( 7 {+} 2\zeta_\pm \right)  
	\right] 
   \omega{\partial\over\partial \omega} \tilde{\mathcal{F}}_\pm( \tau ,\omega) 
\nonumber   \\
 \fl  {+} 
   	\left[ {-}2 \tau ^3\omega^2 {-}  2\left(   \tau {-}1  \right) \omega
     		 \left( {-}2 {+}  \tau \left( \zeta_\pm {-} 1 \right) {-} \zeta_\pm
		 \right) 
	    {+}   2{\left(  \tau {-}1 \right) }^2 \left( 1 {+}  \tau  {+} \zeta_\pm \right) 
	 \right] 
    t{\partial\over\partial t} \tilde{\mathcal{F}}_\pm( \tau ,\omega)
\nonumber   \\ 
\fl      {+} \left( \omega {-} 1 \right)  4\omega^2
    {\partial^2\over\partial \omega^2} \tilde{\mathcal{F}}_\pm( \tau ,\omega) 
{+}4 \left( \tau {-} 1 \right)  \tau \omega {\partial^2\over\partial  \tau \partial \omega}  \tilde{\mathcal{F}}_\pm( \tau ,\omega) {+} \left(   \tau {-}1 \right)^2  \tau^2   {\partial^2\over\partial  \tau ^2}  \tilde{\mathcal{F}}_\pm( \tau ,\omega) =0
\end{eqnarray}

Note that in the scaling variables the particular solution reads
\begin{equation}
F_0(u,v)=-\tau^{-2} (1-\omega)^{-1}\ .
\end{equation}
Although not simple, (\ref{PDE4tF}) implies that its solutions can be expanded in powers of $\tau$. Indeed, let us
expand in $\tau$ the functions $ \tilde{\mathcal{F}}_\pm( \tau ,\omega)$
\begin{equation}
\tilde{\mathcal{F}}_\pm( \tau ,\omega)=\sum_{k=0}^\infty {\tau^k\over k!}\tilde{\mathcal{F}}_\pm^{(k)}(\omega)
\end{equation}
Setting $\tau=0$ in (\ref{PDE4tF}) fixes the equation for the dominant term to
\begin{eqnarray}
2\omega(\zeta_\pm{+}4)\tilde{\mathcal{F}}_\pm^{(0)} (\omega)
{+}
2\left((7{+}2\zeta_\pm)\omega-(3{+}2\zeta_\pm)\right)\omega\,{\partial\over\partial \omega} \tilde{\mathcal{F}}_\pm^{(0)}(\omega)\nonumber\\
{+}
\left( \omega{-} 1 \right) \, 4\,\omega^2\,
    {\partial^2\over\partial \omega^2} \tilde{\mathcal{F}}_\pm^{(0)}(\omega)=0
\end{eqnarray}
This is nothing but the hypergeometric differential equation (\ref{hgeq4F}) for the scaling function $\mathcal{F}(\omega)$ obtained previously. Its solution is thus
\begin{equation}
\tilde{\mathcal{F}}_\pm^{(0)} (\omega)={_2F_1}(1+\zeta_\pm/2,(3+\zeta_\pm)/2,\zeta_\pm+3/2;\omega)
\end{equation}
The expansion in $\tau$ gives a hierarchy of hypergeometric-like differential equations for the correction-to-scaling functions $\tilde{\mathcal{F}}_\pm^{(k)}(\omega)$
with a non-zero r.h.s. involving the previous scaling functions $\tilde{\mathcal{F}}_\pm^{(k')}(\omega)$, $0\le k'<k$. It is easy to check that these equations admit a unique solution $\tilde{\mathcal{F}}_\pm^{(k)}(\omega)$ which is analytic at $\omega=0$ and regular in the domain $\omega\in [0,1[$ (with a singularity at $\omega=1$).

From this analysis, the coefficients $\mathbf{C}_+$ and $\mathbf{C}_-$ in the full scaling expansion
(\ref{GenSolF}) are those already calculated in sect.~\ref{ss-zeta&h}, (\ref{C1&2expl}):
\begin{equation}
\mathbf{C_+}=0.713263\ldots \quad \mbox{and} \quad \mathbf{C_-}=0.519299\ldots
\end{equation} 
The important result is that, using the inverse transformation $F(u,v)\to\langle h_{\mathcal C}(i,n)\rangle$, the average height function takes the general form
\begin{equation}
\label{HScaCor}
\left<h_{\mathcal C}(i,n)\right> = n^{\zeta_+}\left(\sum_{k=0}^\infty n^{-k}\, \mathcal{H}_+^{(k)}(x)\right) + n^{\zeta_-}\left(\sum_{k=0}^\infty n^{-k}\, \mathcal{H}_-^{(k)}(x)\right) - 1
\end{equation}
where the dominant term is the scaling function obtained in (\ref{H1expl})
\begin{equation}
\label{H+H1}
\mathcal{H}_+^{(0)}(x)=\mathcal{H}_1(x)=\mathbf{E}\,x^\zeta(1-x)^{\zeta}
\end{equation}
the leading subdominant term is  the last term  $-1$ in (\ref{HScaCor}). In fact, since we know that the leading order scales like $n^\zeta$ and therefore only grows relatively slowly with $n$, the correction $-1$ turns out to be important, even at $n = \mathcal{O}(10^3)$, a case that we shall consider below. The subleading corrections $\mathcal{H}_\pm^{(k)}(x)$ are distributions on $[0,1]$ and may in principle be computed from the functions $\tilde{\mathcal{F}}_\pm^{(k)} (\omega)$ via inverse Laplace transforms.

%% file: chap5.tex
\section{Pairing probabilities}
\label{sec:pprobs}
Up to now we focused on the height function and its scaling laws. However, for the original RNA problem pairing probabilities constitute more natural objects. In this section, we compute the single-base pairing probability as well as the scaling limit for the pairing probability $P(i,j)$.

\subsection{Single-base pairing probability}
As mentioned above, $h(1,n)$ is the probability that a given base is involved in a pair. We obtain the generating function $g(v)$ of $h(1,n)$ from $F(u,v)$, introduced in (\ref{Fdef}), through differentiation
\begin{equation}
\label{lf54} 
g(v):=v \left.\frac{\partial F(u,v)}{\partial u}\right|_{u=0} 
=\sum_{n=0}^\infty h(1,n)v^n\
\end{equation}
According to (\ref{LPDE4F}), it is solution of the ordinary differential equation
\begin{equation}
(1-v)g''(v) - 2vg'(v) = \frac{2}{1-v}, \qquad g(0)=g'(0)=0 .
\end{equation}
The initial conditions are due to the fact that $h(1,0)=h(1,1)=0$. For our purposes, it is sufficient to solve for $g'(v)$ and compare to the derivative of the series expansion (\ref{lf54}):
\begin{equation}
  g'(v) = \frac{1-e^{-2v}}{(1-v)^2} =\sum_{n=1}^\infty n h(1,n) v^{n-1}
\end{equation}
Comparison of the series development on both sides leads us to the explicit expression
\begin{equation}
h(1,n) = -\sum_{k=0}^{n-1}
   \frac{(-2)^{k+1}}{(k+1)!}+\frac{1}{n}\sum_{k=0}^{n-1}
   \frac{(-2)^{k+1}}{k!}
   \label{eqn:pprob}
\end{equation}
In the limit of large strands $n\to \infty$, the series converges and we obtain
\begin{equation}
  \lim_{n\to \infty} h(1,n) = 1-{1 \over e^2}
  \label{eqn:h1asymp}
\end{equation}
For large $n\gg 1$, the corrections to this result can be determined as follows:
\begin{equation}
 h(1,n) = 1-{1 \over e^2} -\frac{2}{n e^{2}} + r(n)\ , \qquad |r(n)| \le \frac{2^{n}}{n!\, \ln n}\ ,
\end{equation}
where the bound is obtained by approximating the remaining terms in the sum by an integral. 
We shall reconsider this probability later when comparing the arch deposition model to a tree-growth model.

\subsection{Scaling law for the pairing probability $P(i,j)$}
\label{ss-slP}
In this section we compute the scaling function $\mathcal P$ as defined in (\ref{eqn:scaling}). First of all, $P(i,j)$ is indeed only a function of the distance -- despite the fact that we have singled out an origin. Therefore we expect for large $n$ the scaling form (\ref{eqn:scaling}), 
\begin{equation}
P(i,j) \mathop{\sim}_{n\to\infty} n^{-\rho}\;\mathcal{P}\left(\frac{|i-j|}{n}\right)
\end{equation}
With this ansatz,  relation (\ref{eqn:height}) in terms of the scaling functions for the height field and the pairing probability turns into the integral equation
\begin{equation}
 n^{\zeta} \mathcal{H}_1(z) = n^{2-\rho}\int_0^{z-\epsilon}\mathrm{d} s\int_{z+\epsilon}^1\mathrm{d} t\,\mathcal P (t-s)
\end{equation}
This identifies 
\begin{equation}
\rho=2-\zeta\ .
\end{equation}
The dimensionless scaling functions for height and pairing probability are then related by: 
\begin{equation}
  \mathcal{H}_1(z) = \int_0^{z-\epsilon}\mathrm{d} s\int_{z+\epsilon}^1\mathrm{d} t\,\mathcal P (t-s)
\end{equation}
Note that we have introduced a small ultraviolet cutoff $\epsilon$ in order to circumvent possible singularities as $t-s\to 0$. Upon differentiating twice, we find the harmless expression
\begin{equation}
  \mathcal{H}''_1(z) = -(\mathcal P(1-z+\epsilon)+\mathcal{P}(z+\epsilon))
  \mathop{=}_{\epsilon\to 0}-(\mathcal P(1-z)+\mathcal{P}(z))
\end{equation}
It is clear that the limit $\epsilon\to 0$ will not lead to any problems since potentially divergent terms have canceled out.
Since  $\mathcal{P}(z)$ is  invariant with respect to the transformation $z\to 1-z$, it is possible to deduce its exact form
\begin{equation}
  \mathcal{P}(z)=-\mathcal{H}''_1(z)/2 = \mathbf{E}\,\frac{\zeta}{2}\;
  z^{\zeta-2}(1-z)^{\zeta-2}\left[1-\zeta+ 2(2 \zeta -1) z (1-z)\right]\ .
\end{equation}
We see that $\mathcal{P}(z)$ factorises into a beta law with characteristic exponent $\rho = 2-\zeta$ and a polynomial correction. Note that a pure beta law is obtained if and only if $\zeta=1/2$; this corresponds to the RNA homopolymer roughness exponent. Since the amplitude $\mathbf{E}$ is known from (\ref{E1D1expl}), we have entirely characterised the scaling law.

%% file: chap6.tex
\section{Numerical simulations}
\label{sec:numerics}
Numerical simulations not only provide a verification of our analytical results in the scaling limit $n\to\infty$ but are useful in order to quantify finite-size corrections. In this section, we present a simple algorithm for random generation of hierarchical arch structures. Furthermore, we compare the statistics obtained from random sampling to the exact solutions.

\subsection{Outline of the algorithm}
We give a description of the algorithm which we have used to generate hierarchical structures $\mathcal C$: Information about arches is stored in the ``adjacency matrix'' $\Phi_{\mathcal C}(i,j)$. In order to take into account the planarity condition we label each basis $i=1,\dots,n$ with a ``colour'' $c(i)\in \mathbb{Z}$. During the construction process, two bases $i,j$ may be linked by an arch if and only if $c(i)=c(j)$. Furthermore, we introduce two special colours: if a structure $\mathcal{C}$ contains an arch $(i,j)$ we colour its endpoints with $c(i)=1$ and $c(j)=-1$ (what turns out to be convenient).

\begin{figure}[htpb]
  \centering
  \includegraphics[width=0.9\textwidth]{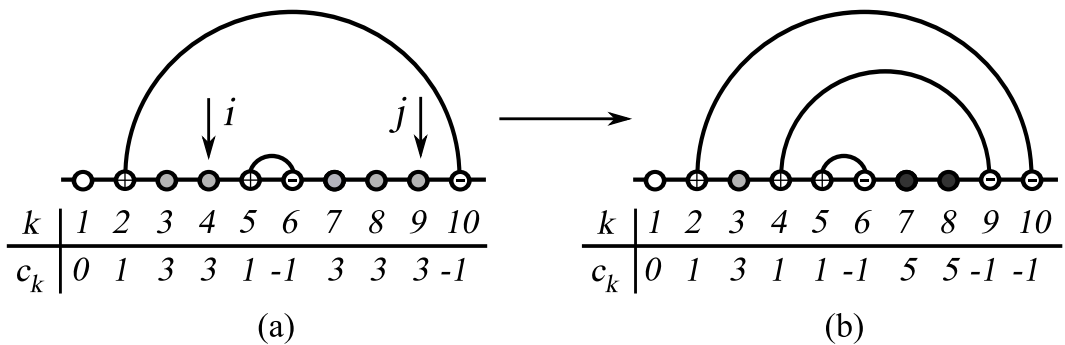}
  \caption{A single step of the construction of hierarchical structures: (a) random choice of the point $i$. Point $j$ is chosen amongst all points of the same colour as $i$ (light gray). (b) Once the arch $(i,j)$ is determined all the points $i<k<j$ are re-coloured in order to mark the new substructure (dark gray).}
  \label{fig:alg}
\end{figure}

The deposition of arches is carried out in the following way: Initially all colours are set to $c(k)=0,\, k=1,\dots,n$ and all entries of the adjacency matrix to $\Phi_{\mathcal C}(i,j)=0,\, i,j =1,\dots,n$.
First, we randomly choose a base $i$ among all unpaired bases. Next we collect all bases $k$ which may be paired to $i$ 
without violation of the planarity constraints, i.e.\ with the same colours $c(k) = c(i),\, k\neq i$, in an ordered list $\ell$. If $\ell$ is empty, the point may be removed from the set of unpaired points and the procedure restarted.
From the list $\ell$ of compatible bases we randomly choose a second base $j$.
For simplicity, let us suppose that $i<j$ 
( the converse case is similar). We store information about the so-created arch $(i,j)$ by setting $\Phi_{\mathcal{C}}(i,j)=1$.
Moreover we label the starting point and the endpoint of the arch with colours $c(i)=1$ and $c(j)=-1$. Finally, in order to mark the new substructure due to the insertion of this arch, we set $c(k) = i+1$ for all $i < k < j$. We repeat the procedure until no more points can be paired without violation of planarity. See figure\ \ref{fig:alg} for illustration of a single cycle.

Once this procedure is finished, the matrix $\Phi_{\mathcal{C}}(i,j)$ contains all information about the structure. To compute the height field, the colours $c(i)$ may be used as well: if we set $c(i)=0$ for all $i$ such that $c(i)>1$, then we obtain a sequence $\{c(i)\}_{i=1}^n$ with entries $0,\pm 1$. It precisely corresponds to the discrete derivative of the height function $c(i) = h_{\mathcal C}(i,n)-h_{\mathcal C}(i-1,n)$. Therefore, we can reconstruct
\begin{equation}
  h_{\mathcal C}(i,n)=\sum_{k=1}^i c(k)
\end{equation}
For a given strand of length $n$, we perform this construction $N$ times in order to  average over the samples. This algorithm is a variant of the point process to be discussed in section \ref{sec:growth}. Though not being dynamically equivalent to the arch deposition model, the key feature of partitioning into independent sub-structures leads to the same final probability law in configuration space. 

\subsection{Results}
We have constructed structures with up to $n=6500$ bases in order to test our theoretical predictions. For $n\leq 200$ bases, we have sampled $10^6$ structures whereas for $n>200$ bases, $10^5$ structures per data point were sampled. 

\begin{figure}[htpb]
\centering
\begin{tabular}{cc}
\includegraphics[width=0.47\textwidth]{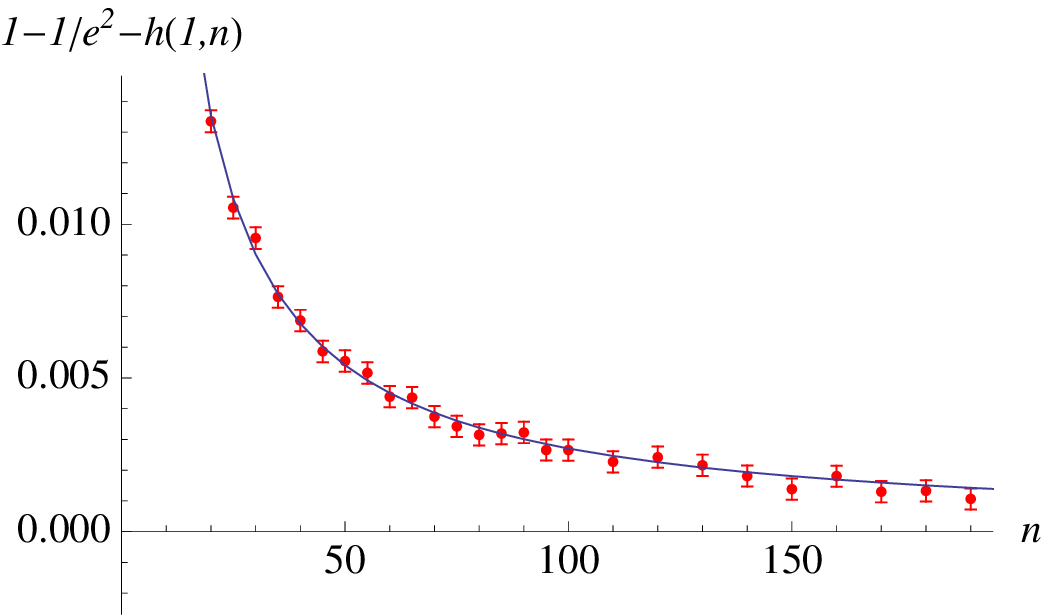}&
\includegraphics[width=0.47\textwidth]{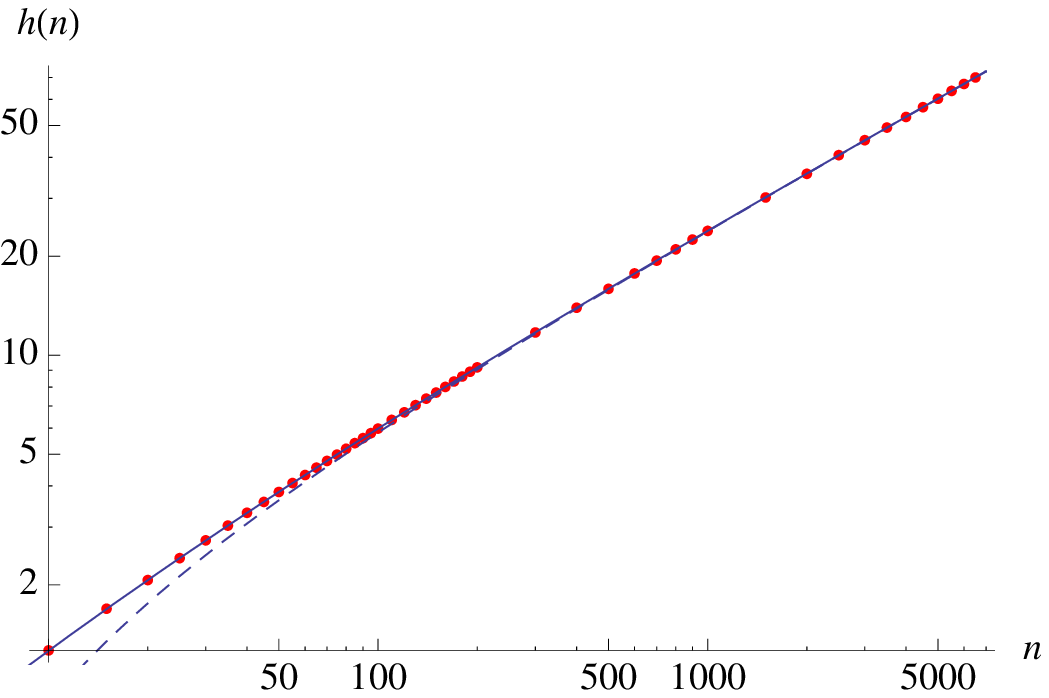}\\
(a) & (b)
\end{tabular}
\caption{(a) Deviation of the single base pairing probability $h(1,n)$ from $1-e^{-2}$ as a function of $n$. The dashed line is the theoretical prediction (\ref{eqn:pprob}) (b) Log-log-plot of the average mean height $h(n)$ as a function of the number of bases for $5\leq n\leq 6500$. The straight line corresponds to the theoretical prediction from (\ref{eqn:hasymp}), the dashed line indicates the scaling limit. (The error bar is of point size.)}
\label{fig:ppdata}
\end{figure}

\paragraph{Single-base pairing probability $h(1,n)$:} Results for the probability that a base is paired, which equals the height $h(1,n)$, is presented in figure\ \ref{fig:ppdata}a. We find agreement with the theoretical prediction from (\ref{eqn:pprob}) within errorbars.

\medskip

\paragraph{Averaged mean height $h(n)$, and its fluctuations:} In figure\ \ref{fig:ppdata}b we compare results for the averaged mean height $h(n)$ to the theoretical prediction in the limit $n\to \infty$. 
\begin{figure}[htpb]
\centering
\begin{tabular}{cc}
\includegraphics[width=0.47\textwidth]{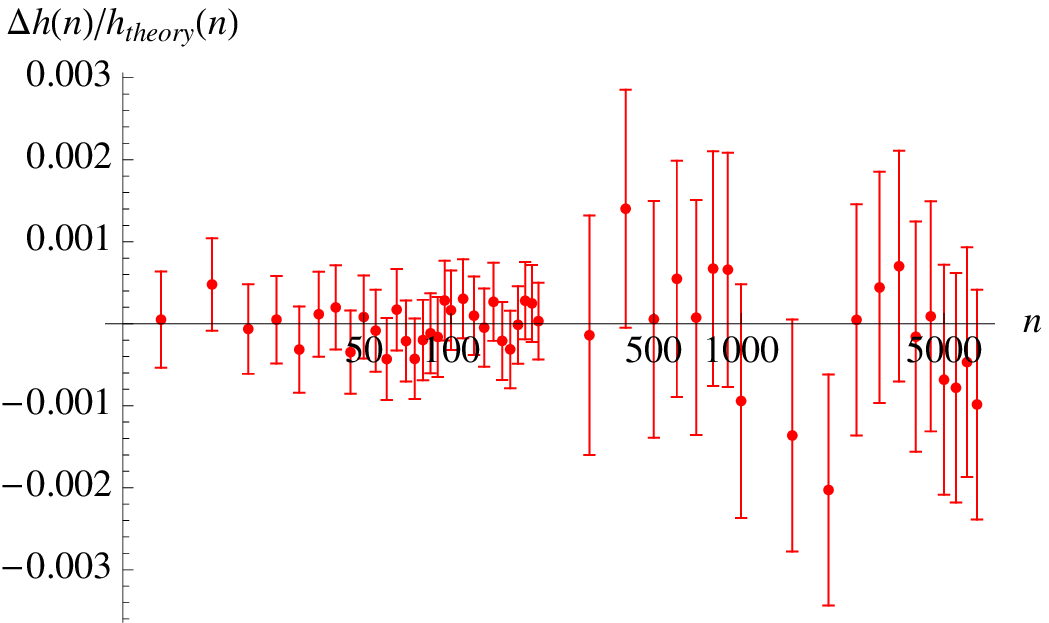}&
\includegraphics[width=0.47\textwidth]{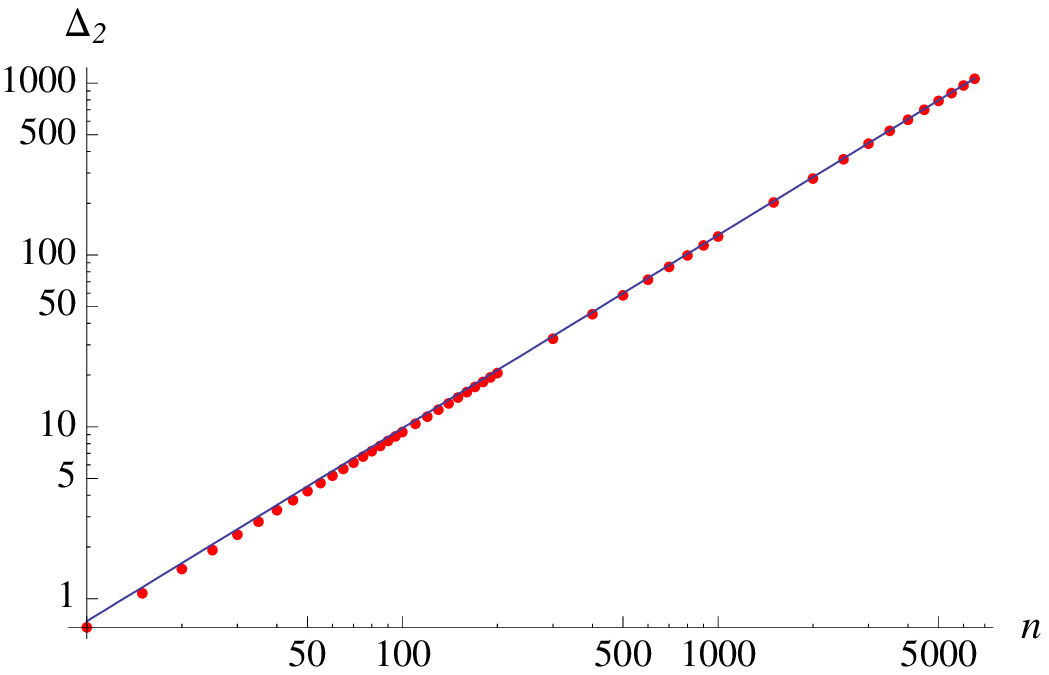}\\
(a) & (b)
\end{tabular}
\caption{(a) Deviations of the averaged mean height $h(n)$ from the theoretical prediction, normalised by the result of (\ref{eqn:hasymp}). Clearly, the deviations are of smaller than the errorbars. (b) Log-log-plot of the averaged second moment $\Delta_2$. (The error bar is of point size.) The straight line presents the scaling limit.}
\label{fig:mhdata}
\end{figure}
Taking into account all terms of (\ref{eqn:hasymp}) is sufficient to show that the difference between numerical results and theory is of the order of the statistical error, see figure \ref{fig:mhdata}a.

In order to compare our data to the results of section \ref{sss-akm} on averaged $k$-moments, we evaluate the height fluctuations via $\Delta_2 =\sum_{k=1}^n(\langle h_{\mathcal C}(k,n)^2\rangle-\langle h_{\mathcal C}(k,n)\rangle^2)/n$ in figure\ \ref{fig:mhdata}b. The data shows good agreement with the prediction from (\ref{eqn:delta2}).

\begin{figure}[htpb]
\centering
\begin{tabular}{cc}
\includegraphics[width=0.47\textwidth]{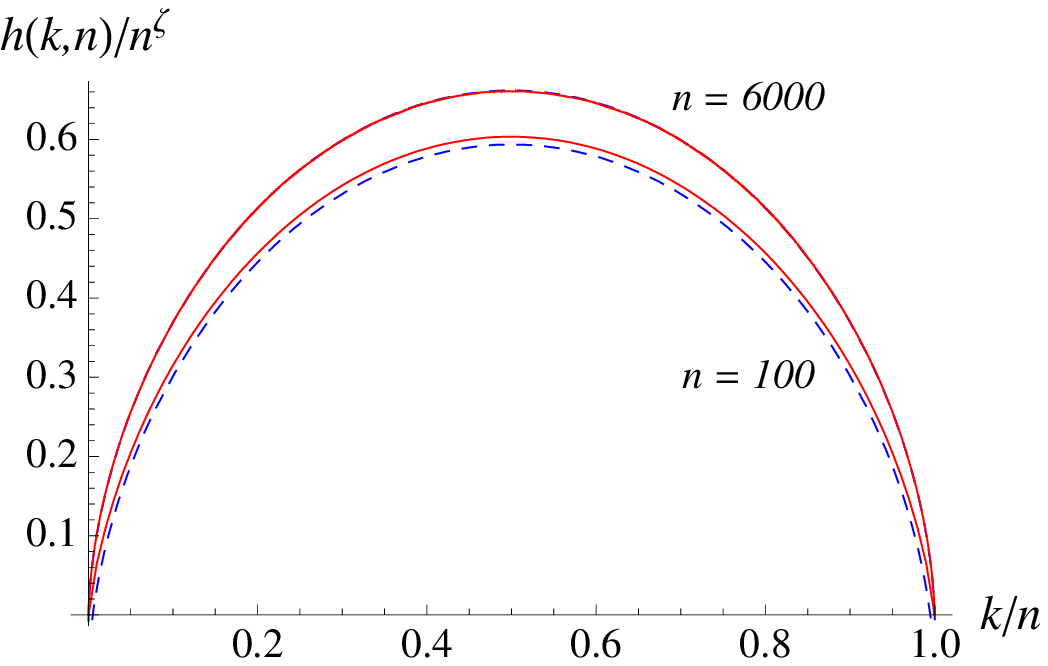}&
\includegraphics[width=0.47\textwidth]{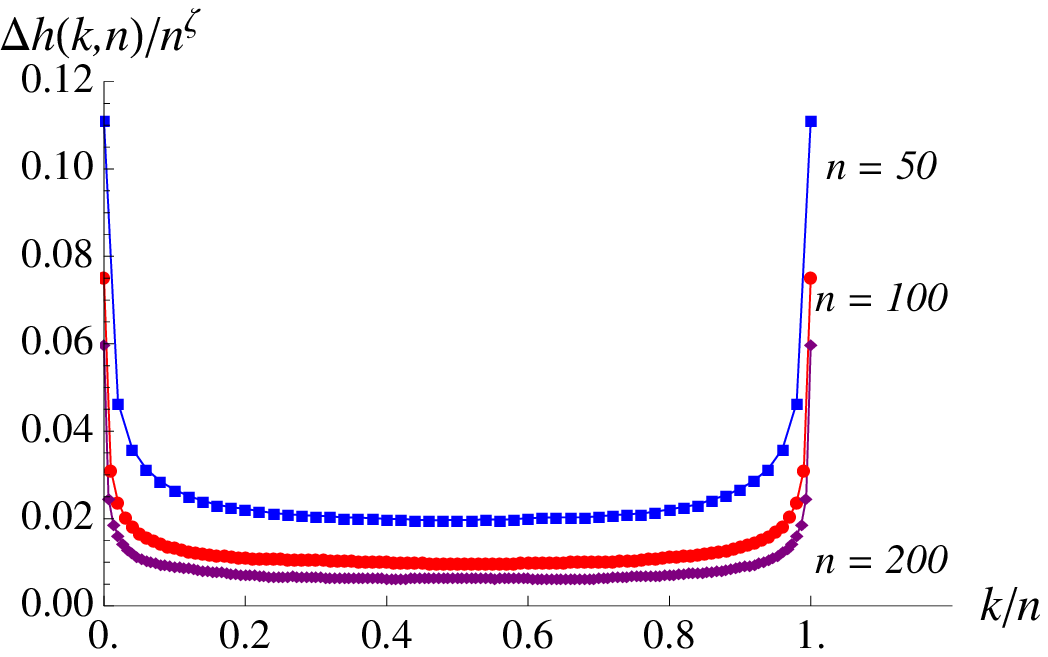}\\
(a)&(b)
\end{tabular}
\caption{(a) Scaling plot for the height function $\mathcal{H}(k,n)$ for two strand lengths $n=100$ and $n=6000$. The dashed lines correspond to the scaling function plus the first finite-size corrections to the scaling limit $n^{-\zeta}(n^\zeta\mathcal{H}(k/n)-1)$. (b) Residual scaled deviations  $\Delta h(k,n)/n^{\zeta}$ for different strand lengths $n=50,100,200$.}
\label{fig:avhf}
\end{figure}

\paragraph{Averaged height function ${h}(k,n)$.} Results for the averaged height functions are shown in figure  \ref{fig:avhf}a,b. In order to point out universal behaviour we plot ${h}(k,n)/n^\zeta$ as a function of $x=k/n$. In figure\ \ref{fig:avhf}b, we compare the data to the first-order corrected scaling limit $n^{-\zeta}\mathcal{H}(k/n)-1$ where $\mathcal{H}(x)$ denotes the scaling function from (\ref{H1expl}). The deviations $\Delta h(k,n) = h(k,n)-n^{-\zeta}\mathcal{H}(k/n)+1$ are large at the ends $k=1$ and $k=n$. 

\begin{figure}[htpb]
  \centering
  \includegraphics[width=0.5\textwidth]{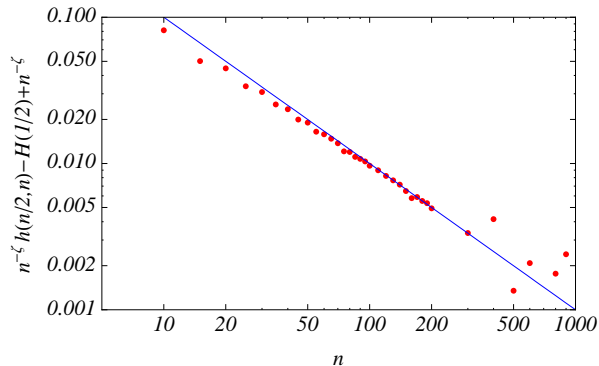}
  \caption{Log-log-plot of deviations of the height function $h(k,n)$ at $k=n/2$ as a function of $n$. The straight line corresponds to the function to $n^{-1}$.}
  \label{fig:hhalf}
\end{figure}

We examine the deviation of the height function $h(k,n)$ from the scaling limit by evaluation of its value at $k = n/2$. 
Figure \ref{fig:hhalf} shows the rescaled deviation $n^{-\zeta}(h(n/2,n)-n^\zeta\mathcal{H}(k/n)+1)$.
Numerically, we find
\begin{equation}
    n^{-\zeta}[h(n/2,n)-n^\zeta\mathcal{H}(k/n)+1] = {\mathcal O}( n^{-1})\ , 
\end{equation}
in agreement with the scaling form (\ref{HScaCor}). 

%% file: chap7.tex
\section{Growth model}
\label{sec:growth}
The model considered in the previous sections (\textbf{model A}) is a \emph{deposition model}. The size of the system is fixed; to study the folding of a strand with $n$ bases, we
start from a set of $n$ unoccupied points $\{1,\cdots, n\}$ on the line
and successively deposit arches in a planar way until the system
is full (no deposition possible). Systems with different size
$n$ and $n'$ are a priori different.

In sections~\ref{ss-Gmodel} and \ref{ss-G'=A}, we show that this arch deposition model \textbf{A} is
equivalent to a \emph{stochastic growth model} \textbf{G} for arch systems, where we start from a system with no points. At each
time step $t$ we deposit a new point according to a simple stochastic process, and create a new arch whenever
it is possible. We show that in the growth model \textbf{G} the
statistics for the arches at time $t$ is the same as the statistics
of arches of the deposition model \textbf{A} for a system of $N=t$ points.

In section~\ref{ss-treegrowth}, we shall also show that this stochastic growth process can be reformulated (by a simple geometric duality) as a tree growth process \textbf{T}.
In section~\ref{ss-MFcalc}, as an application, we compute in a simple way local observables of these models, such as the asymptotic (at large time) distribution of the number of branches for a vertex of the growing tree, which is related to the asymptotic distribution of substructures (maximal arches) in the arch model.
Finally, in section~\ref{ss-Dyncor}, we study the dynamics of this growth model, and compute time-dependent pairing correlation functions.

\subsection{Arch growth models via point deposition}
\label{ss-Gmodel}

\subsubsection{Closed-strand growth model \textbf{G}} 
\label{sss-G}
\begin{figure}[h]
\begin{center}
\includegraphics[width=0.9\textwidth]{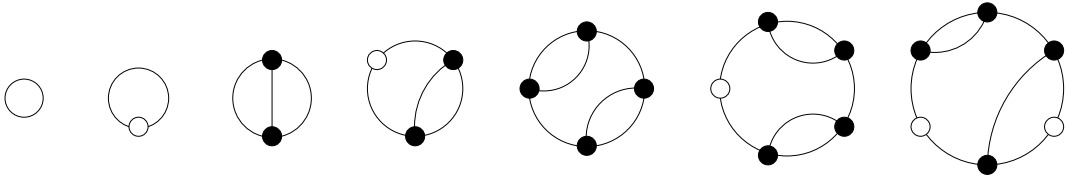}
\caption{Model \textbf{G}: successive deposition of points on a
circle. Unlinked points are marked in white, linked points in black.} \label{k1}
\end{center}
\end{figure}

Let us first define the model \textbf{G} (closed model), illustrated on figure \ref{k1}.

At time $t=0$ we start from a closed line (a circle) with no point.
At time $t=1$ we deposit a point on the circle.
Assume that at time $t$ we have already deposited $t$ points and constructed a
maximal planar arch system between these points. Namely there are
$n_a(t)$ arches and $n_f(t)=t-2n_a(t)$ free points such that it is
impossible {to construct a new arch linking 2 free points without crossing an already constructed arch (planarity condition).}

At time $t+1$ we deposit a $(t+1)^{\mathrm{th}}$ point, with
equiprobability $1/t$ on the $t$ intervals separating the $t$
already deposited points. If it is possible to draw a planar arch
between this last point and one of the free points (i.e. an arch
which does not intersect one of the existing arches) we add this
arch (It is clear that this arch is unique, otherwise the existing
planar arch system at time $t$ would not be maximal). Otherwise the
new point stays free.

\begin{figure}[htpb]
\centering
\fig{0.6}{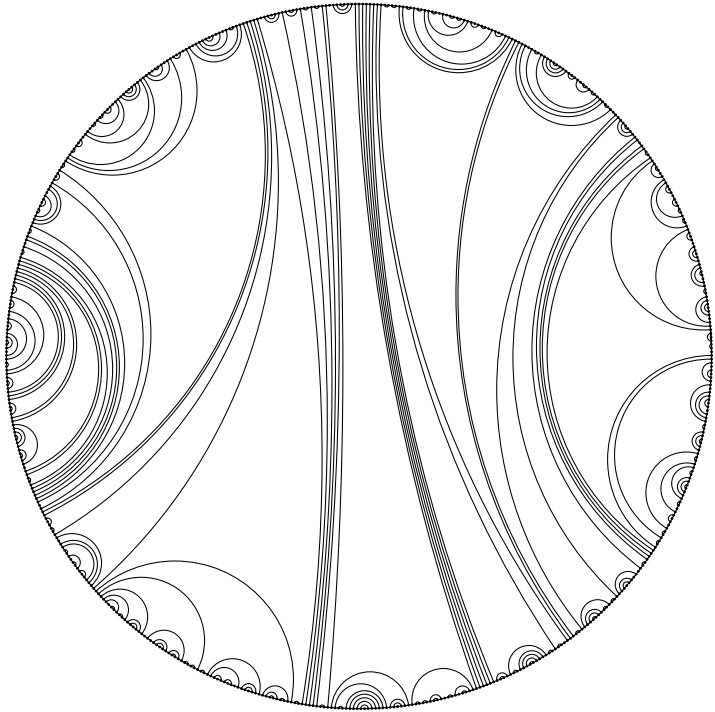}
\caption{An explicit example of a structure with $n=600$ points.}
\label{f:600points}
\end{figure}

Here we give all possible configurations from $n=2$ up to $n=9$ vertices, with
their pr	obabilities. Furthermore figure \ref{f:600points} shows a sample for $n=600$ points.
\begin{eqnarray}
\fl{\cal C}_{2}=
\left\{
1\diagram{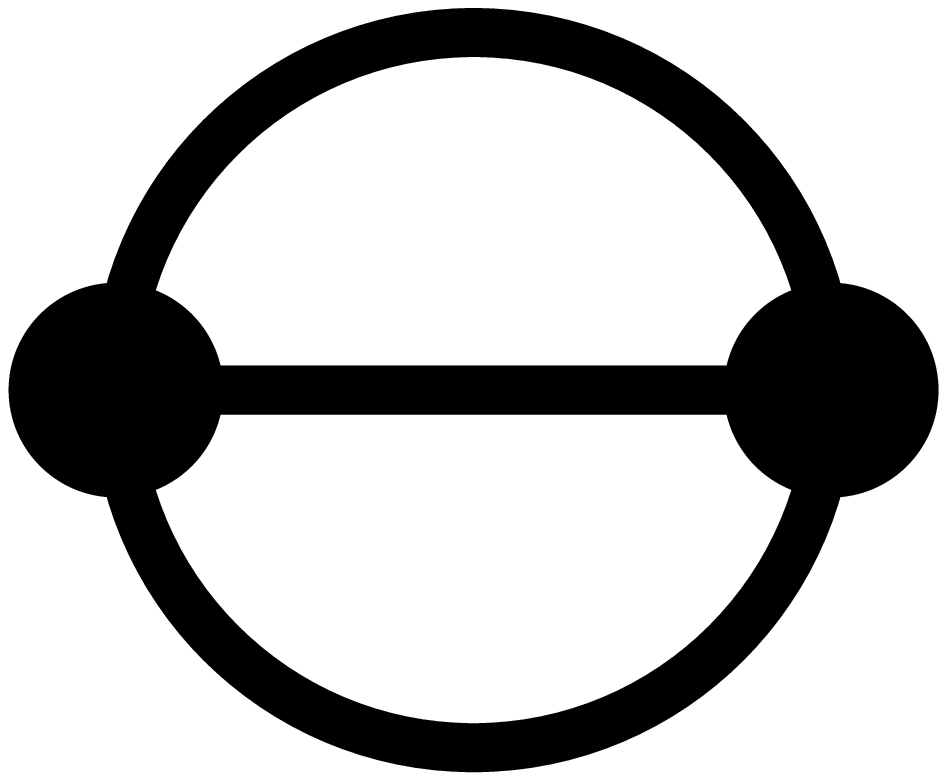} 
\right\}
\quad \qquad
{\cal C}_{3}=
\left\{
1\diagram{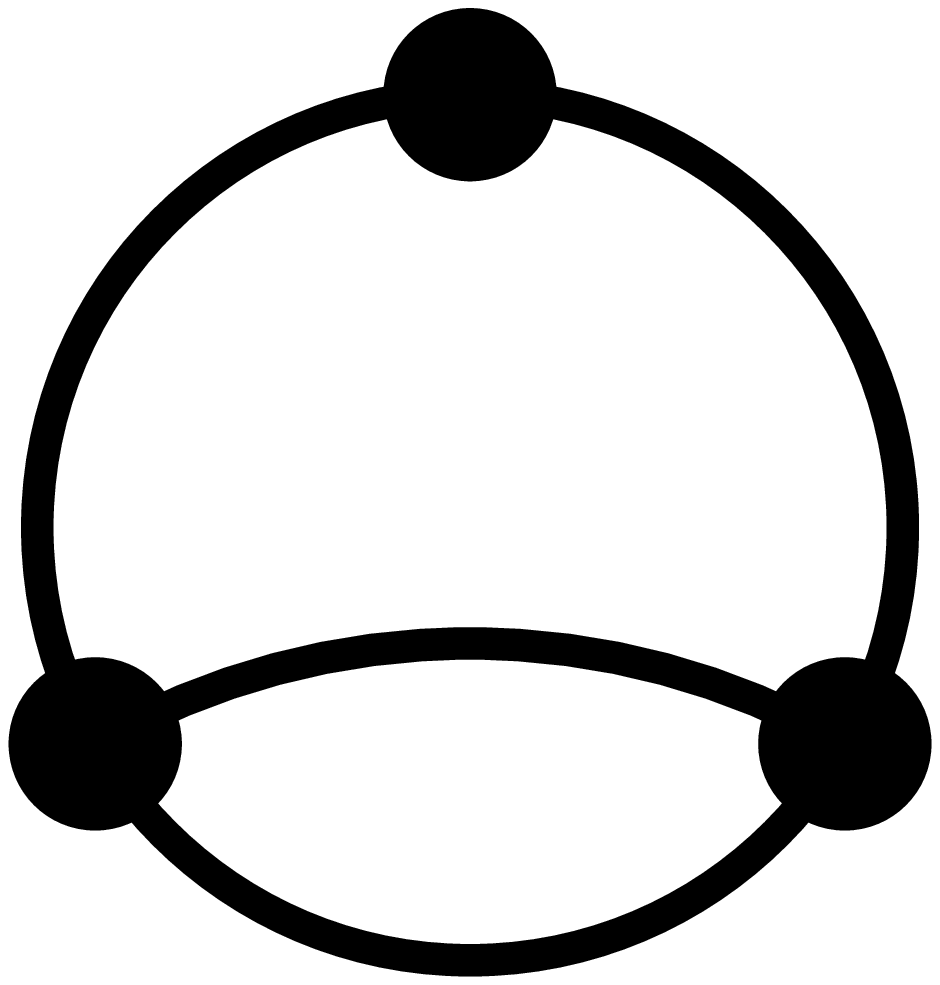}
\right\}
\end{eqnarray}
\begin{eqnarray}
\fl{\cal C}_{4}=
\left\{
\frac23\diagram{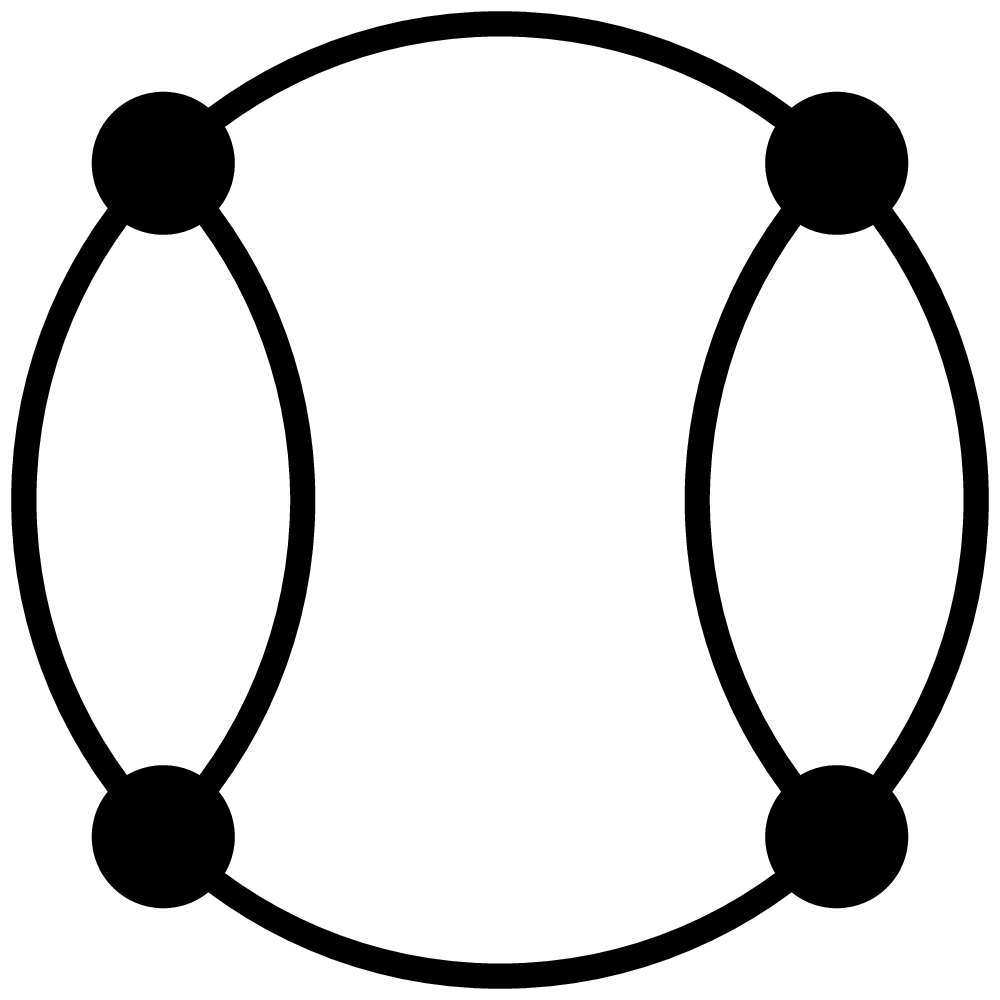}{~,\ }\frac13\diagram{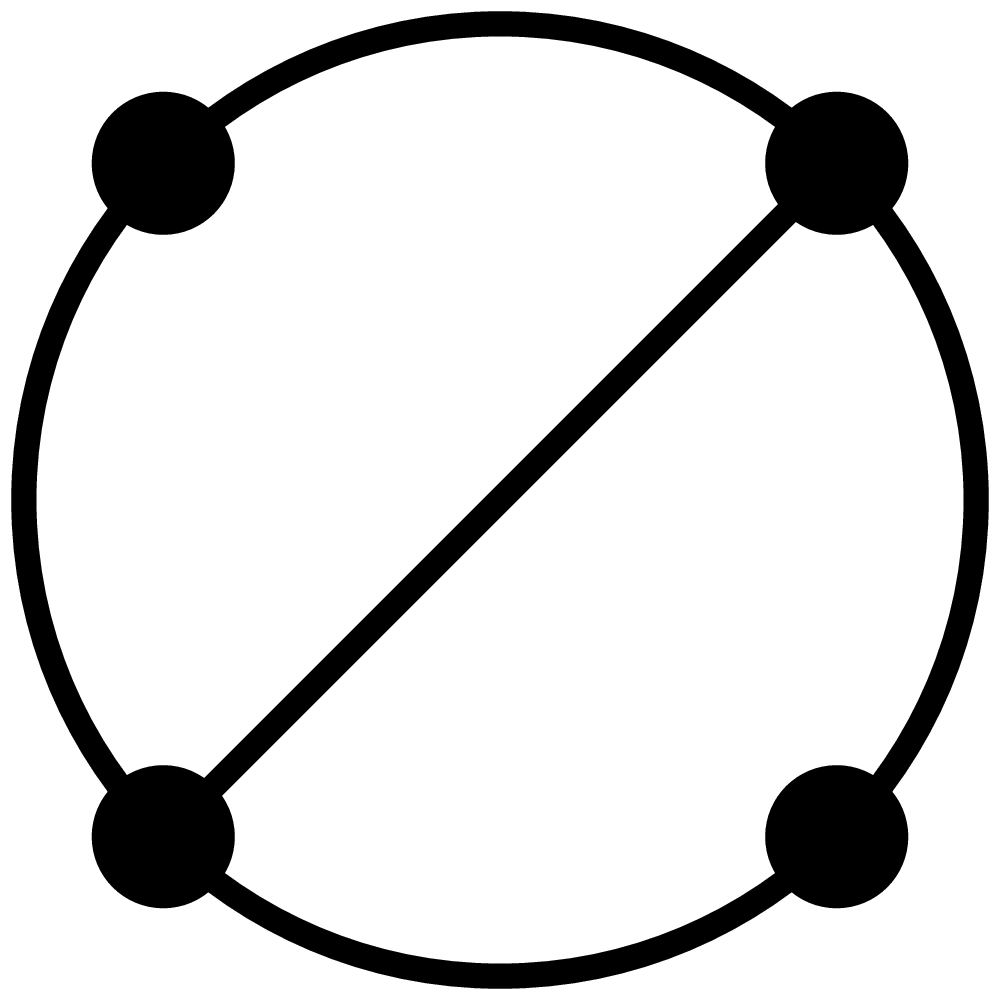}
\right\}\quad\qquad
{\cal C}_{5}=\left\{\frac13\diagram{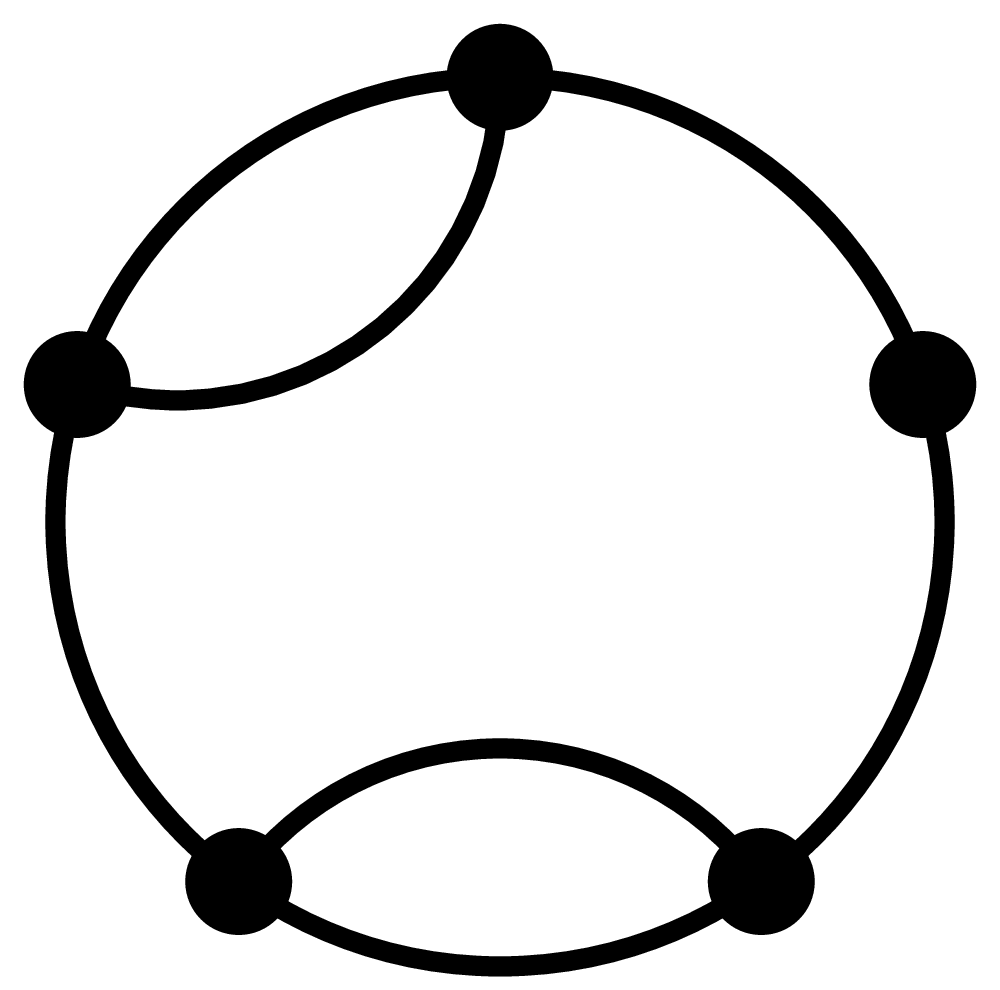}{~,\ }\frac23\diagram{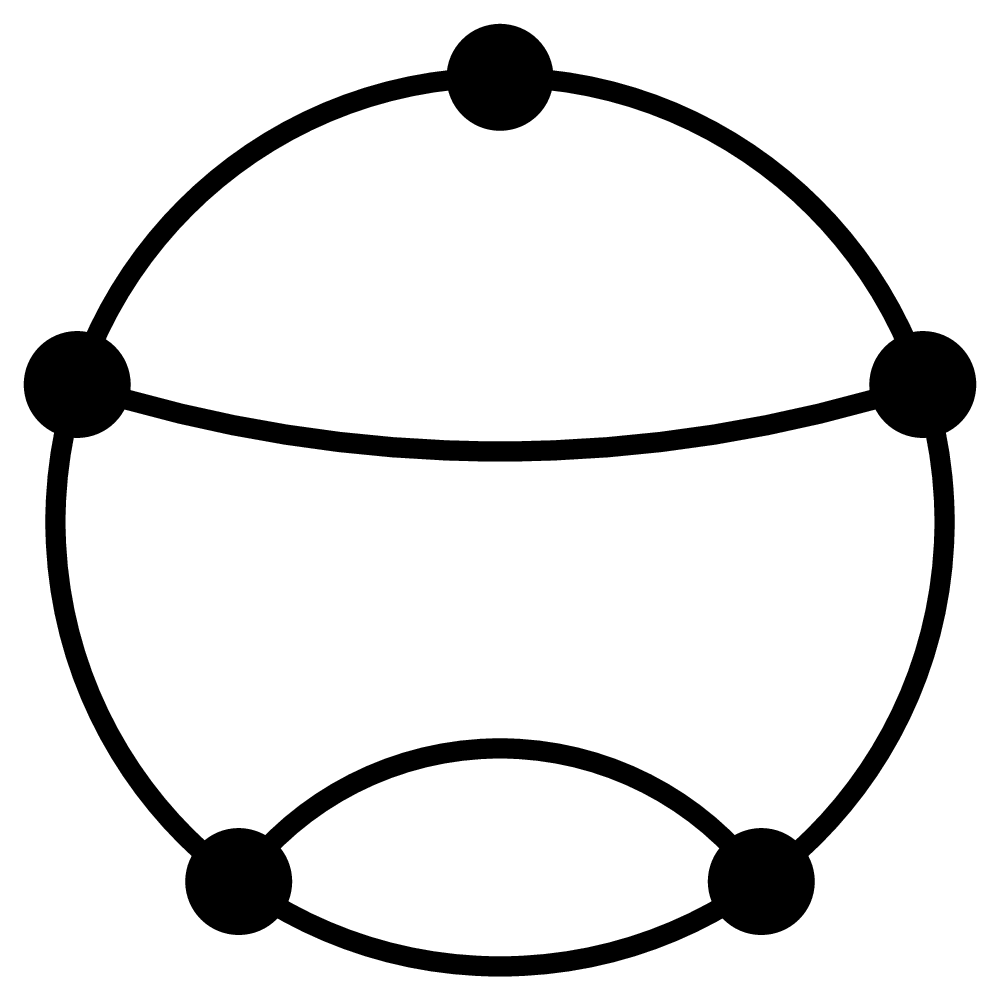}
\right\}
\end{eqnarray}
\begin{eqnarray}
\fl{\cal C}_{6}=%{~,\ }
\left\{
\frac13\diagram{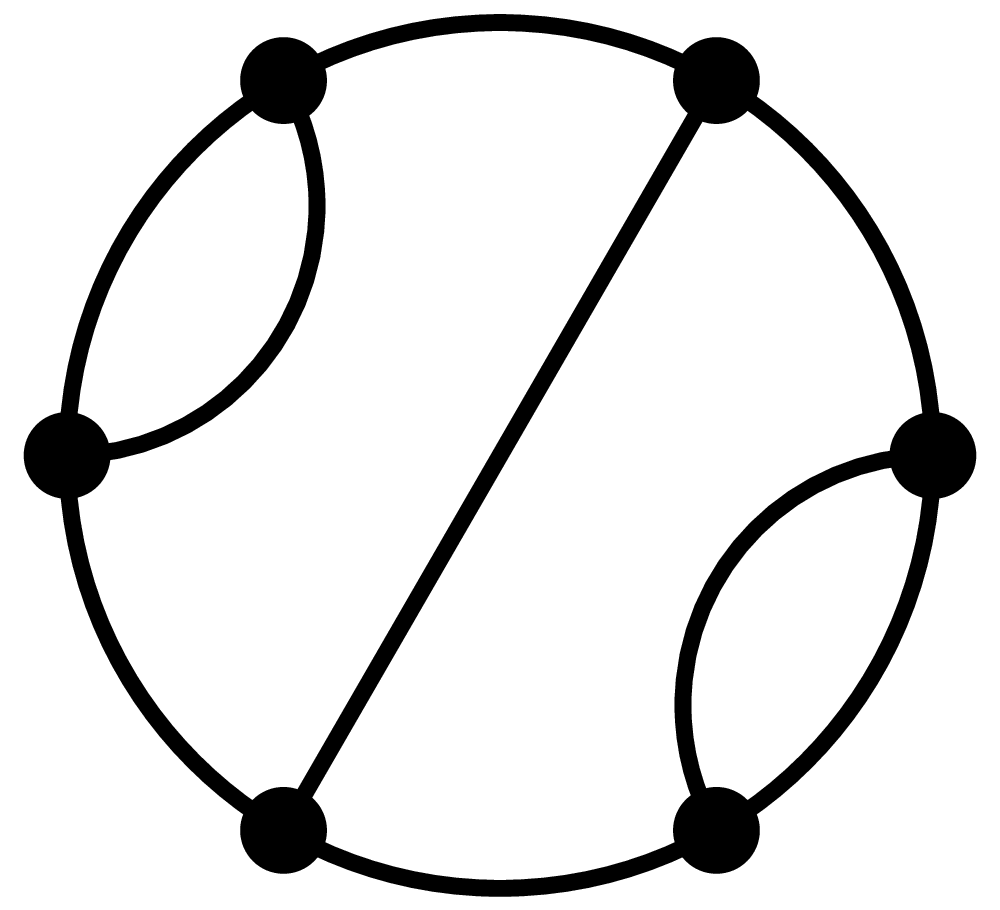}{~,\ }\frac2{15}\diagram{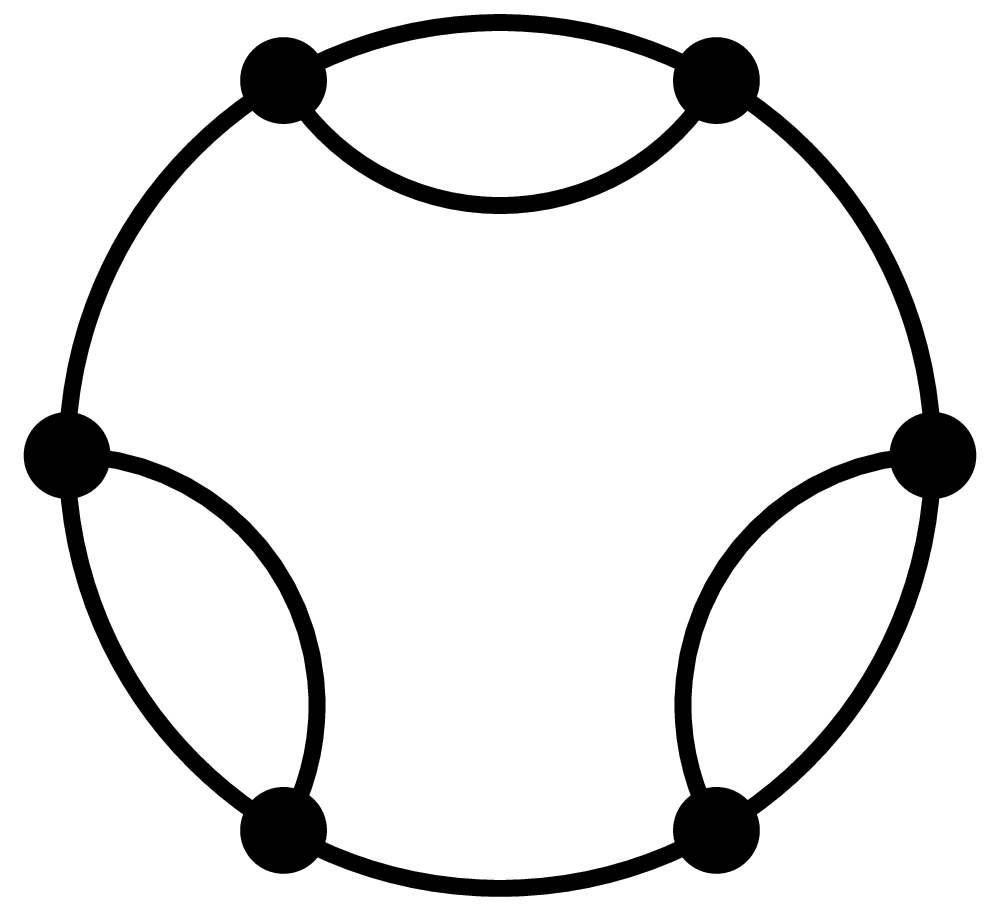}{~,\ }
\frac15\diagram{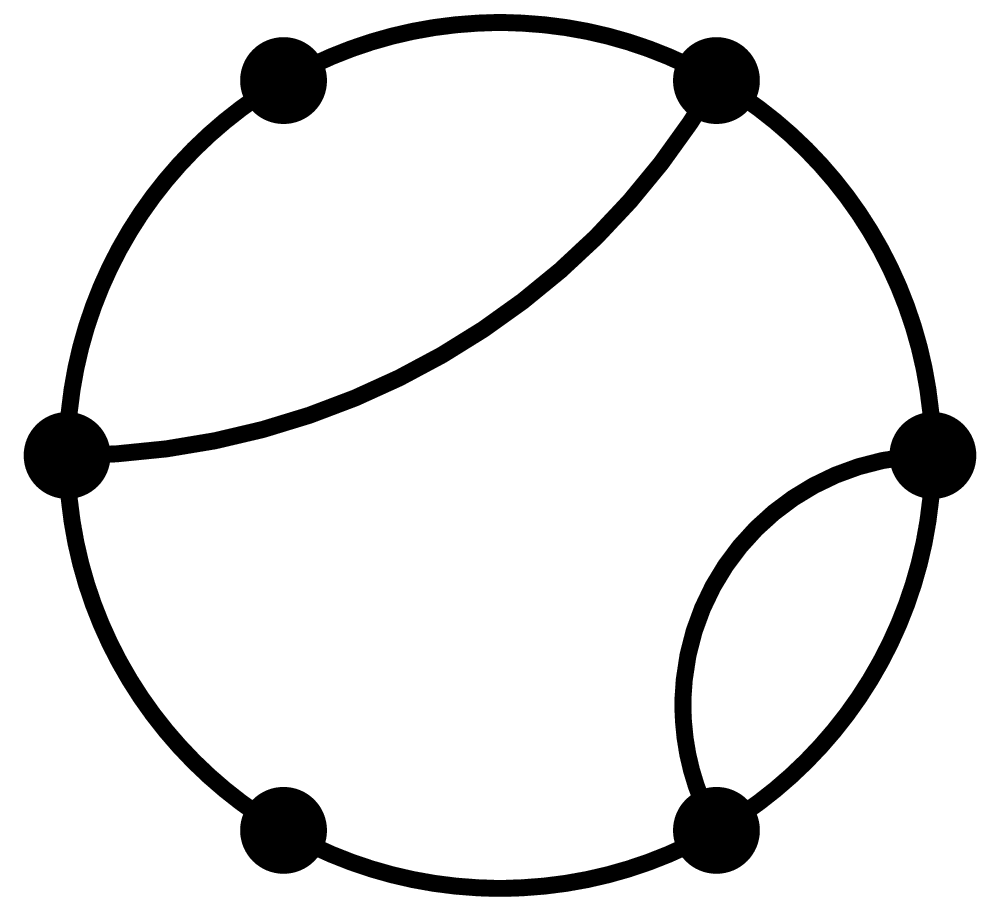}{~,\ }\frac2{15}\diagram{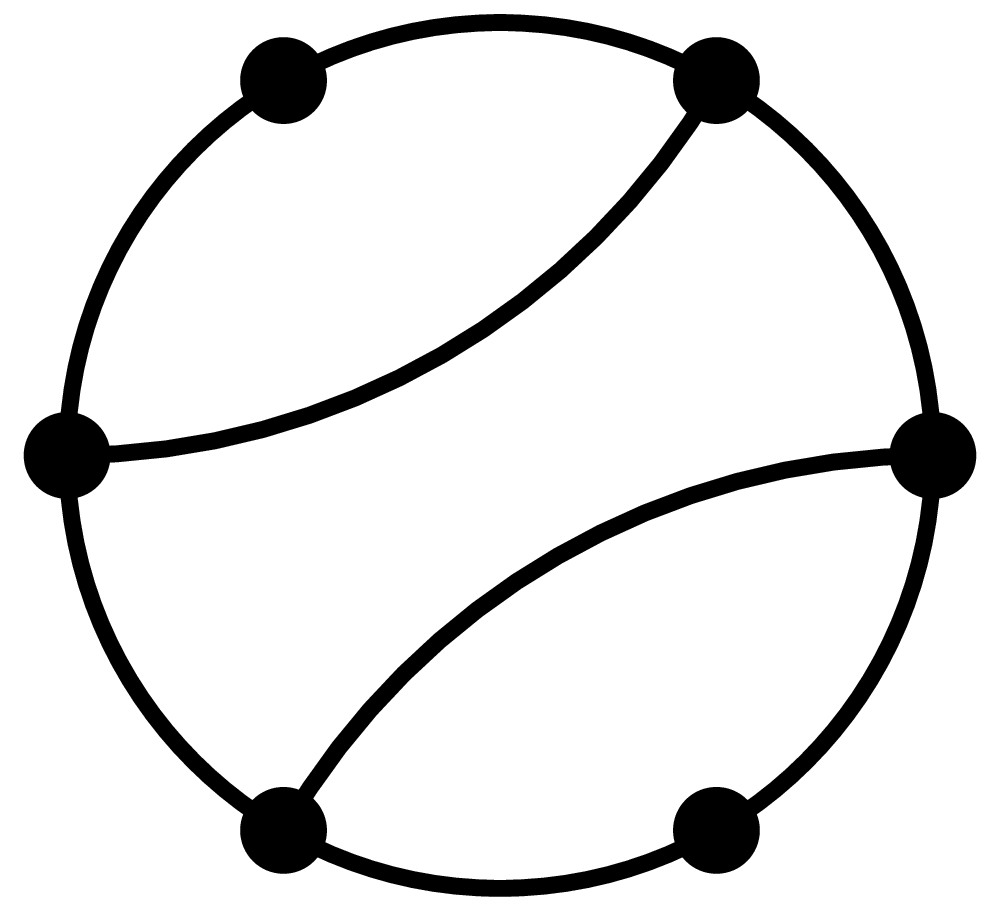}{~,\ }\frac15\diagram{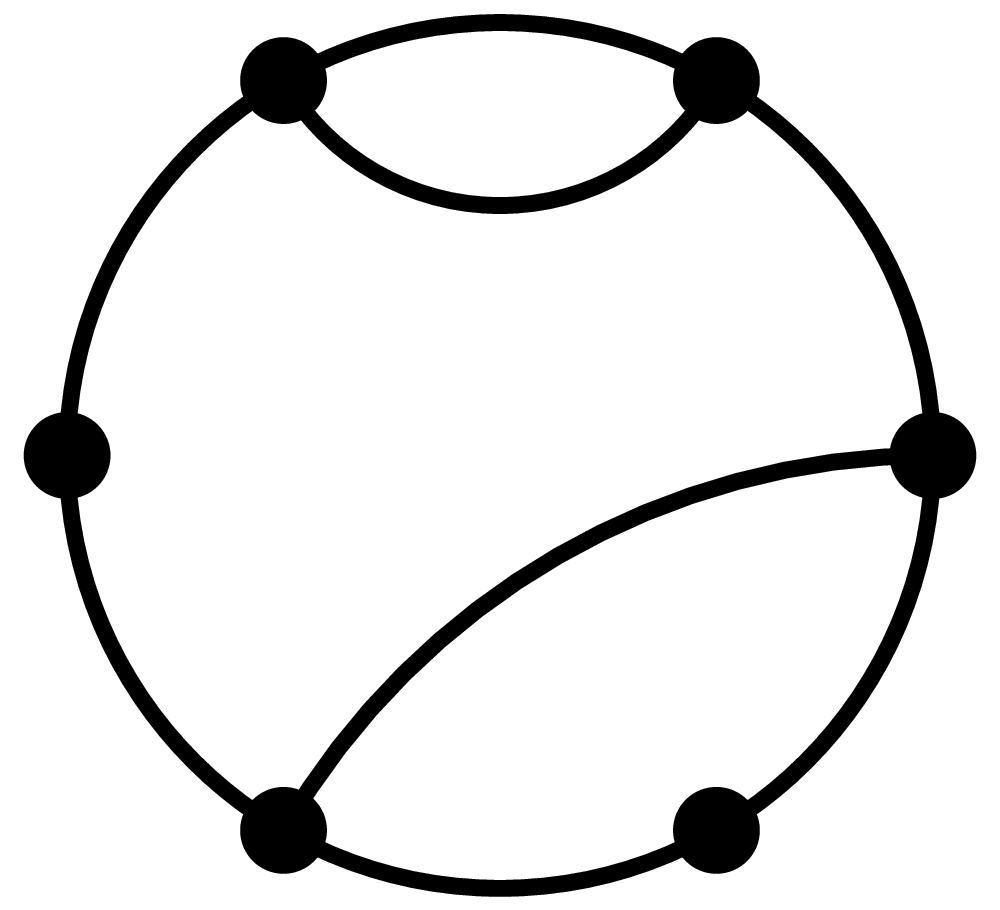}
\right\}
\end{eqnarray}
\begin{eqnarray}
\fl{\cal C}_{7}=%{~,\ }
\Bigg\{
\frac8{45}\diagram{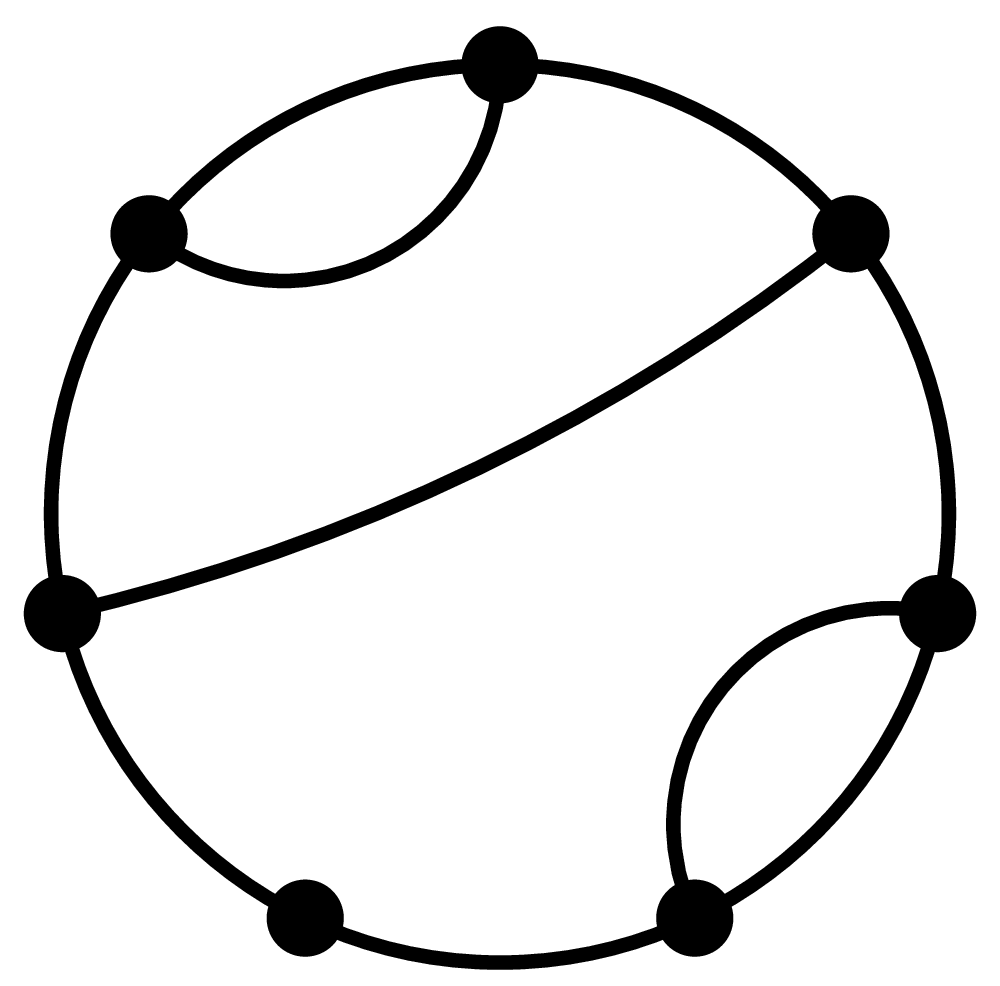}{~,\ }\frac4{15}\diagram{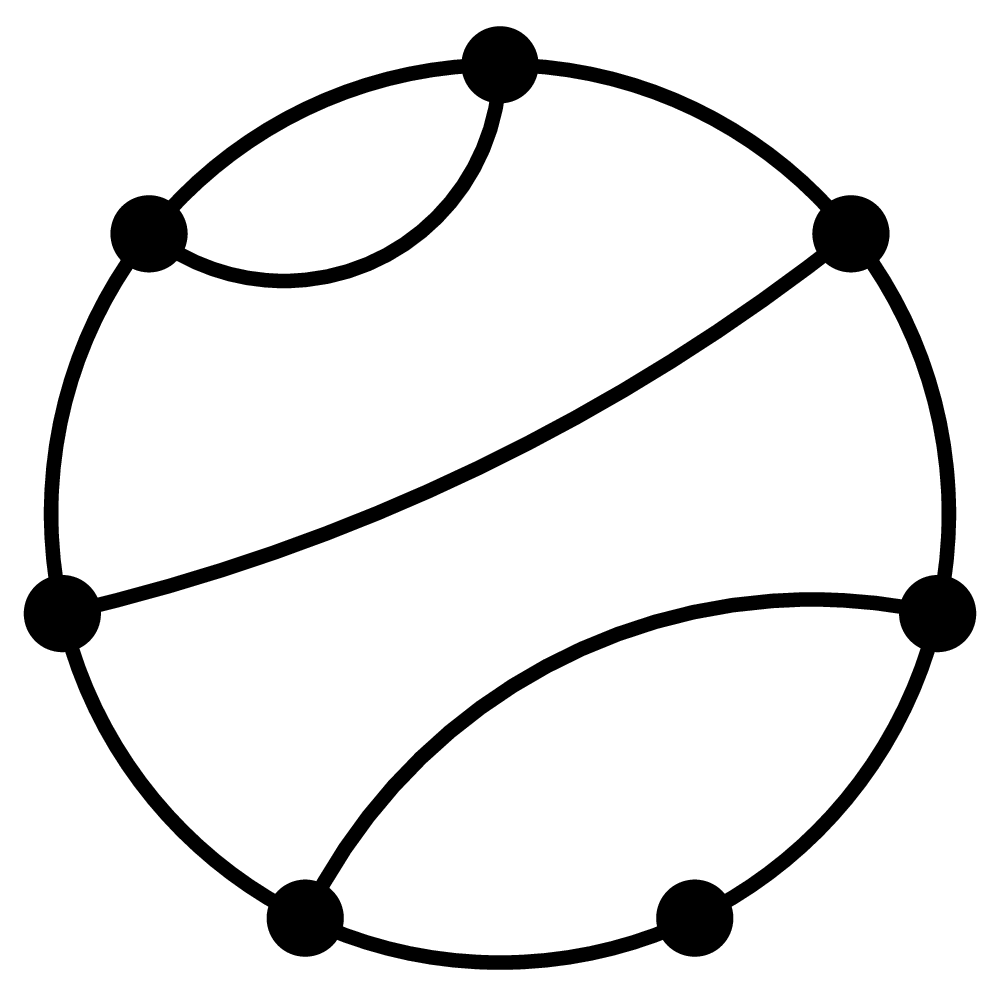}{~,\ }
\frac1{15}\diagram{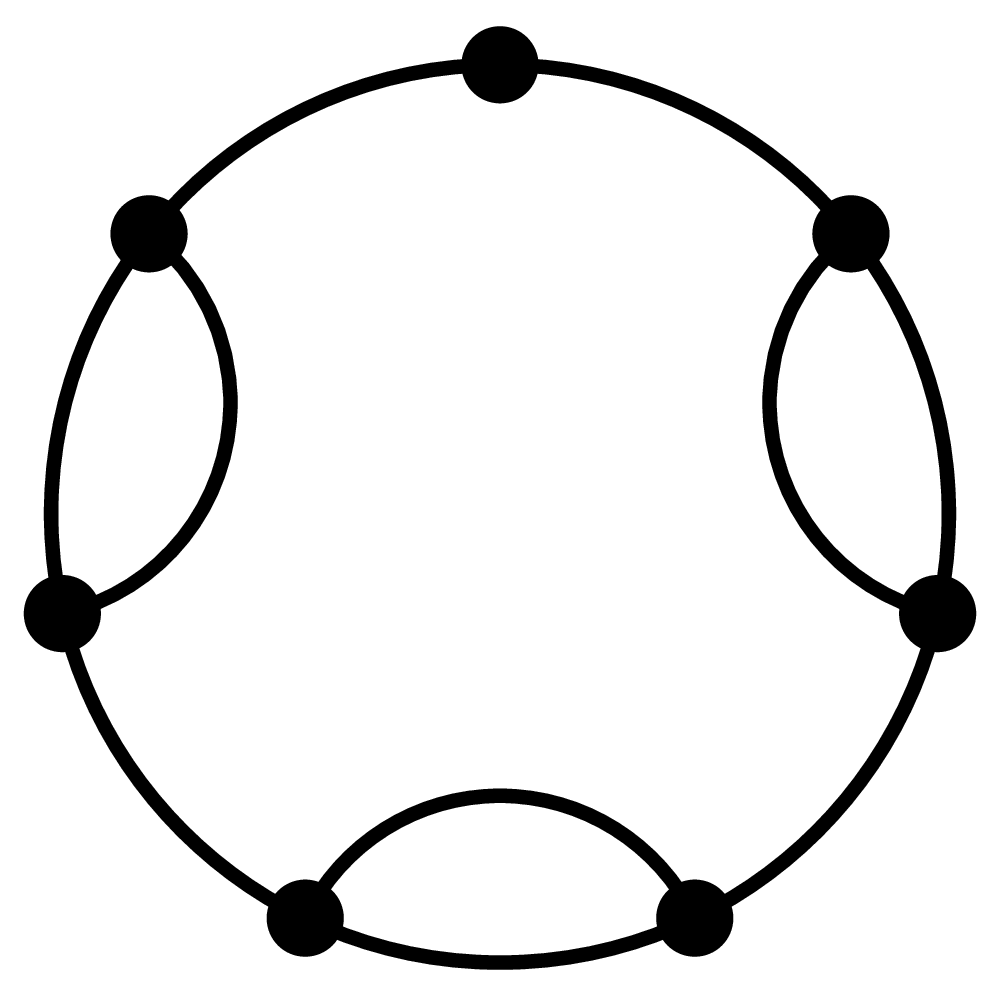}{~,\ }\frac1{5}\diagram{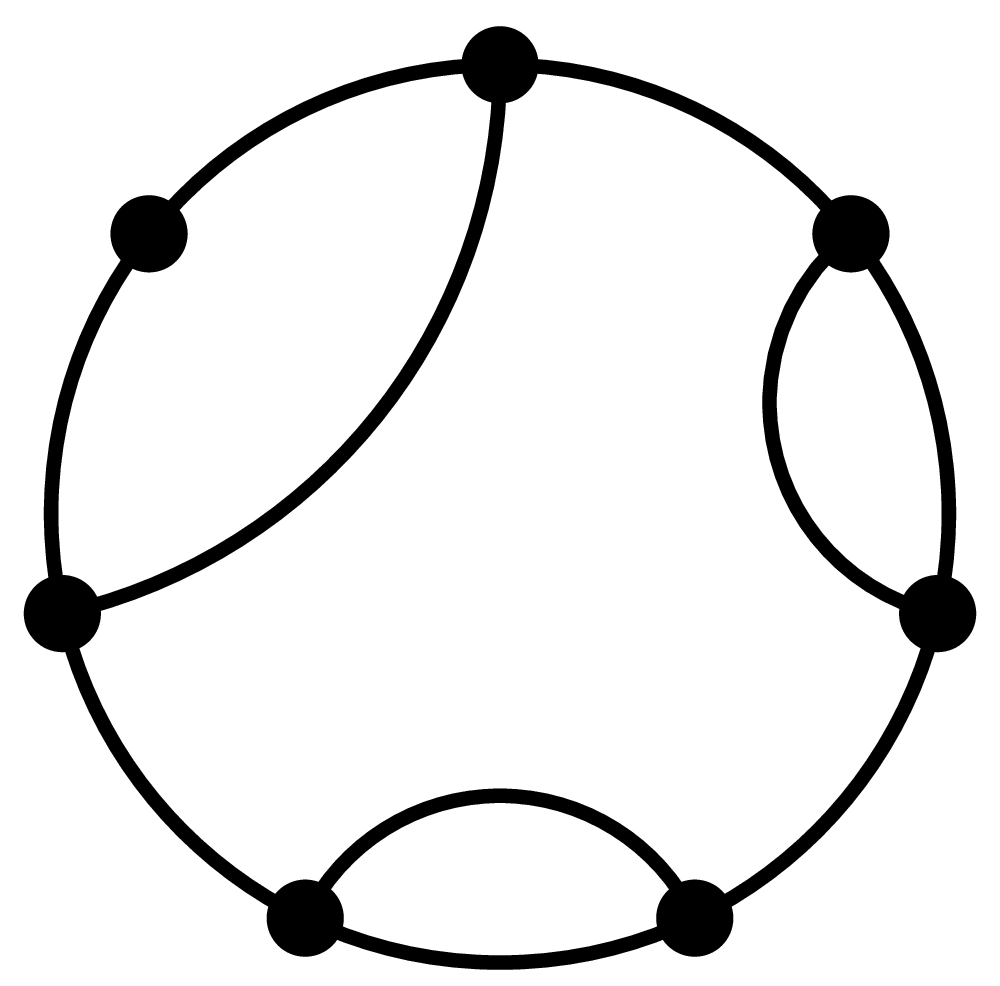}{~,\ }\frac8{45}
\diagram{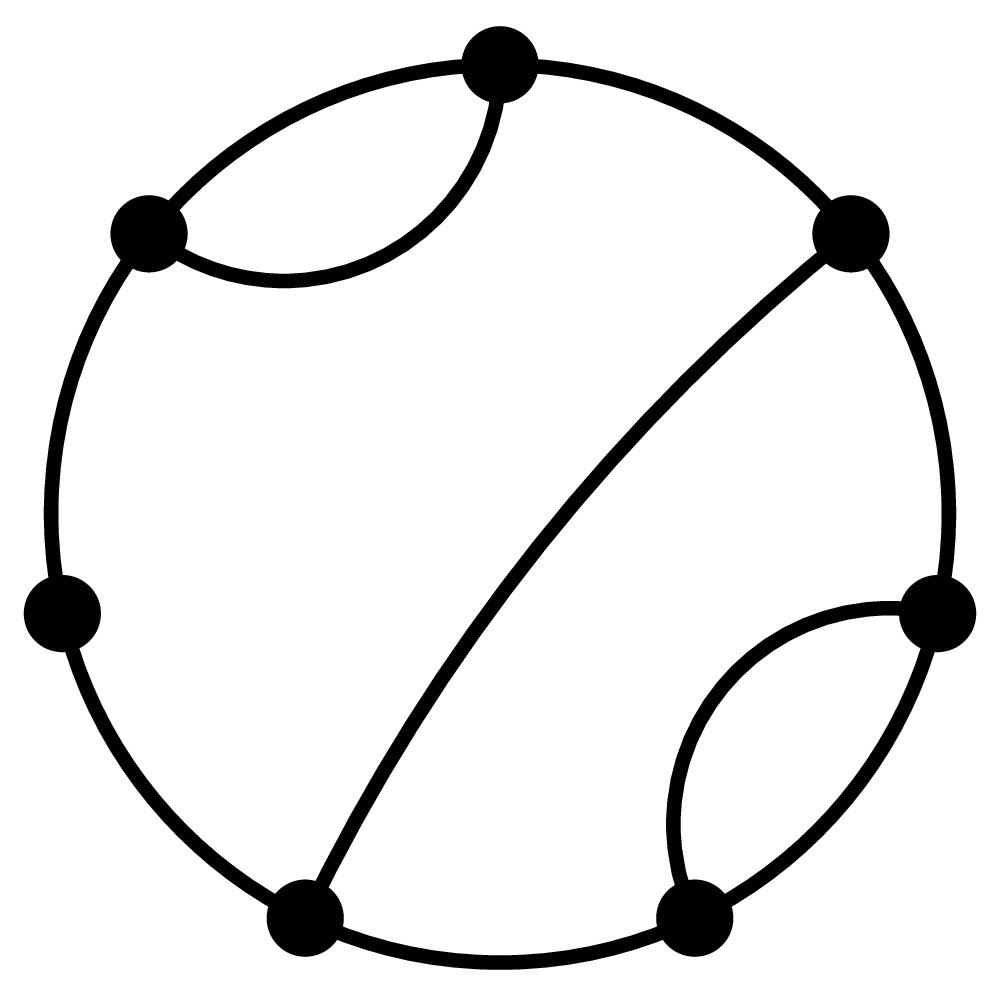}{~,\ }\nonumber\\ \frac19\diagram{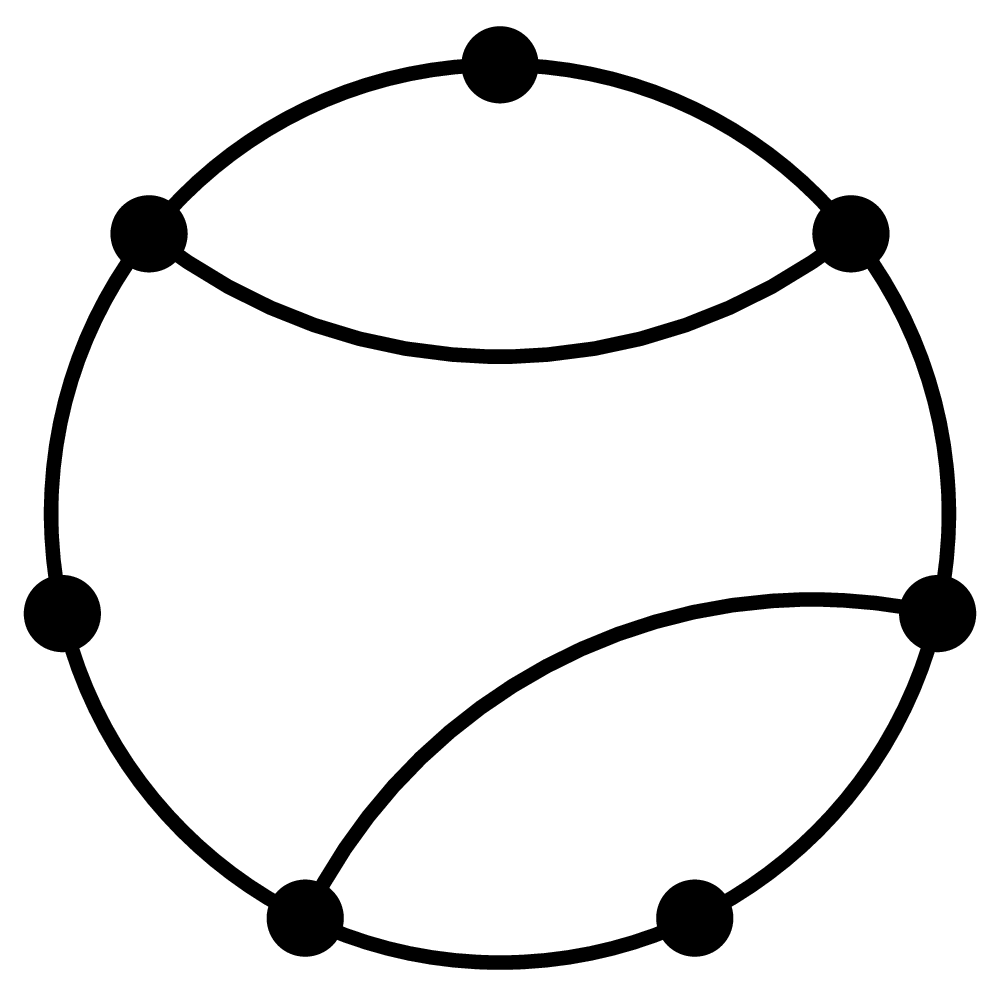}
\Bigg\}
\end{eqnarray}
\begin{eqnarray}
\fl{\cal C}_{8}=&
\Bigg\{
\frac2{105}\diagram{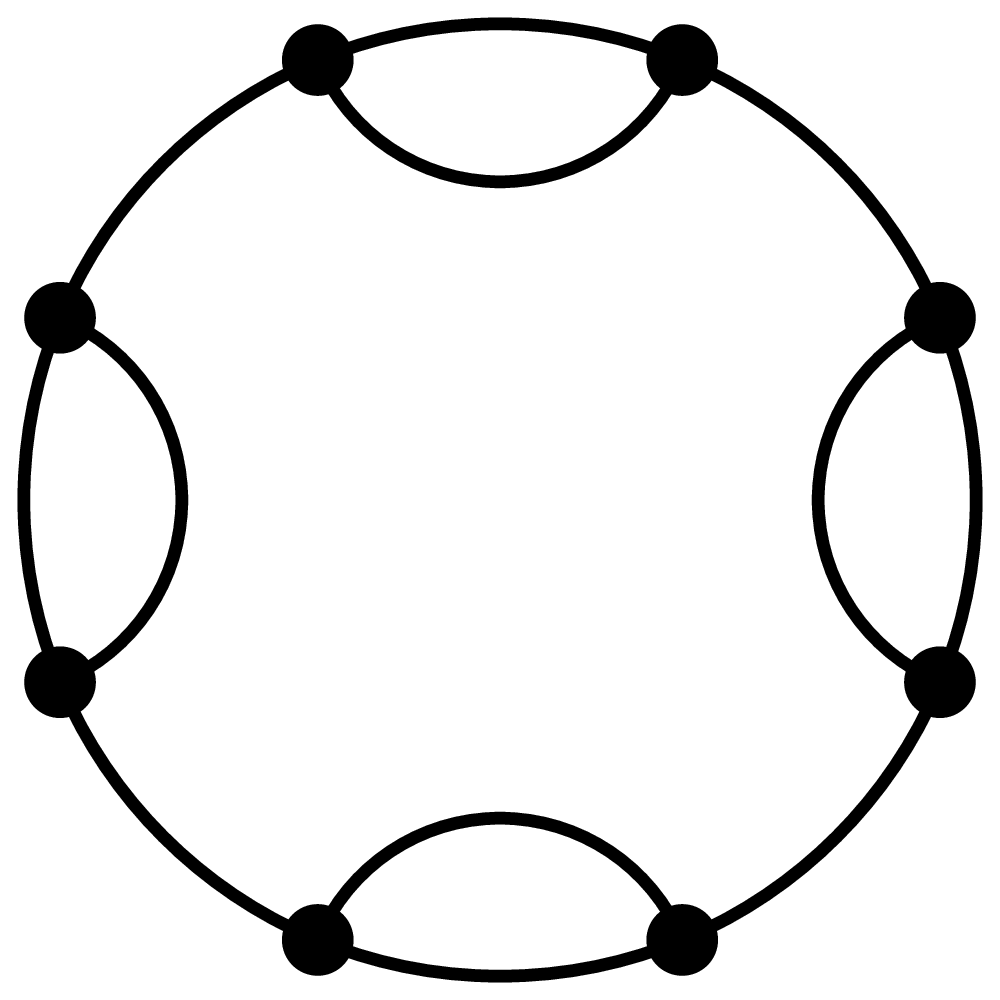}{,\ }\frac8{45}\diagram{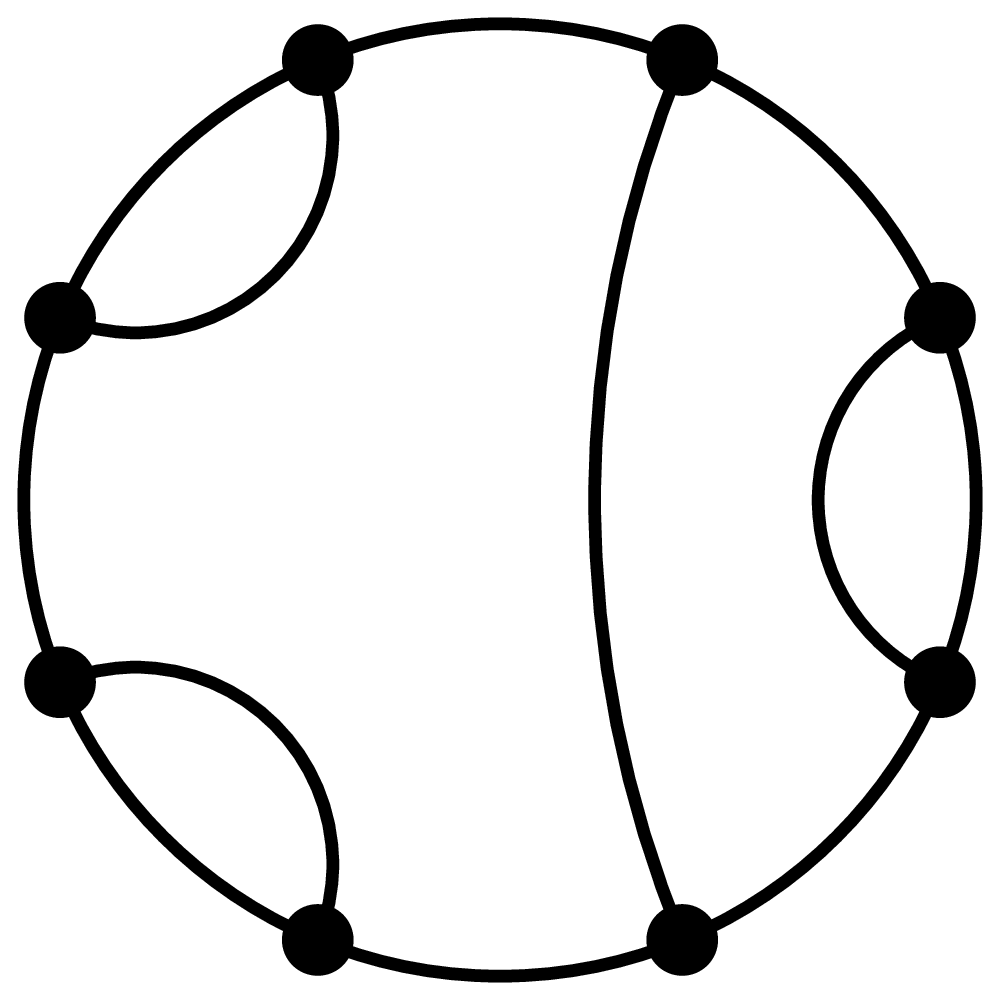}{,\ }
\frac8{63}\diagram{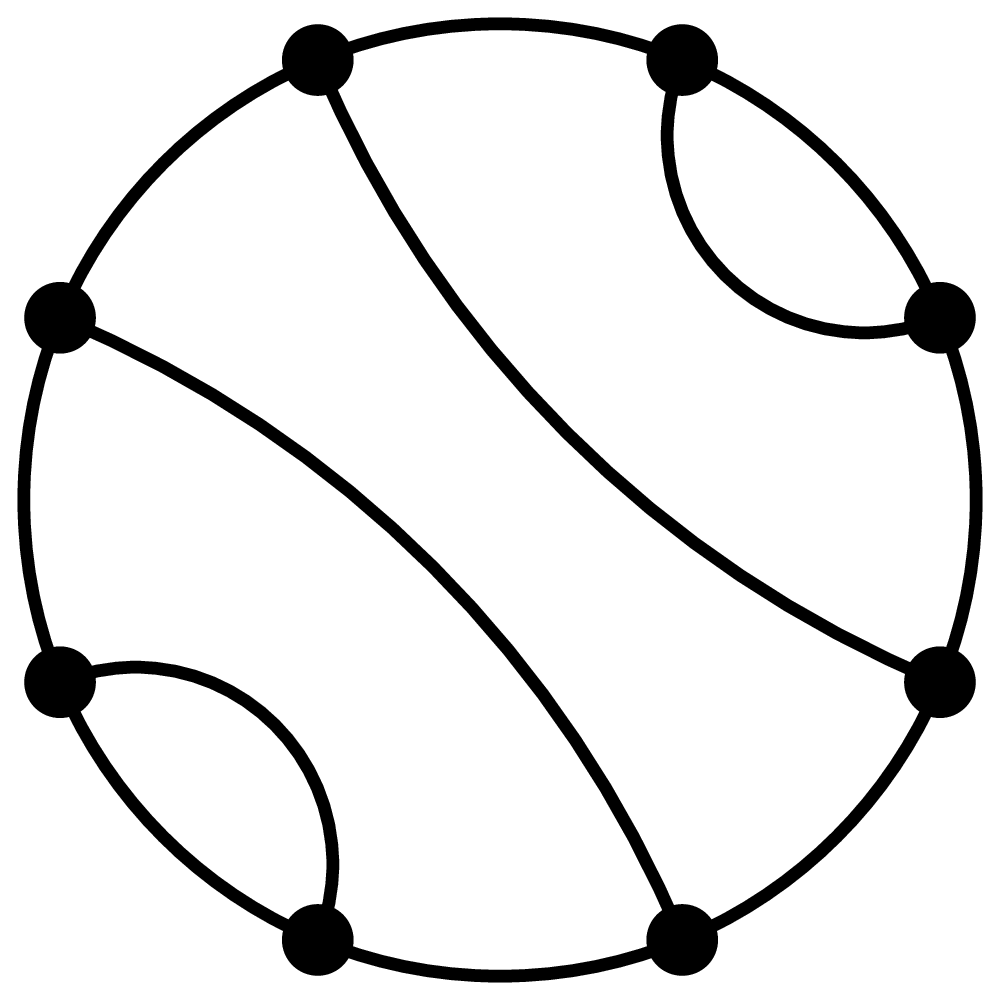}{,\ }\frac{17}{315}\diagram{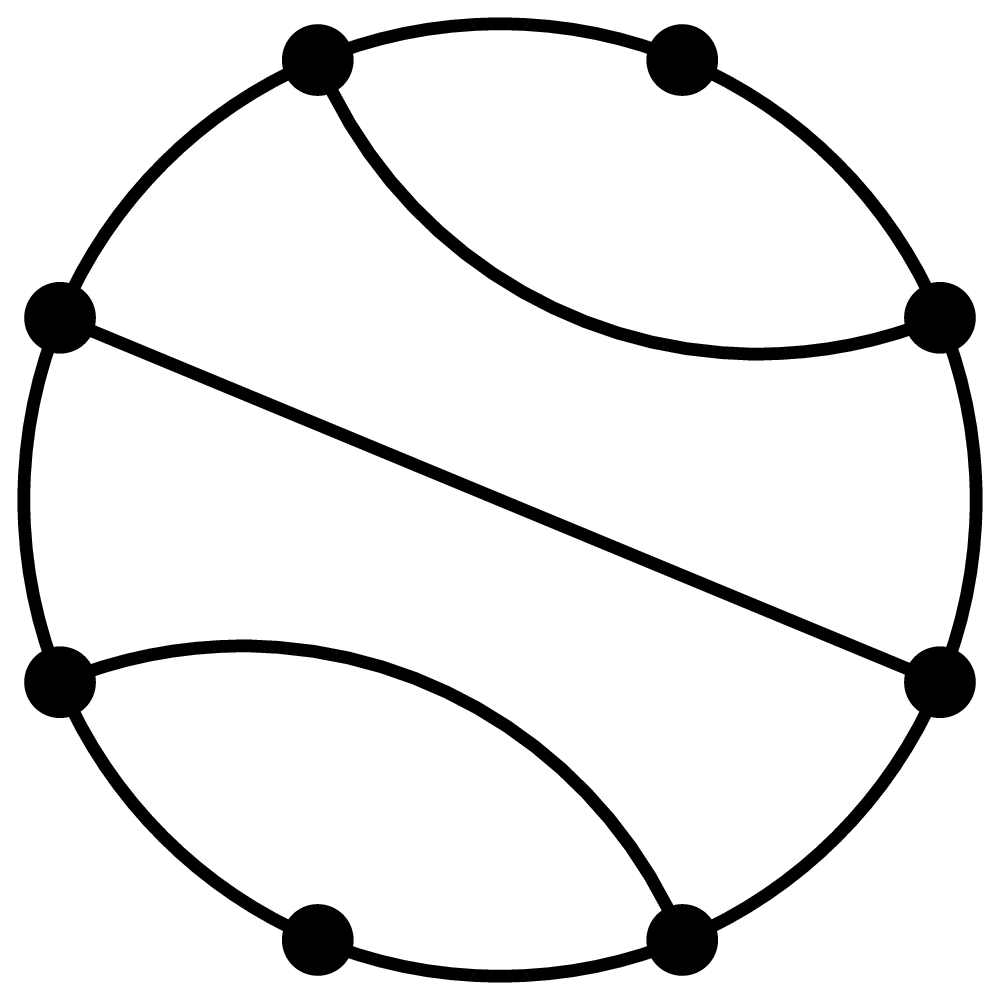}{,\ }\frac4{45}\diagram{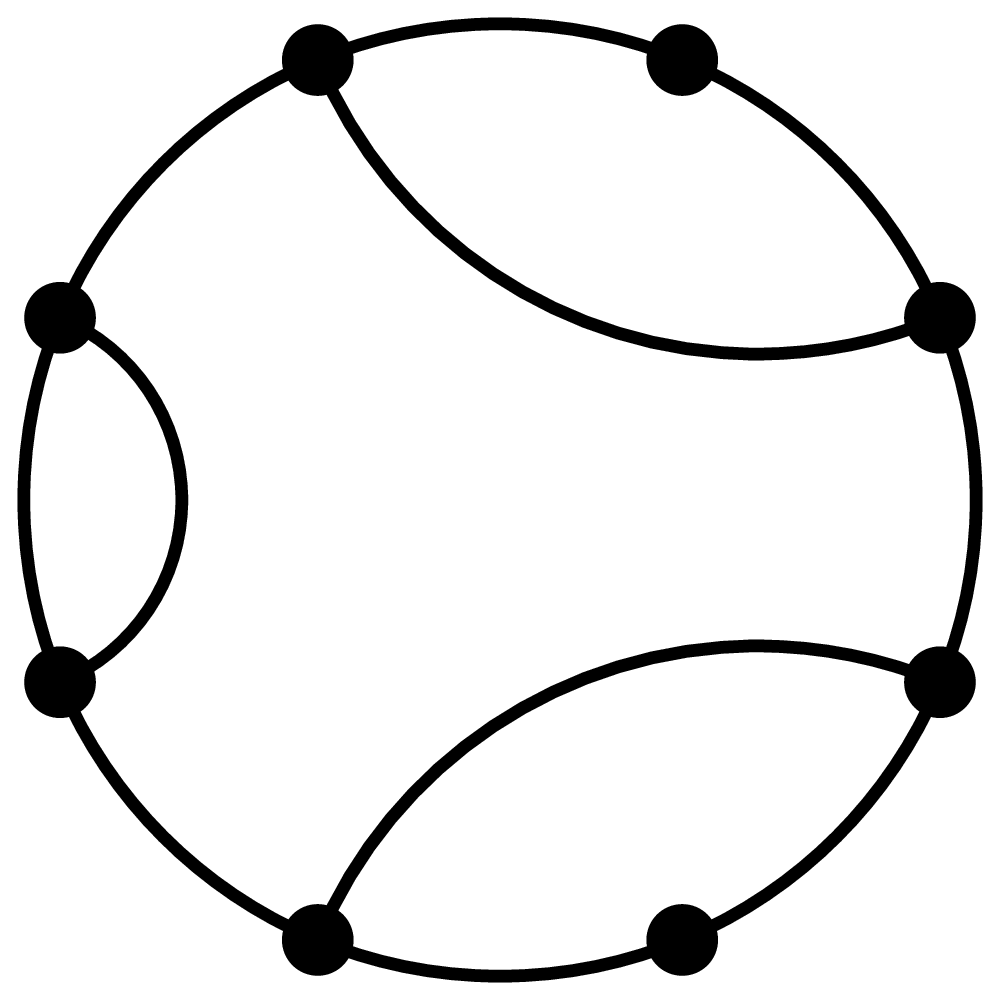}{,\ }\\
&\hphantom{,\ }
\frac4{63}\diagram{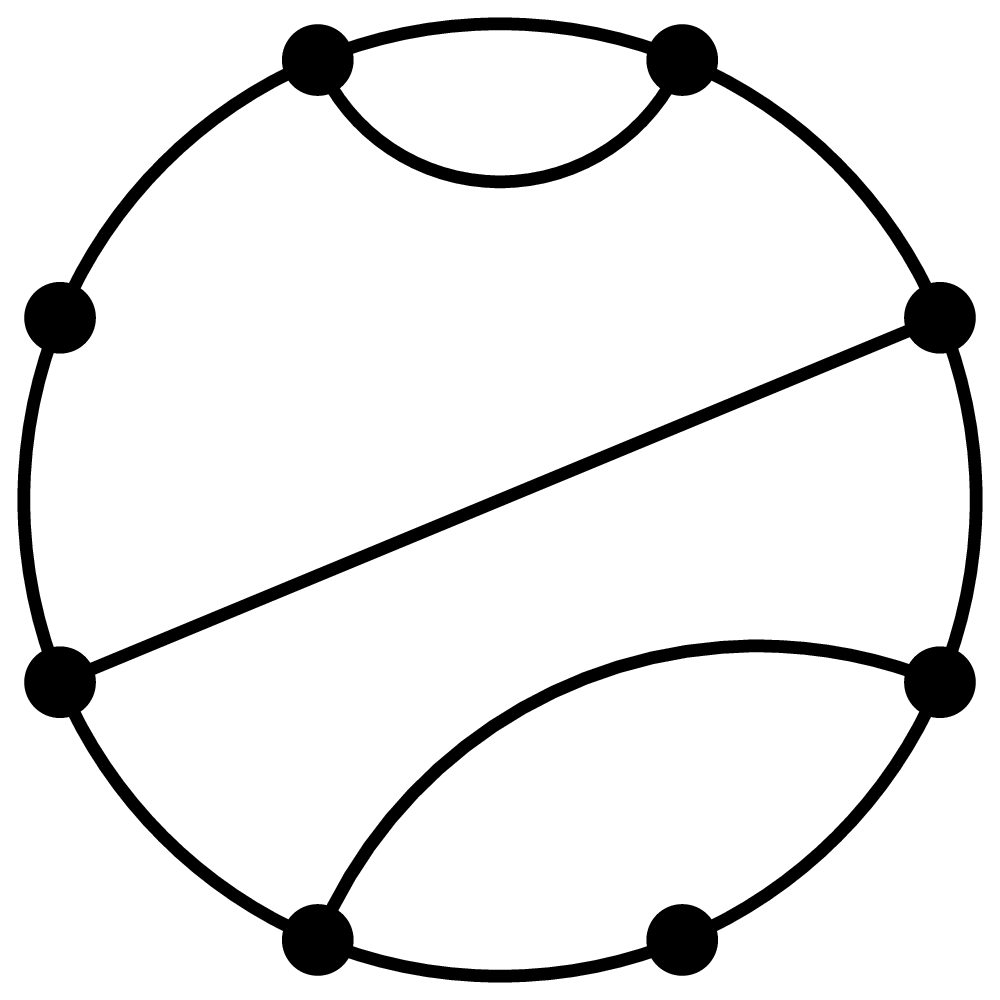} {~,\ }\frac{4}{105}\diagram{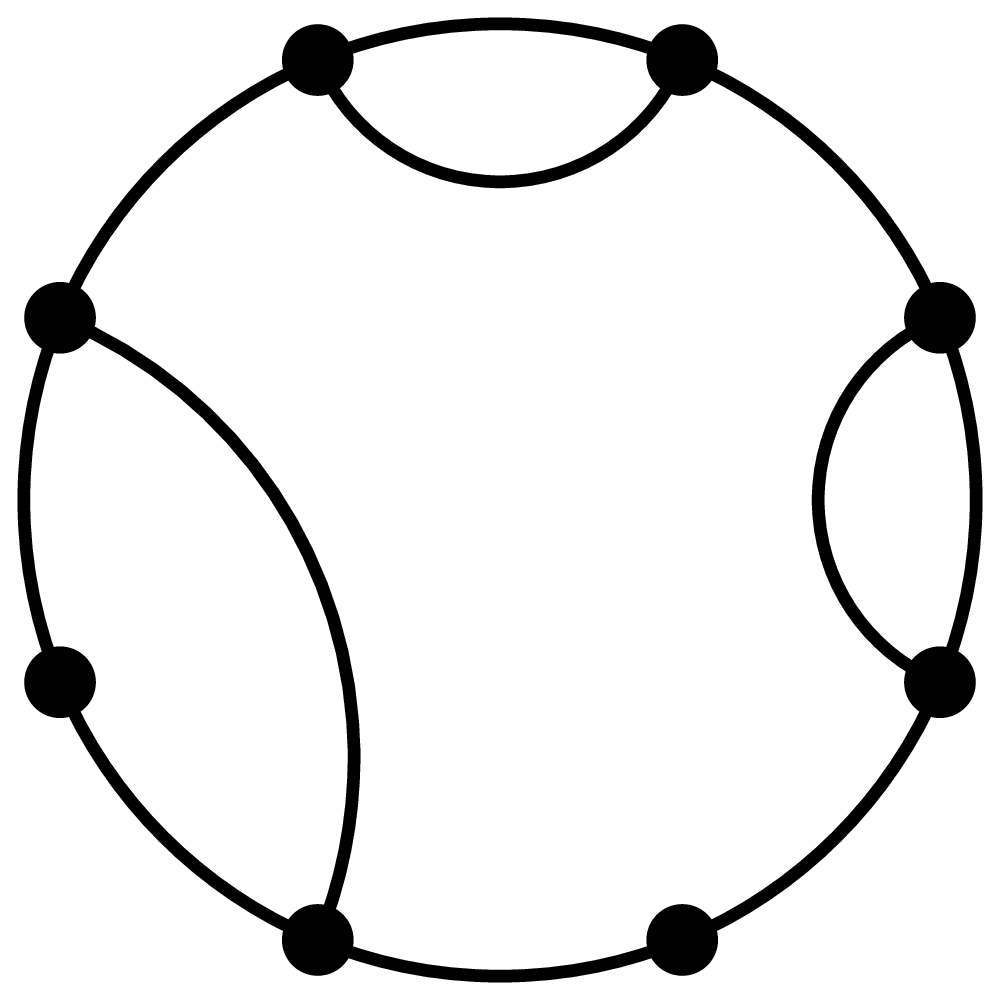}{~,\ }
\frac8{315}\diagram{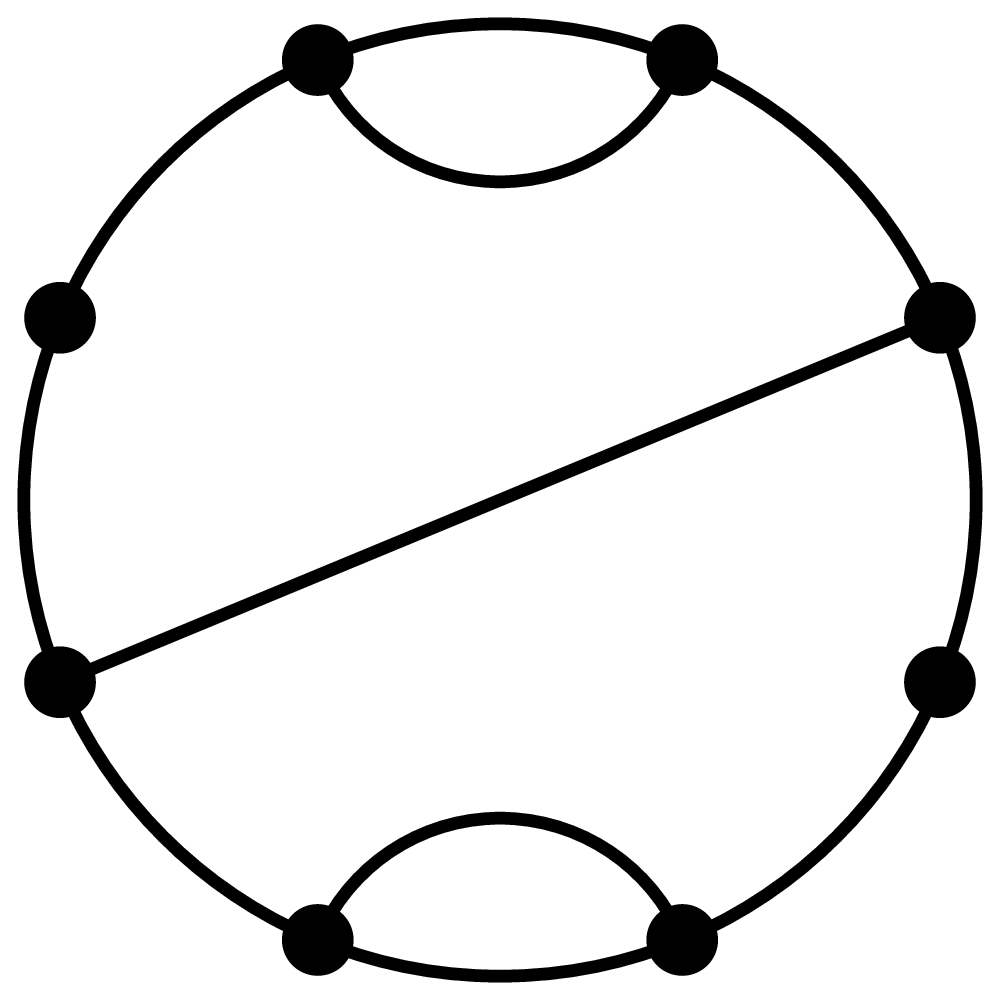}
{~,\ }\frac4{105}\diagram{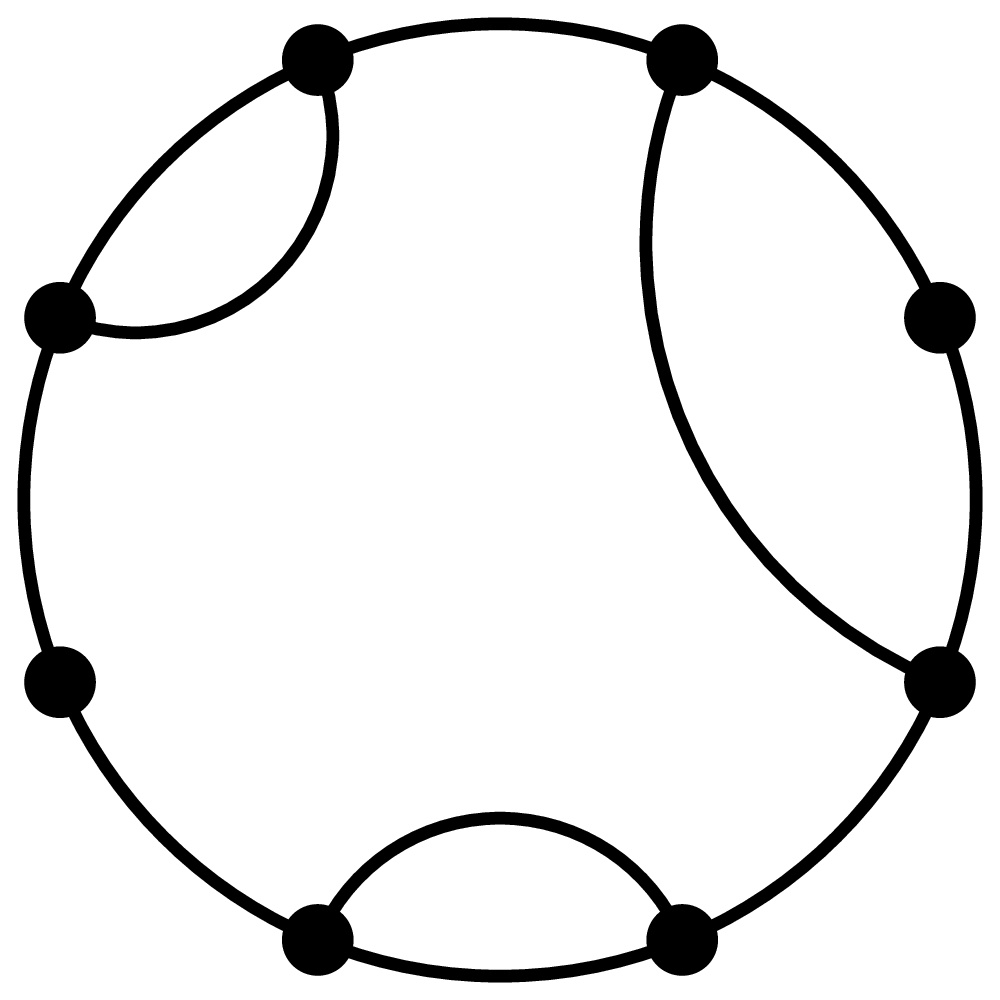}{~,\ }\frac{8}{315}\diagram{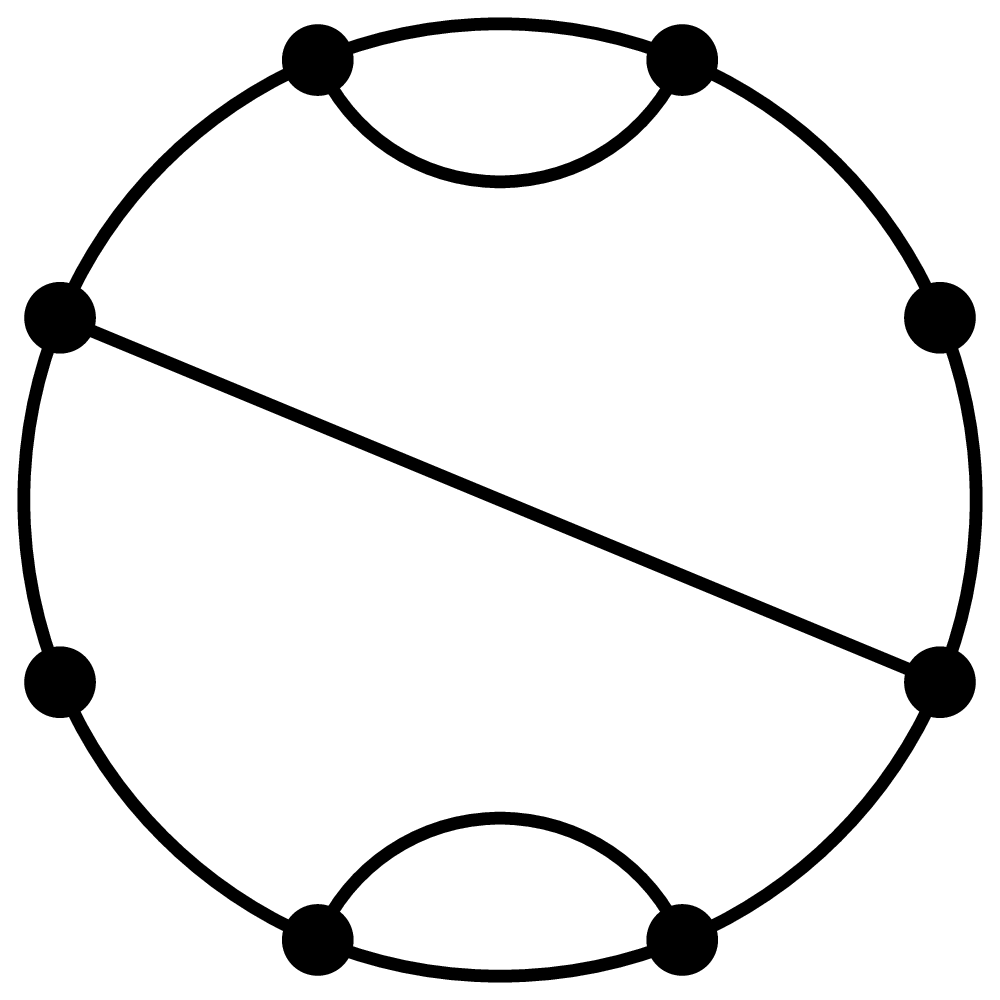}{~,\ }\nonumber\\
&\hphantom{~,\ }\frac2{21}\diagram{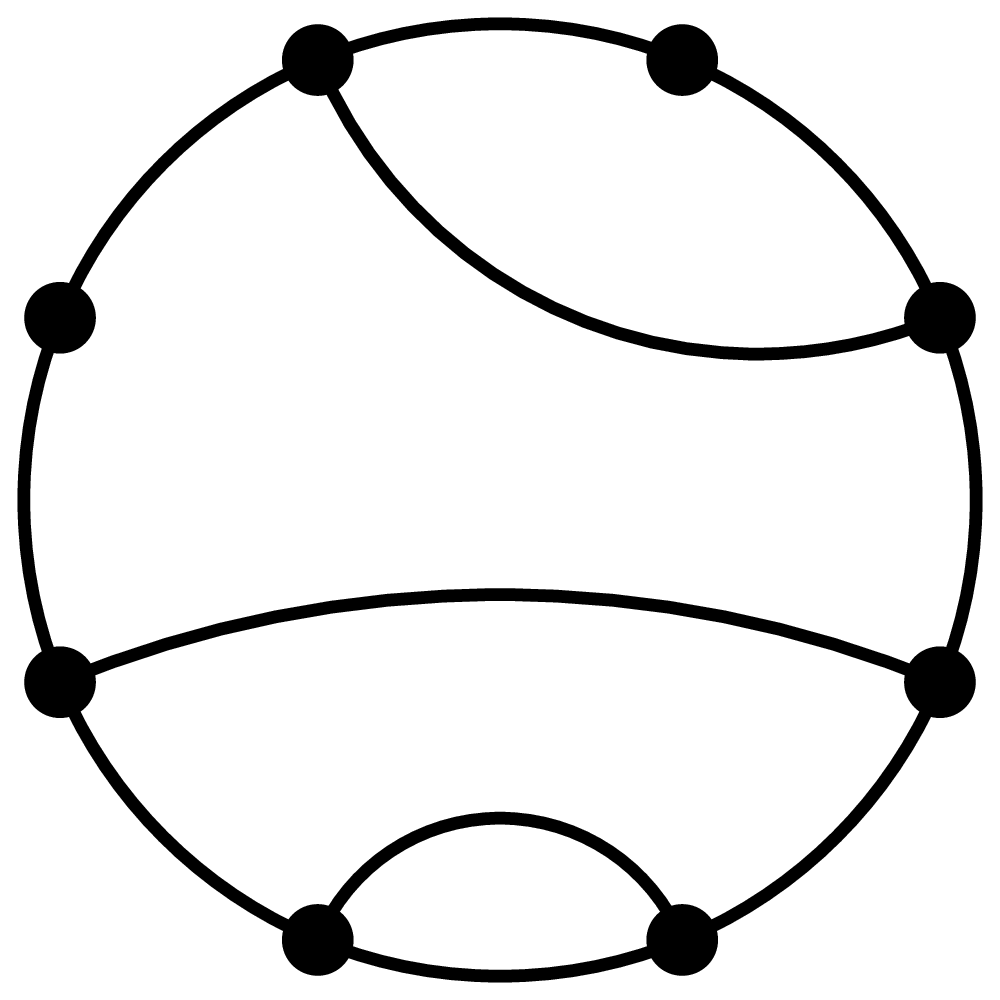}
{~,\ }\frac2{21}\diagram{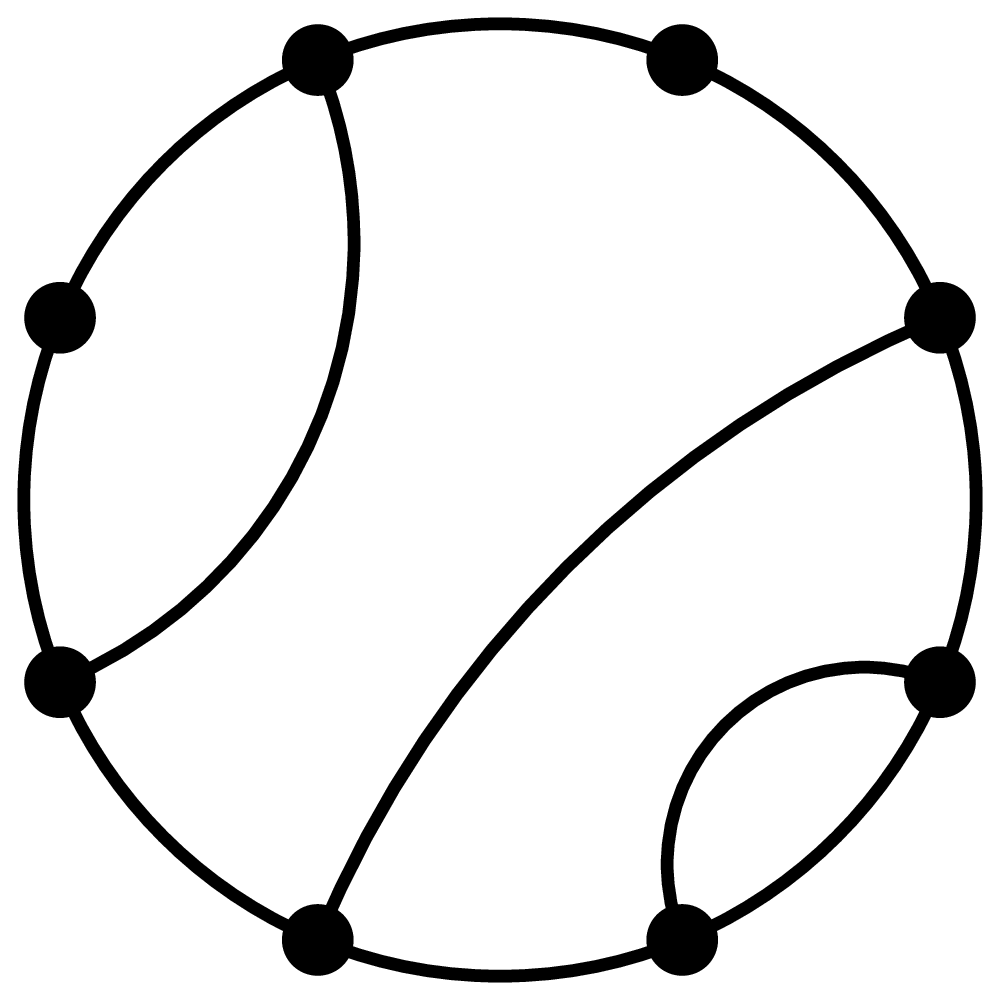}~,\frac{16}{315}\diagram{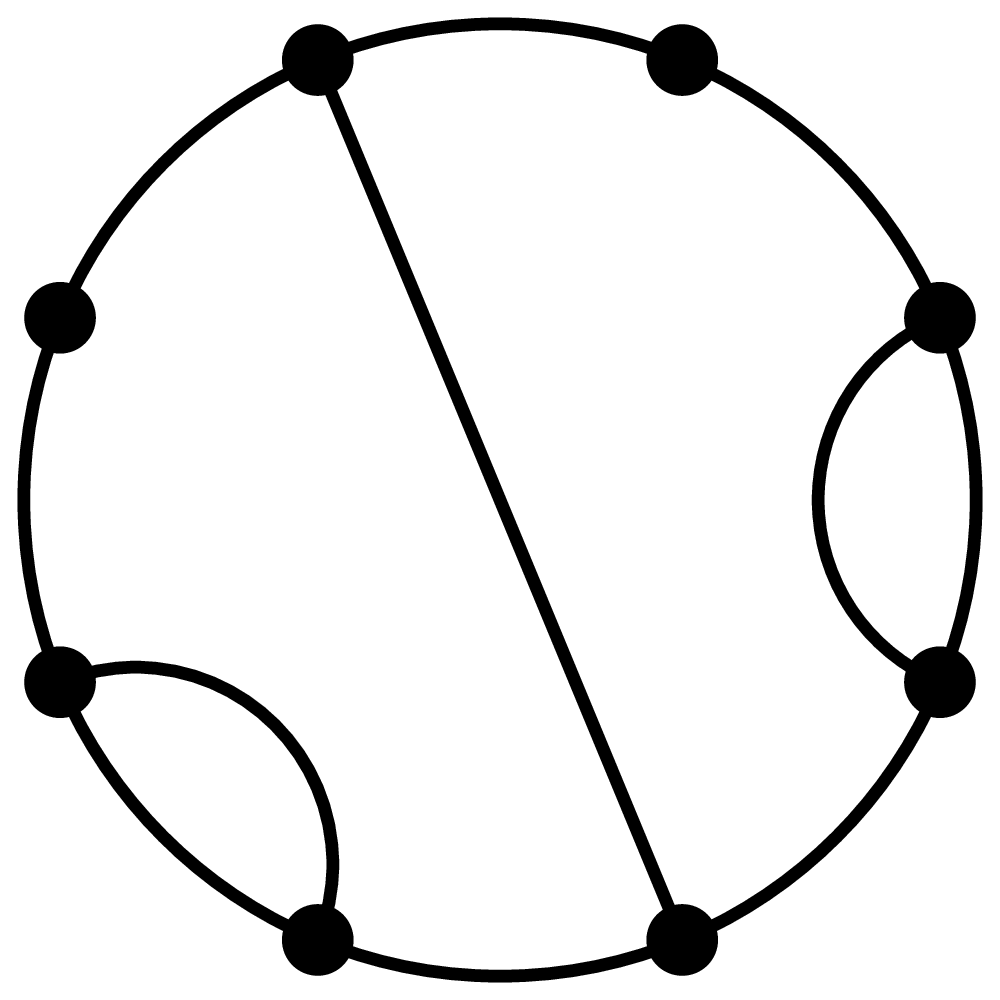} {~,\ }\frac4{63}\diagram{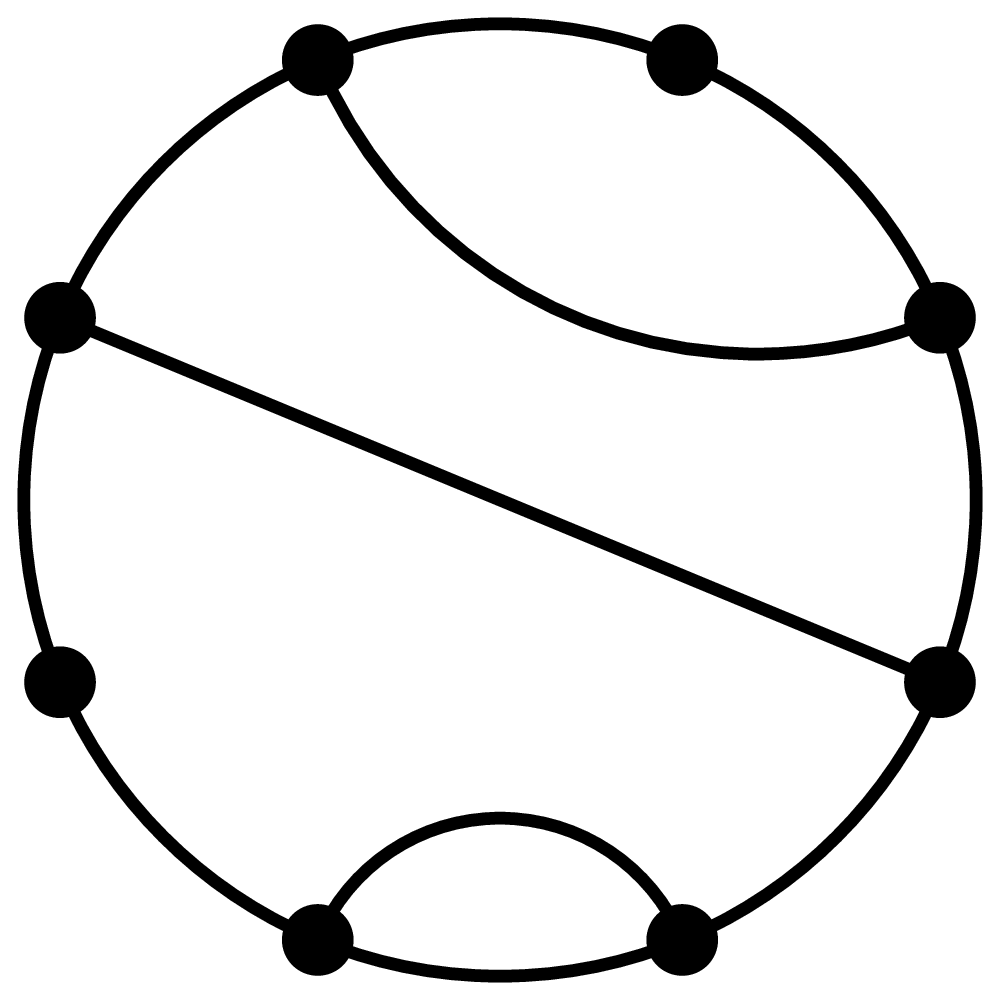}
{~,\ }\frac4{105}\diagram{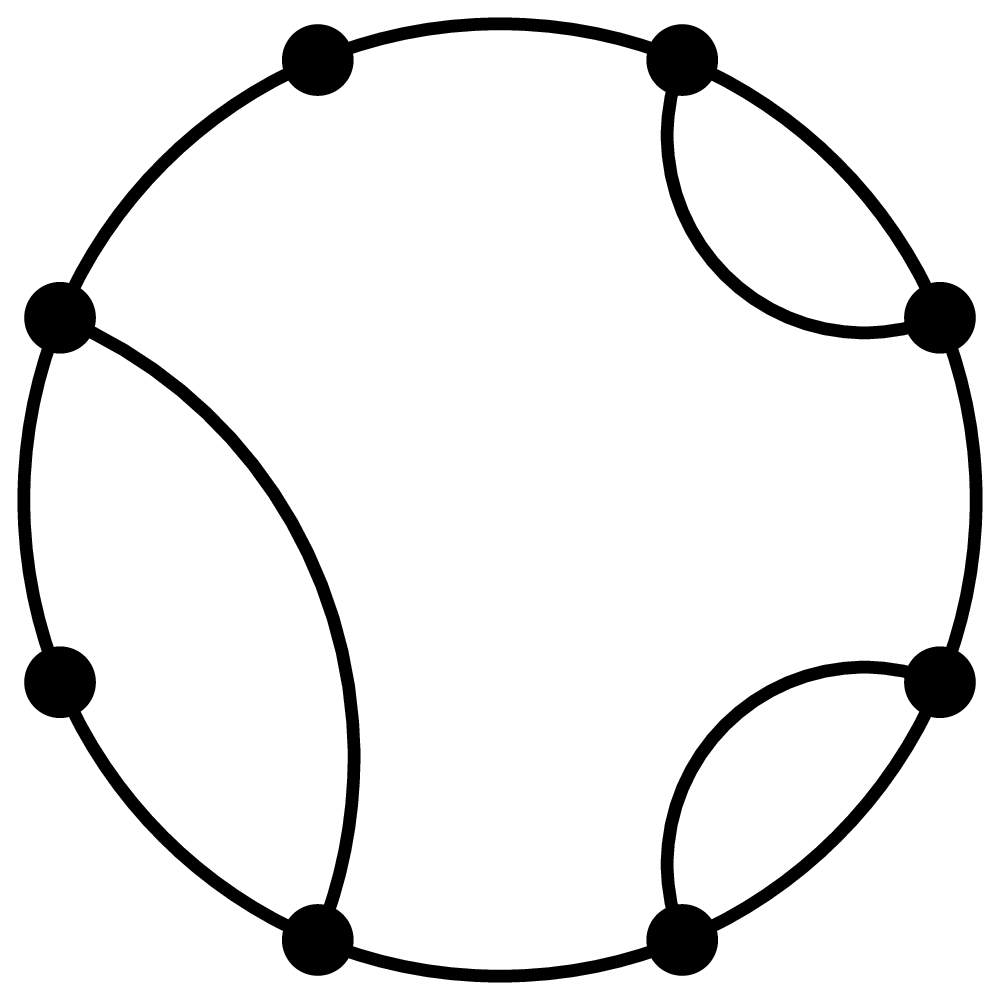}
\Bigg\}\nonumber
\end{eqnarray}
\begin{eqnarray}
\fl{\cal C}_{9}=&%{~,\ }
\Bigg\{
\frac{5}{63}\diagram{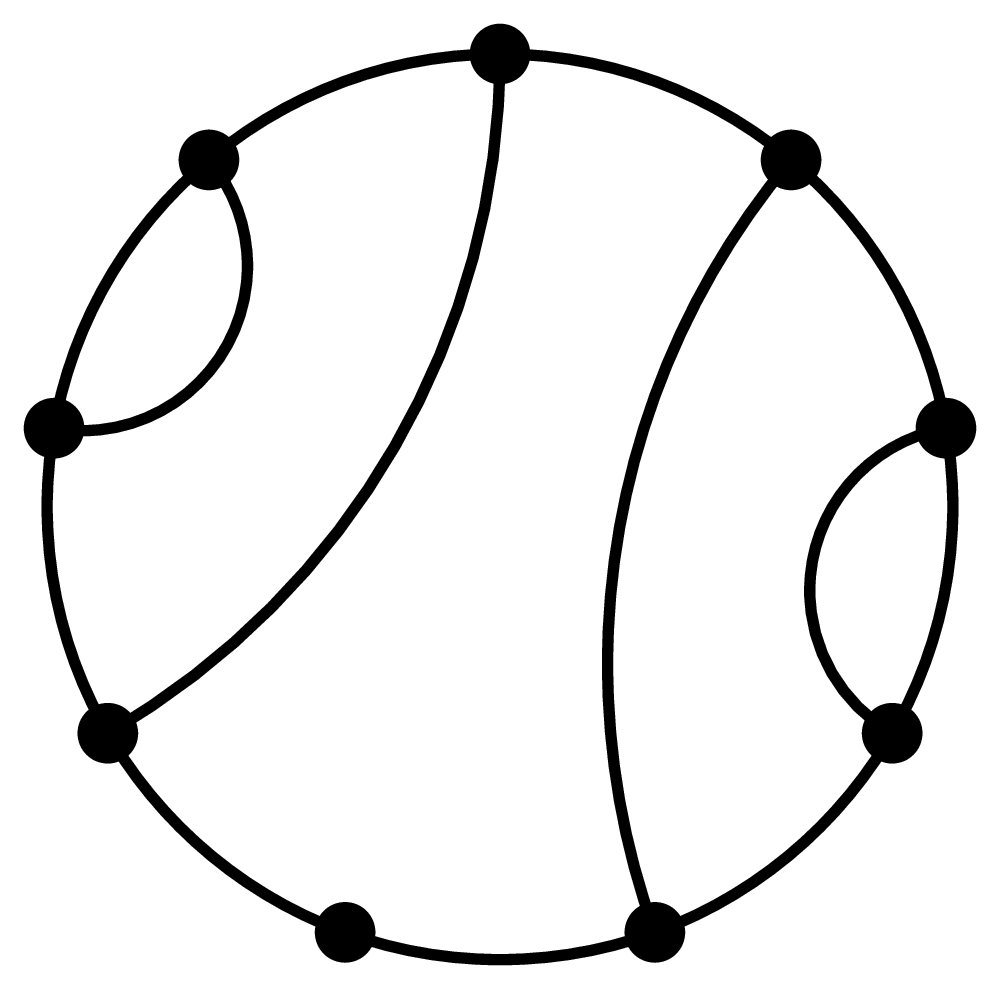}{~,\ }\frac{7}{90}\diagram{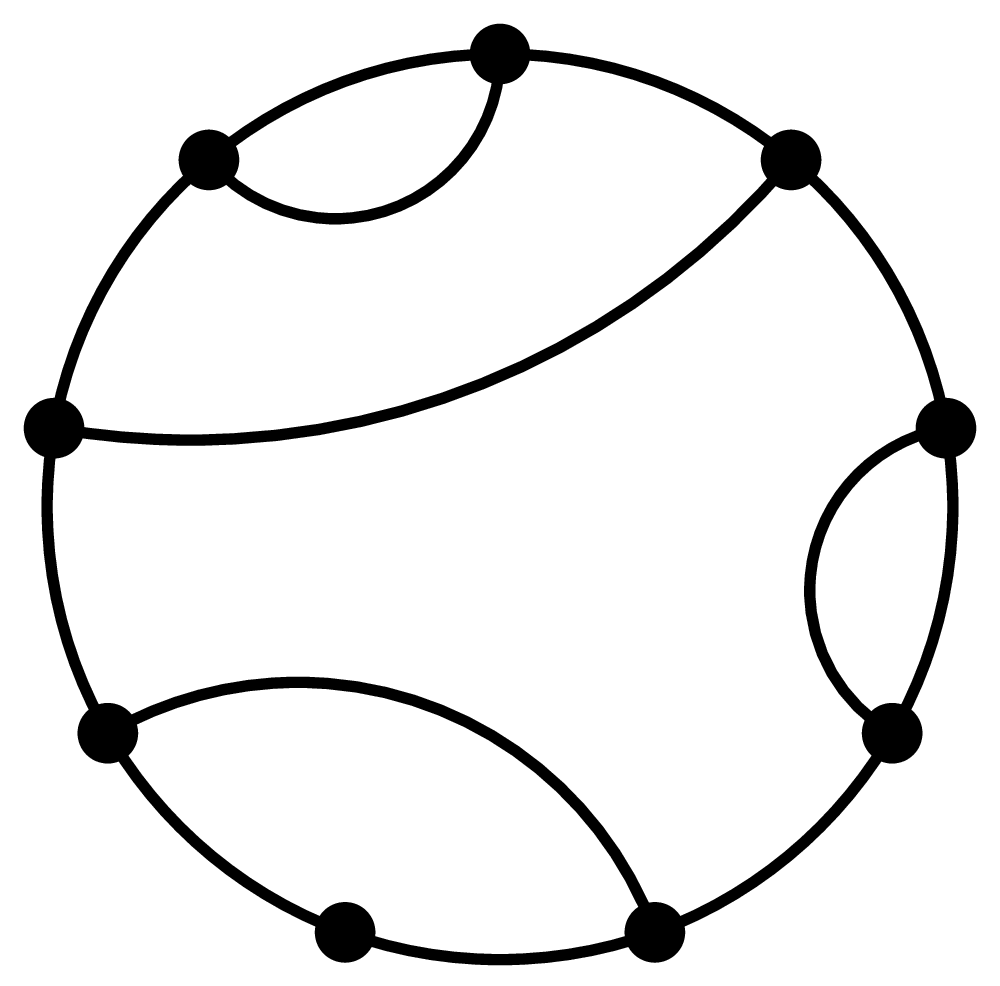}
{~,\ }\frac{2}{63}\diagram{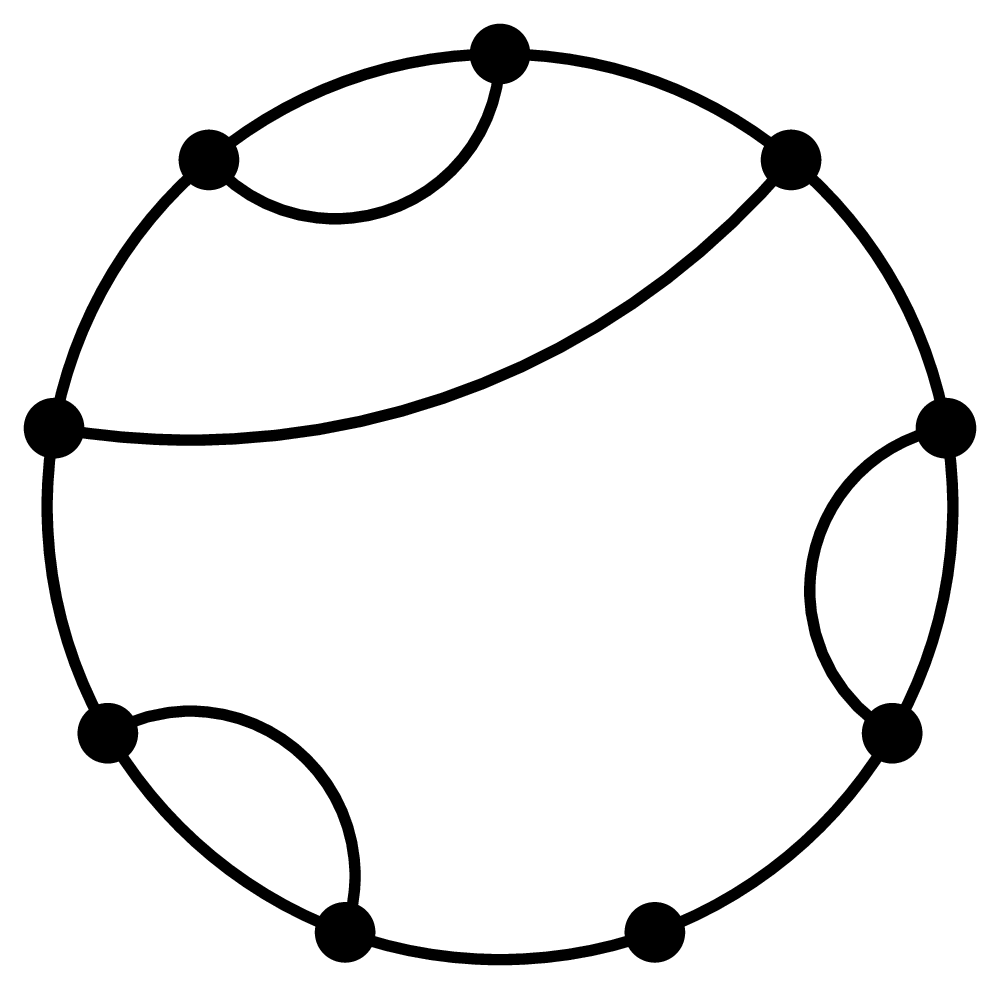}{~,\ }\frac{1}{105}\diagram{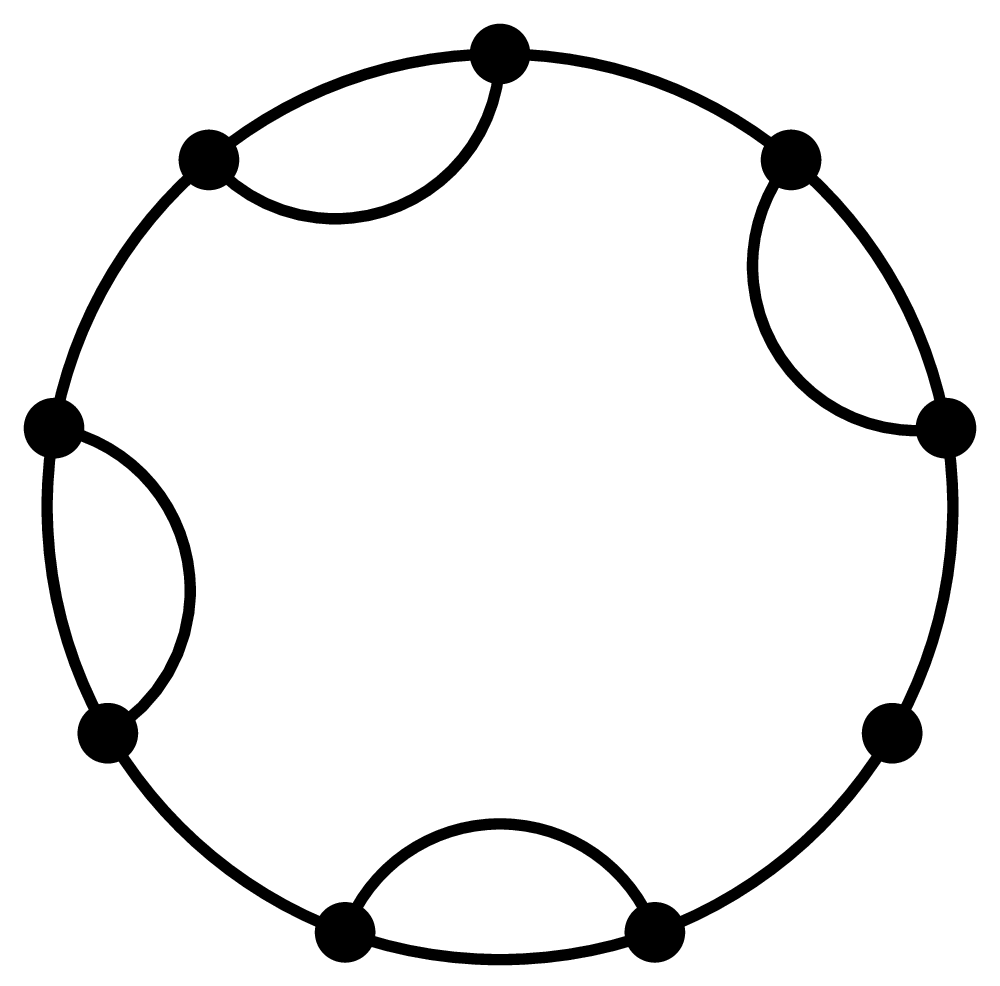}
{~,\ }\frac{4}{63}\diagram{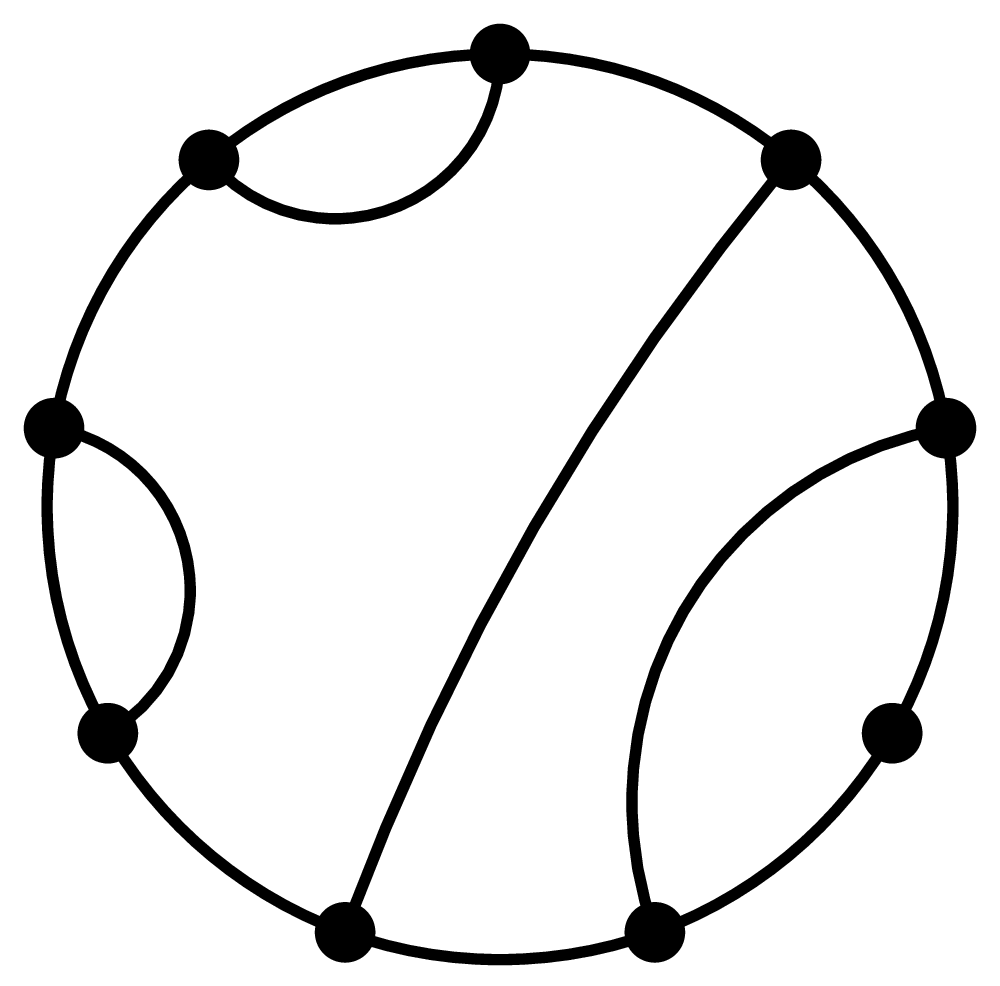}{~,\ }\nonumber\\
&\frac{7}{90}\diagram{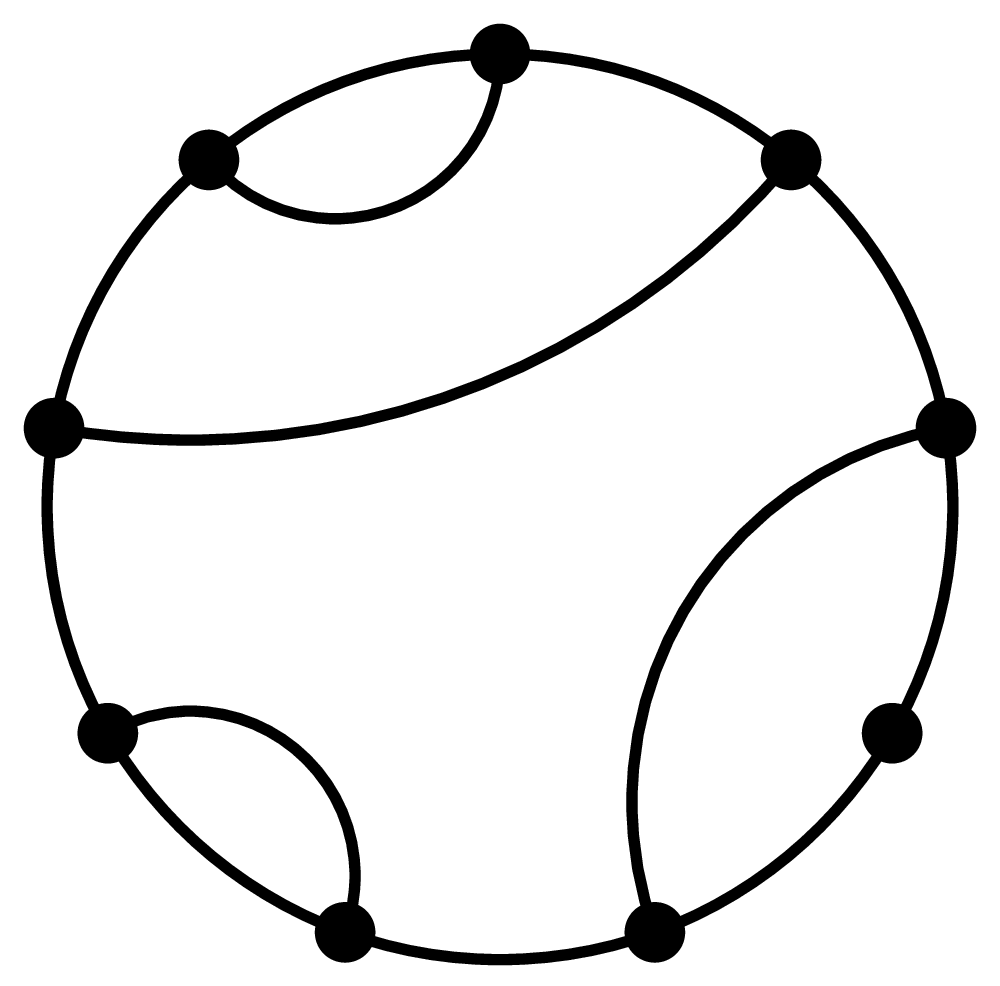}~,
\frac{1}{21}\diagram{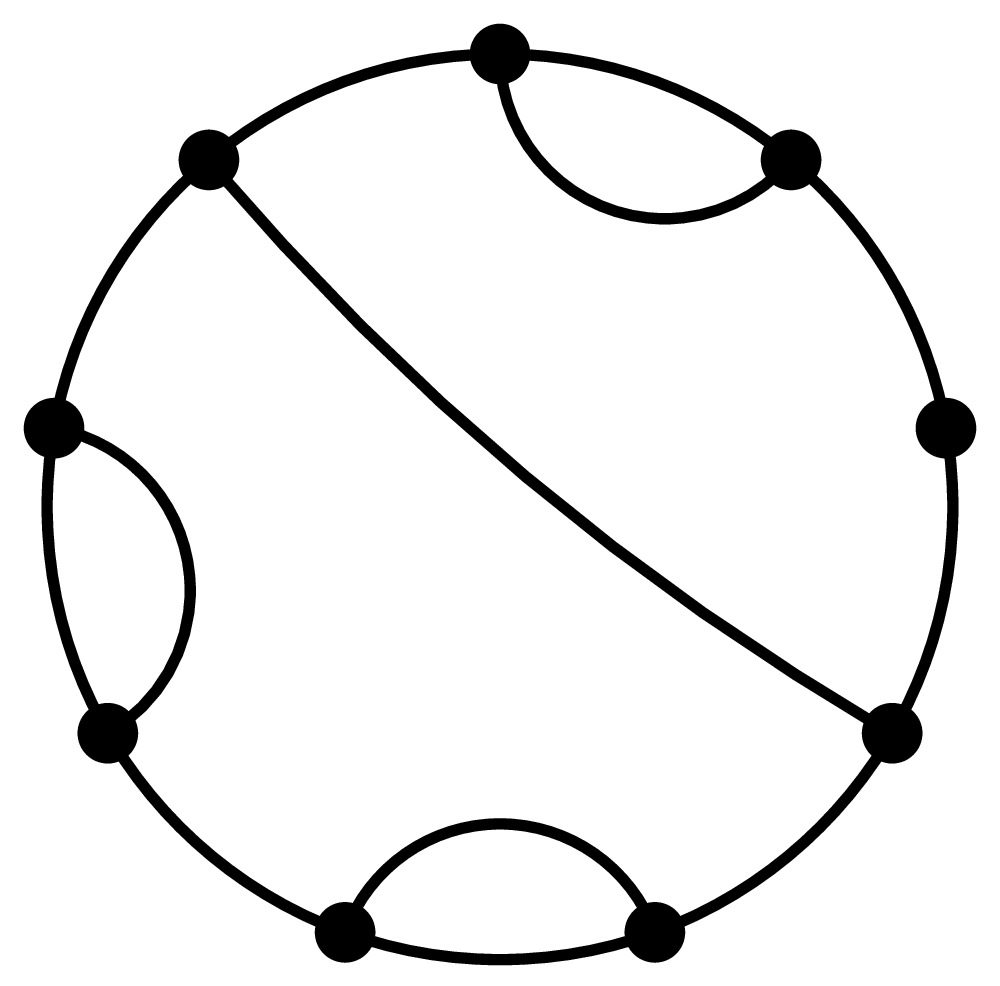}{~,\ }\frac{31}{315}\diagram{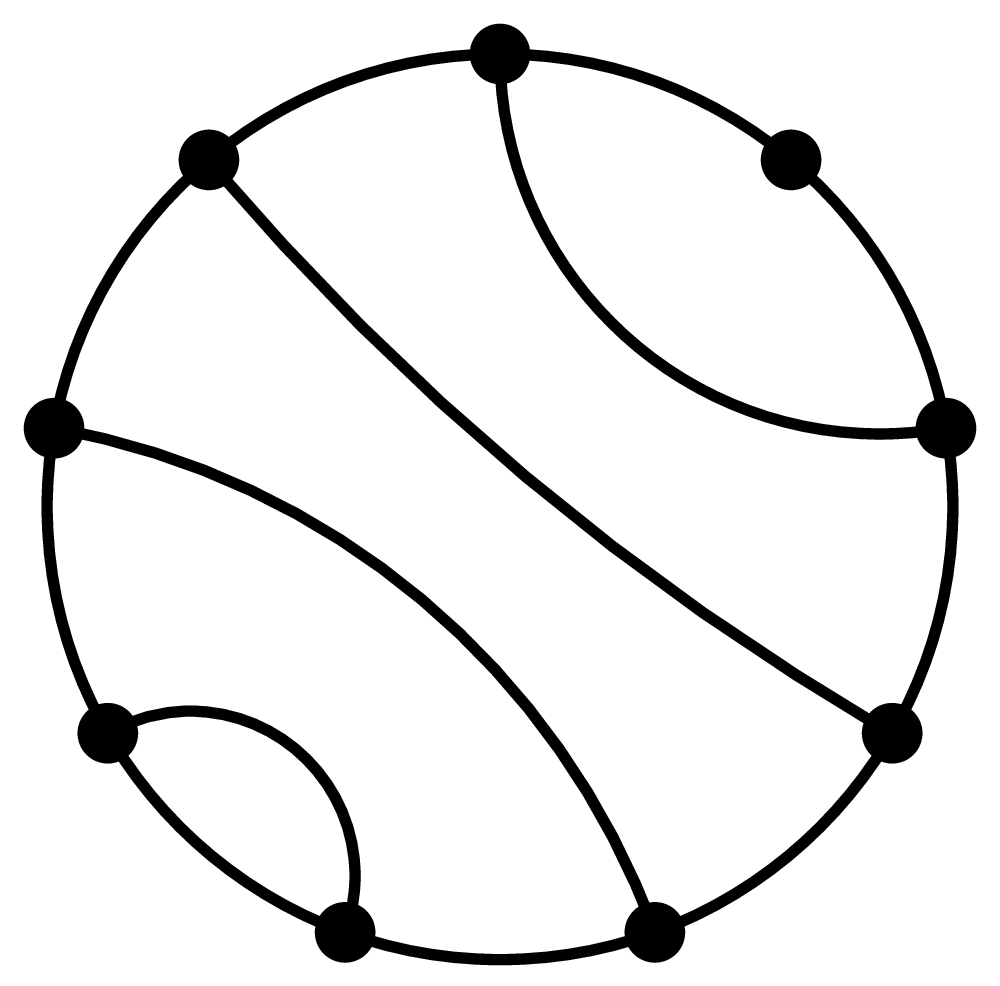}
{~,\ }\frac{1}{21}\diagram{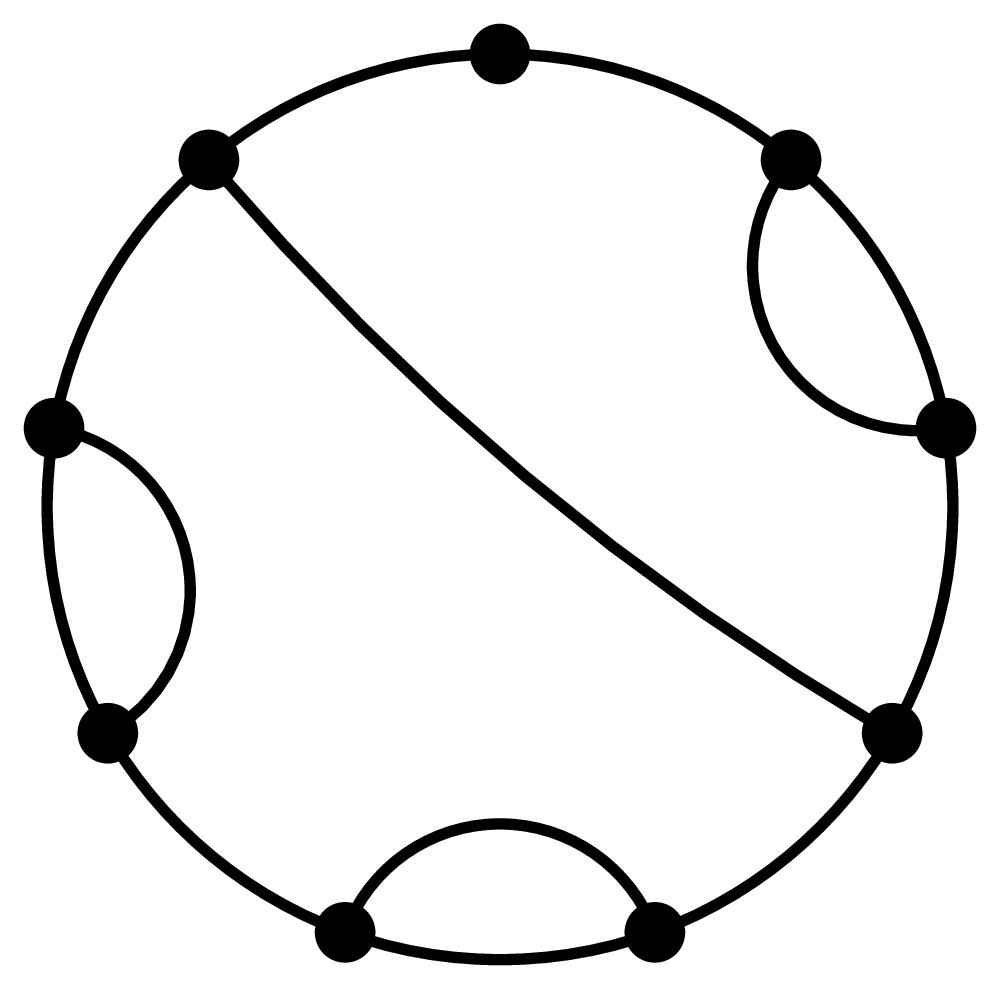}{~,\ }\frac{19}{315}\diagram{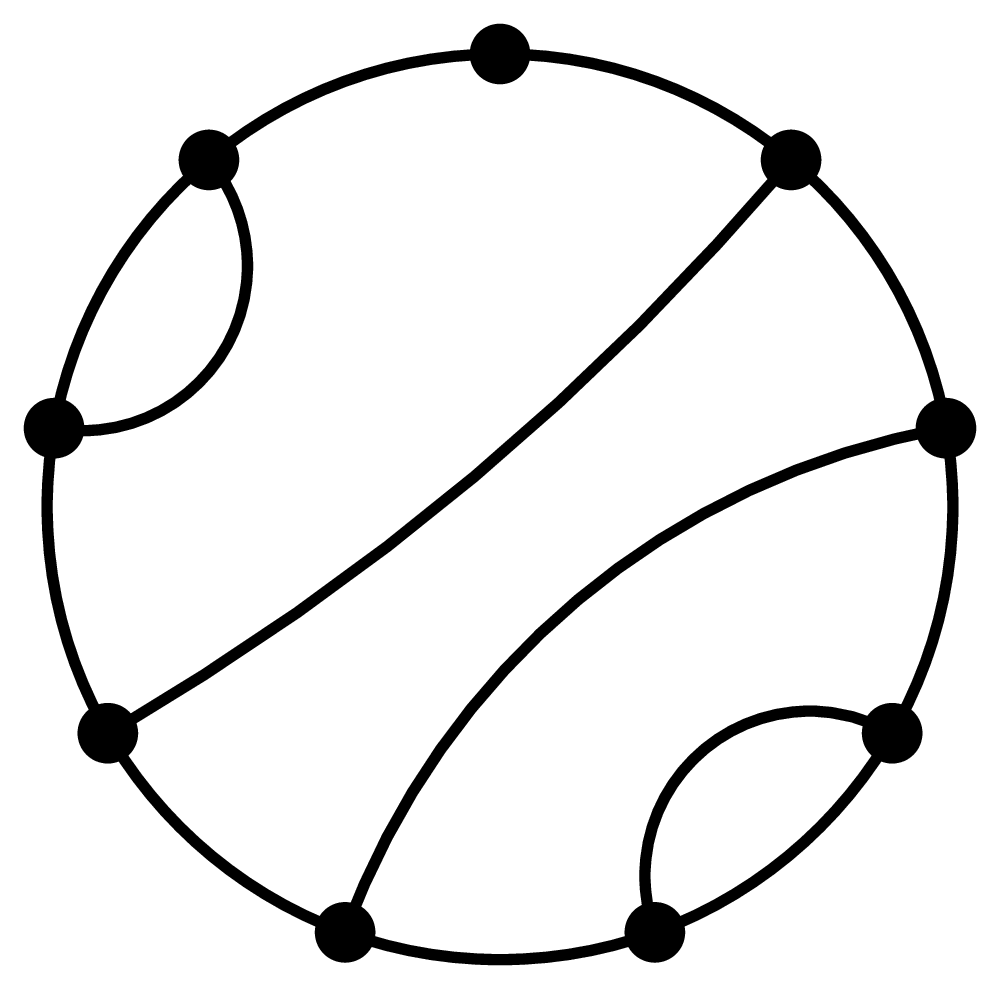}
{~,\ }\nonumber\\
&
\hphantom{~,\ }\frac{2}{63}\diagram{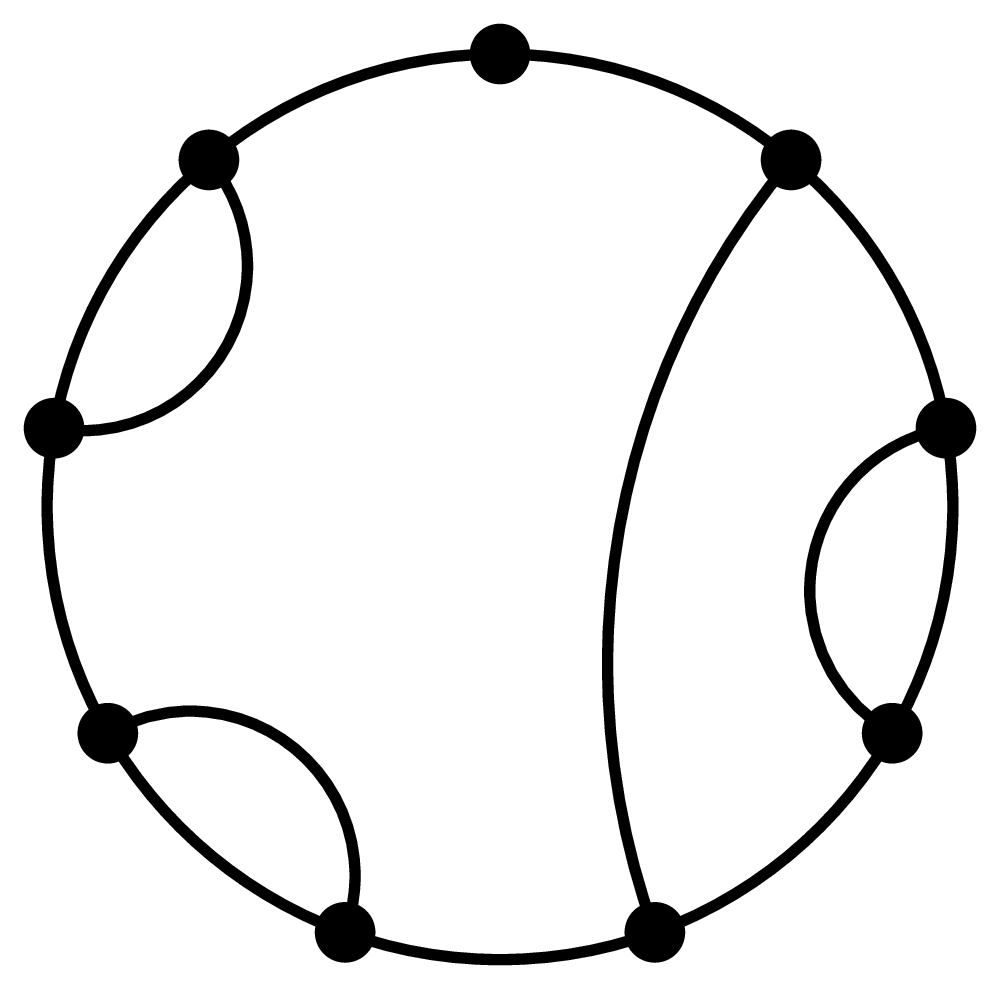}{~,\ }\frac{19}{315}\diagram{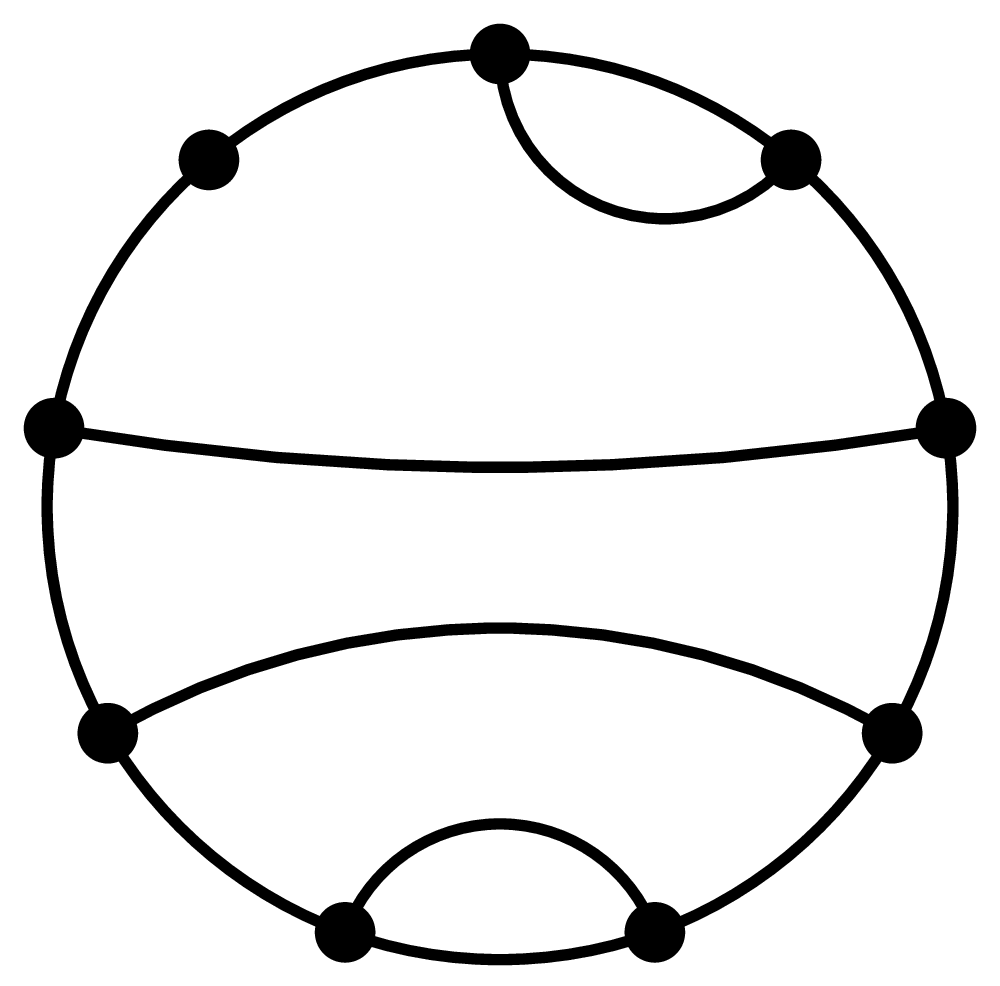}~,\frac{4}{105}\diagram{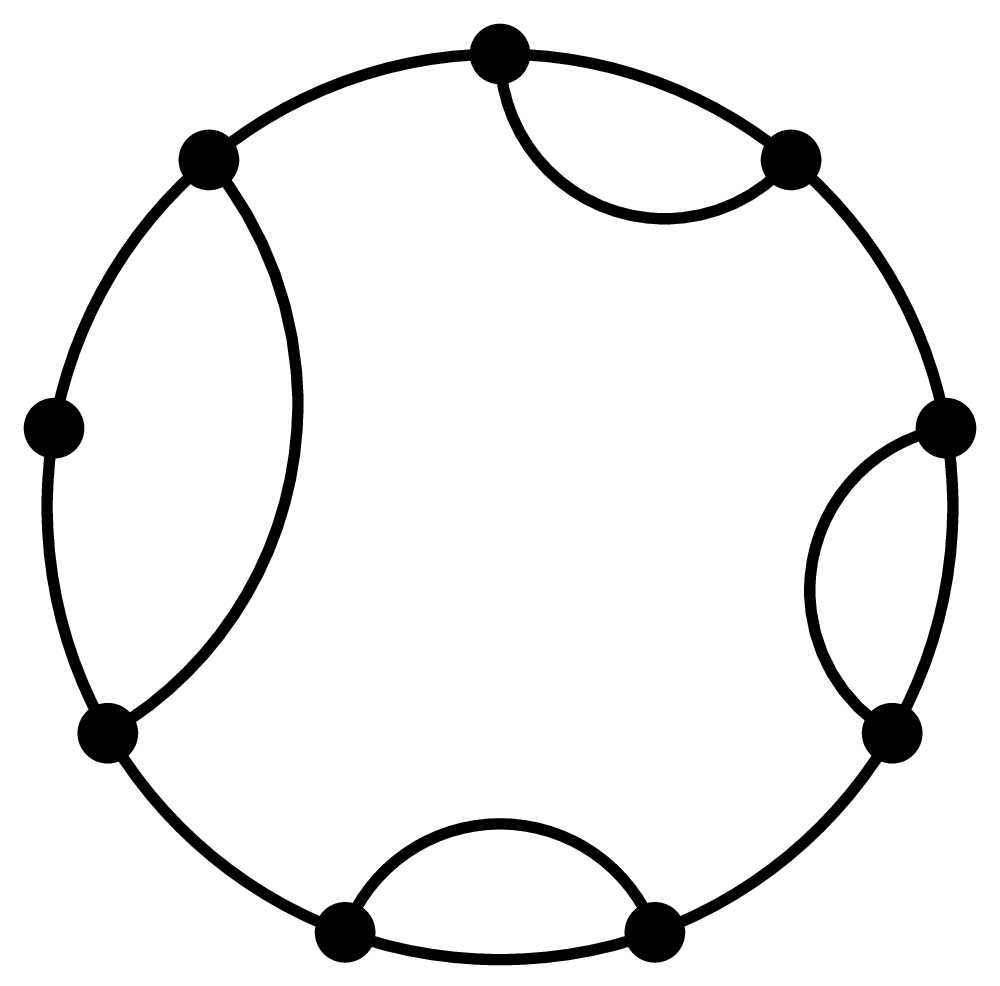}{~,\ }\frac{2}{63}\diagram{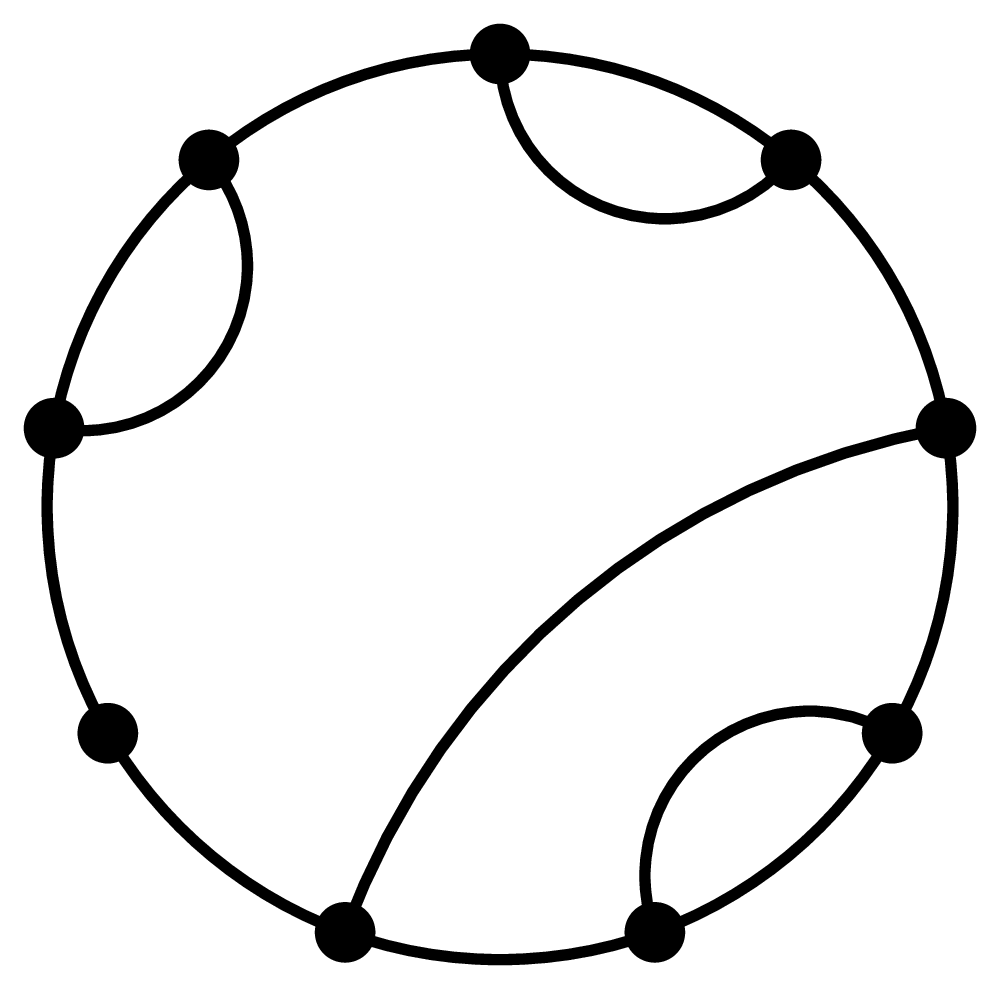}
{~,\ }\frac{13}{630}\diagram{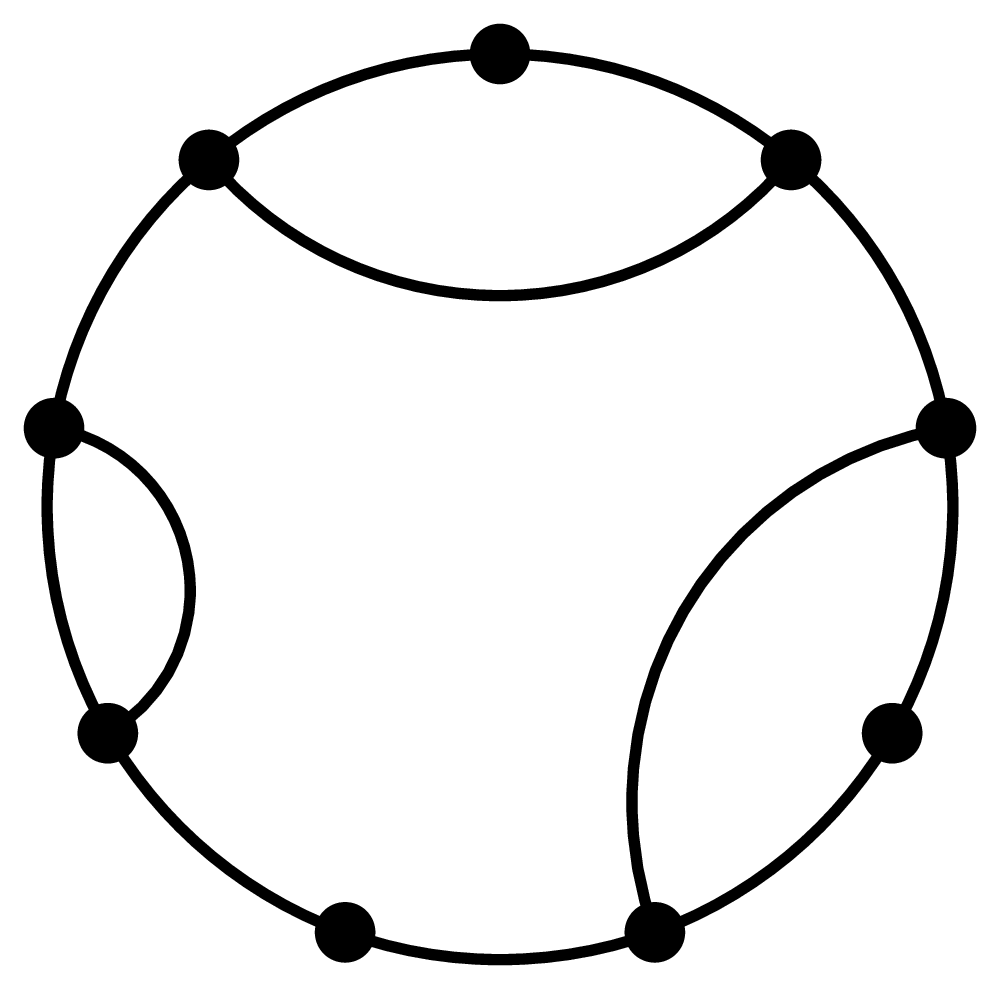}{~,\ }\nonumber\\
&\hphantom{~,\ }\frac{11}{420}\diagram{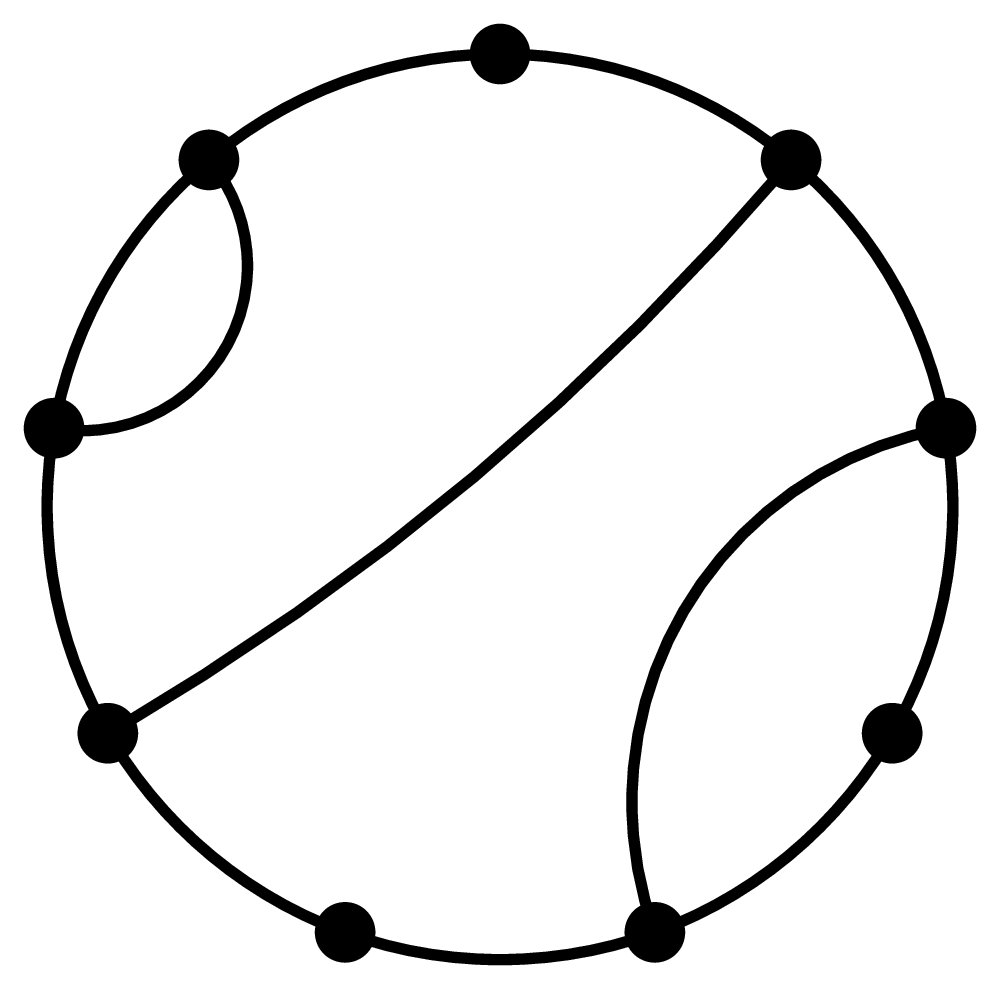}
{~,\ }\frac{13}{630}\diagram{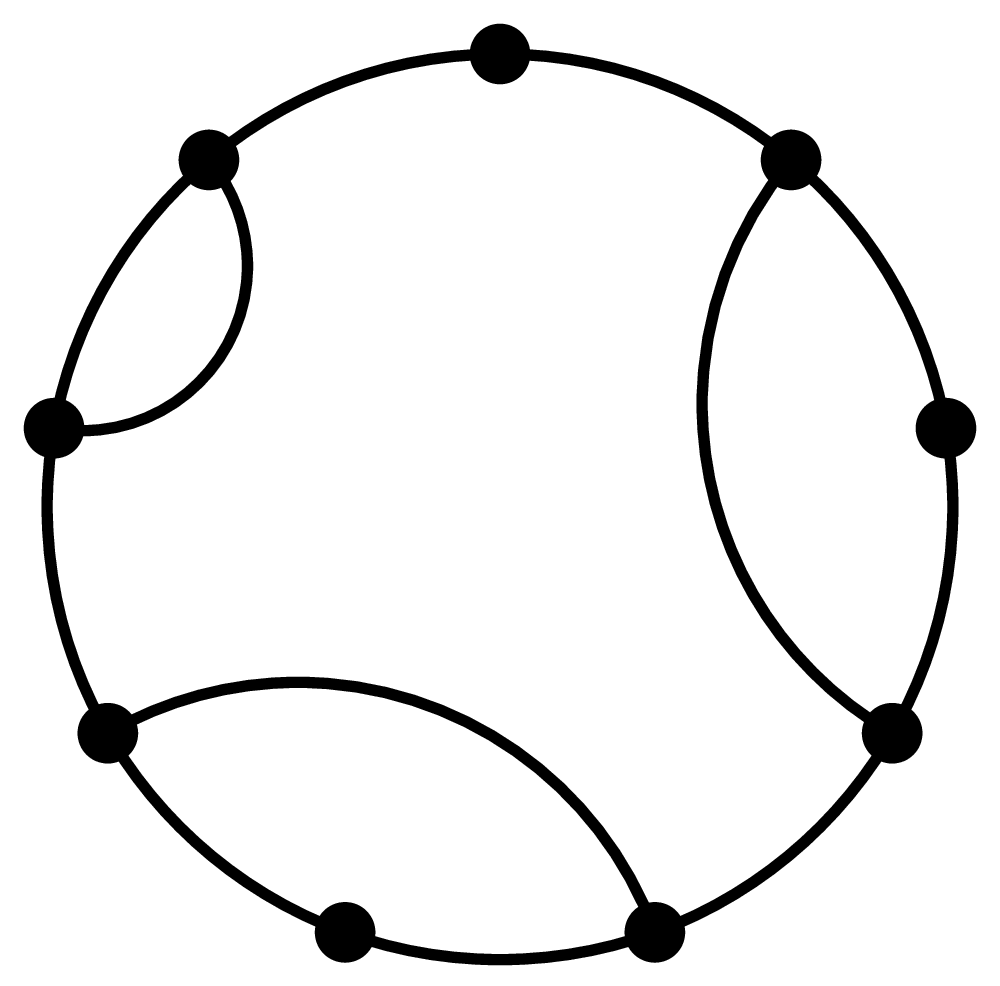}{~,\ }\frac{1}{30}\diagram{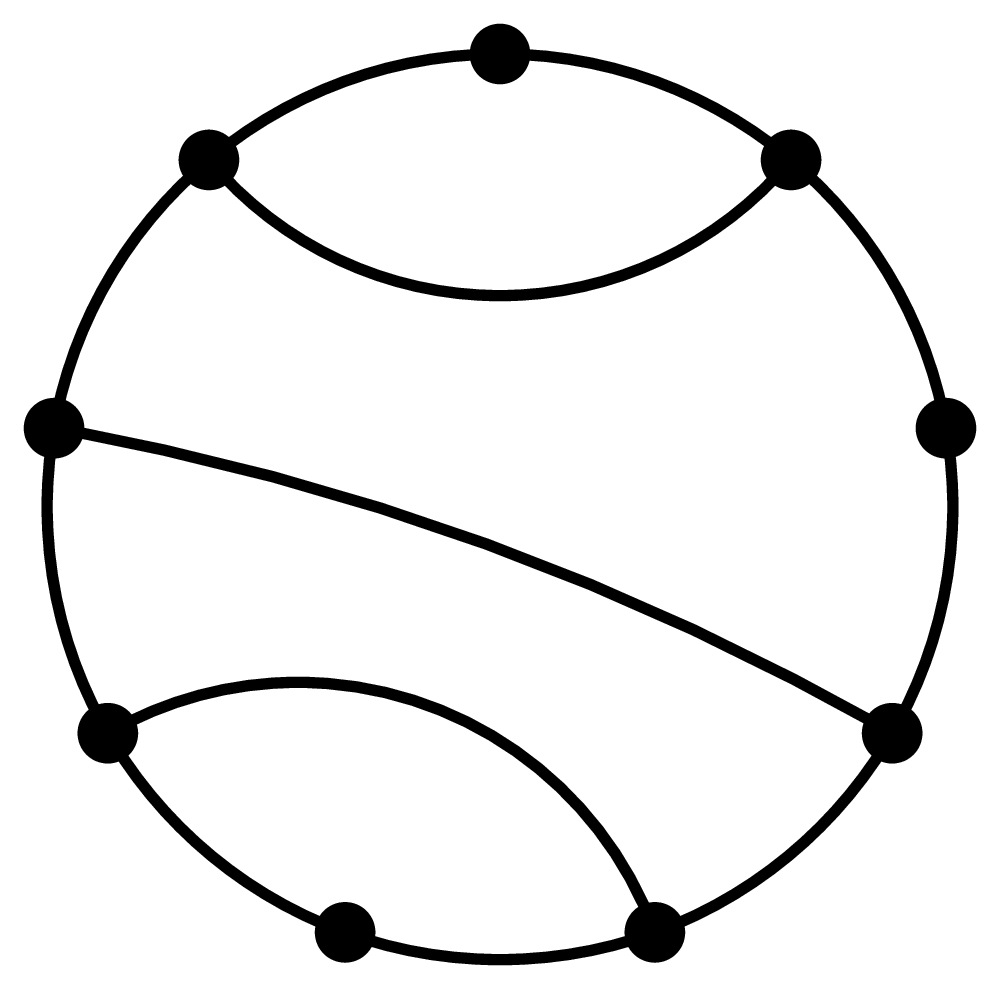}{~,\ }\frac{1}{90}\diagram{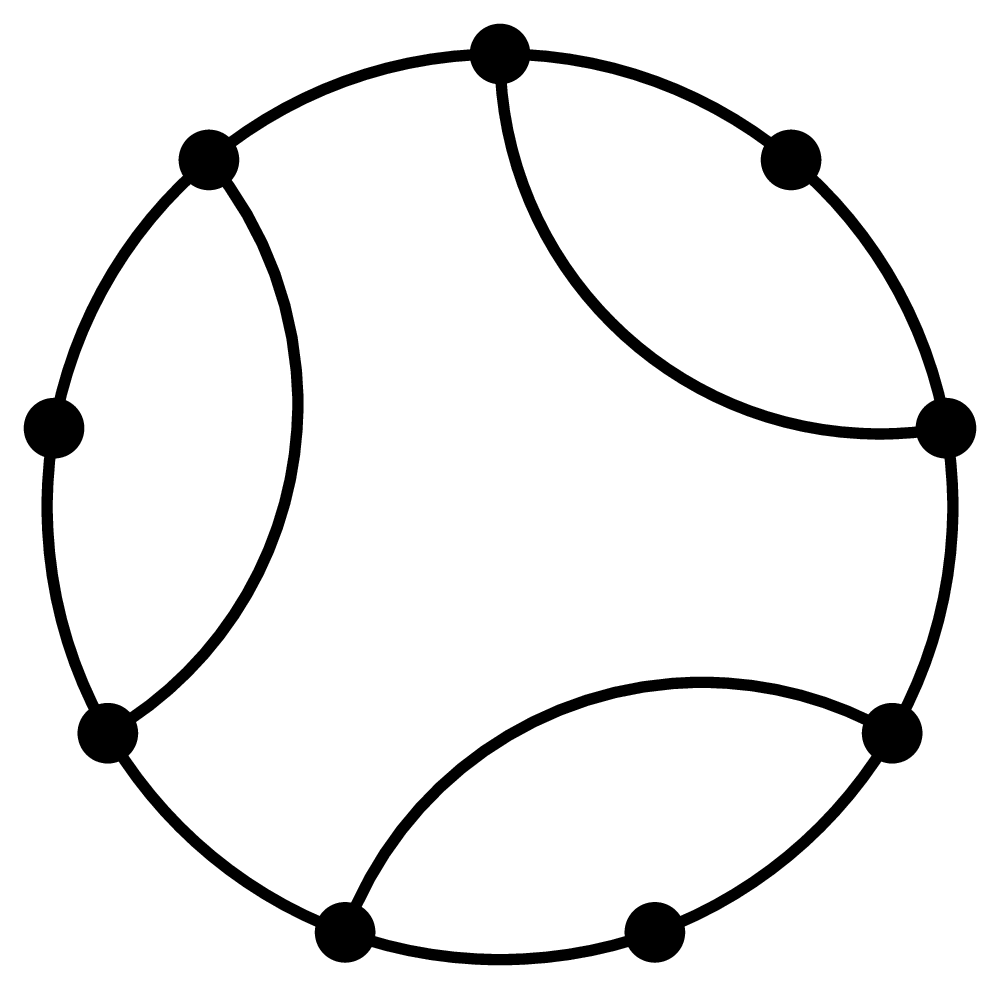}{~,\ }\frac{1}{30}\diagram{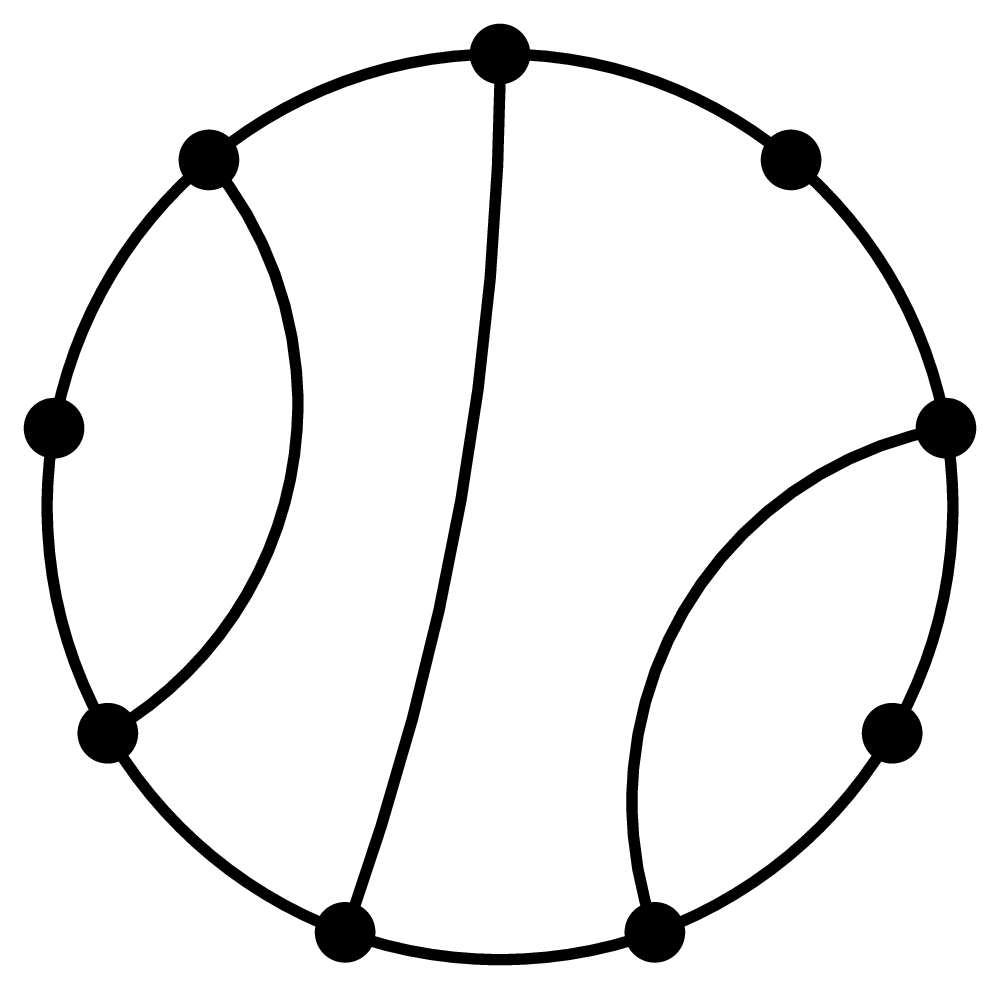}{~,\ }\nonumber\\
&\hphantom{~,\ }
\frac{11}{420}\diagram{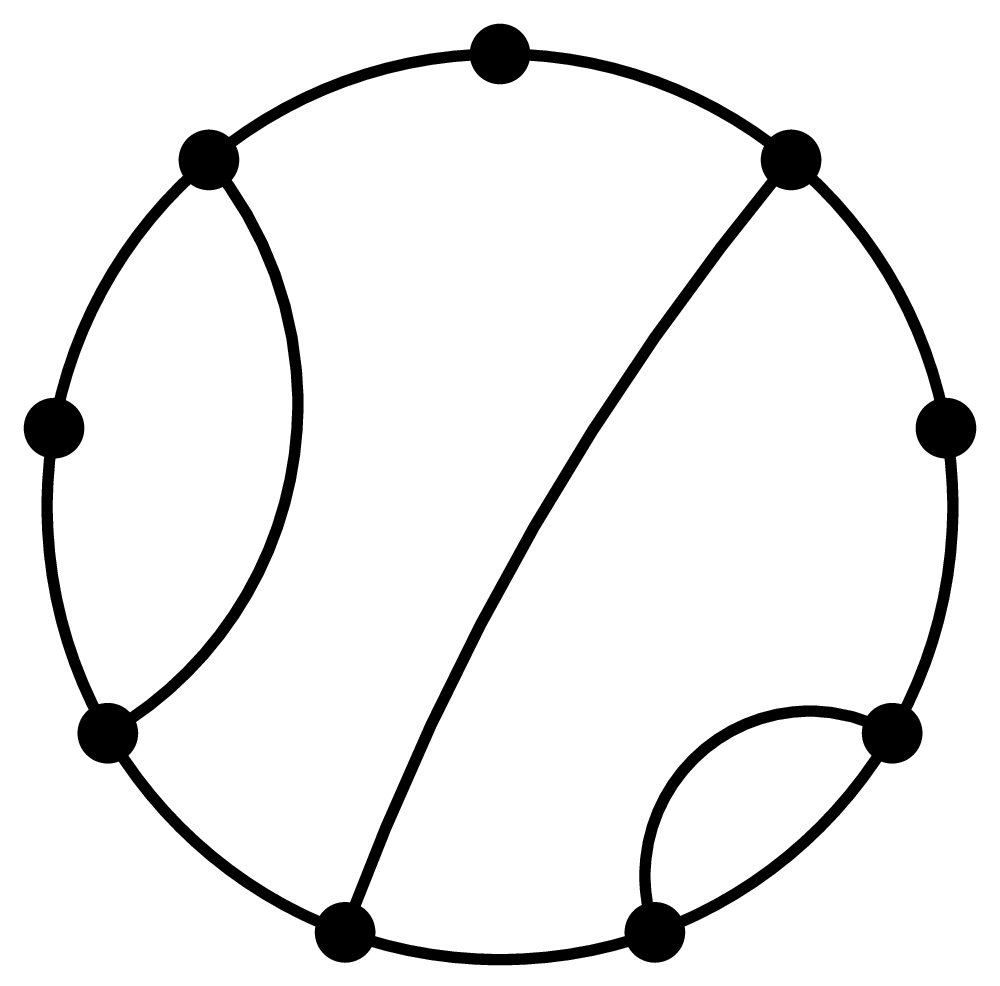}{~,\ }\frac{11}{420}\diagram{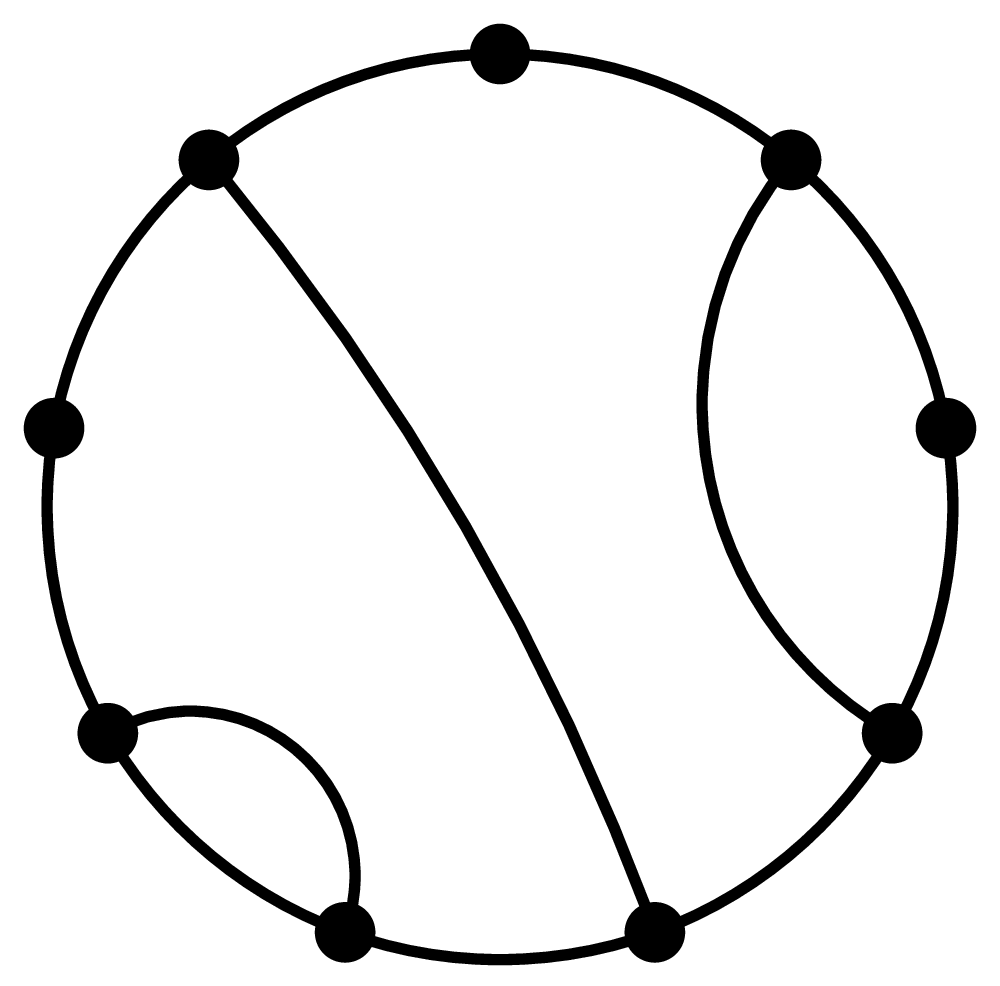}
{~,\ }\frac{11}{420}\diagram{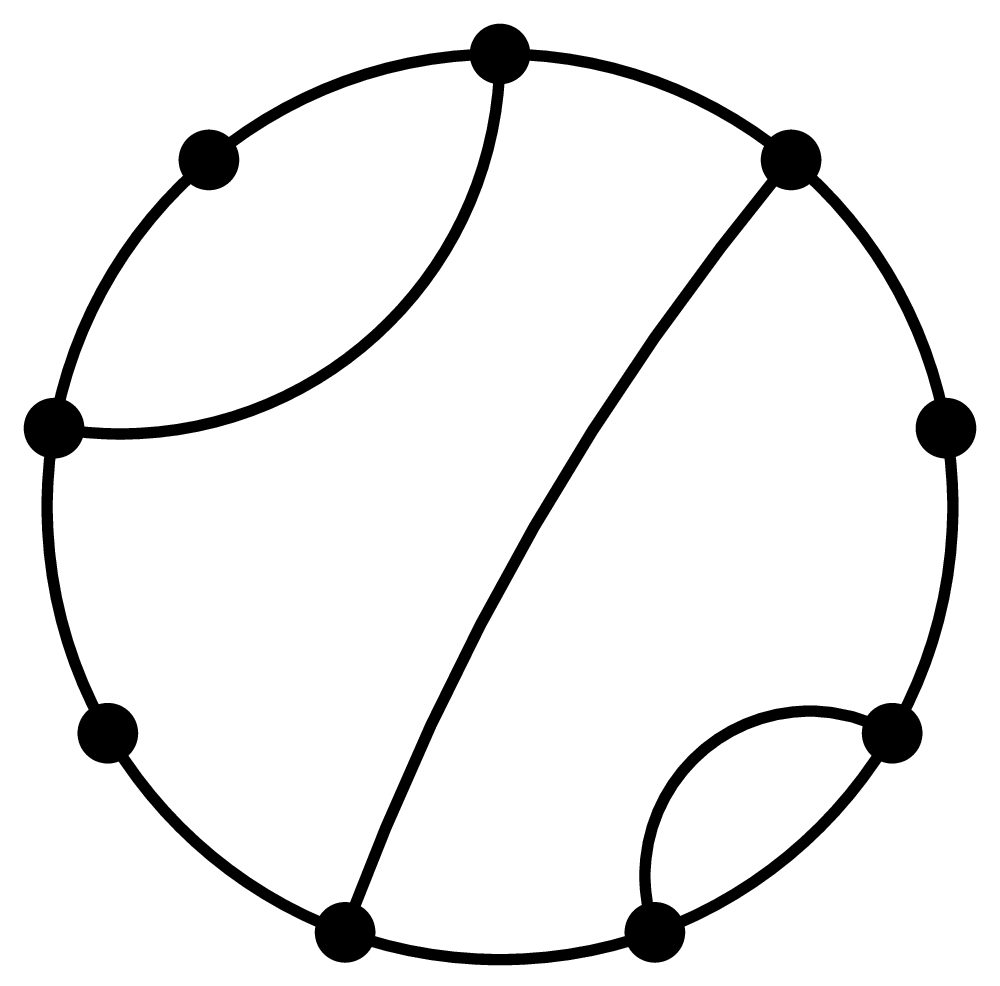}{~,\ }\frac{13}{630} \diagram{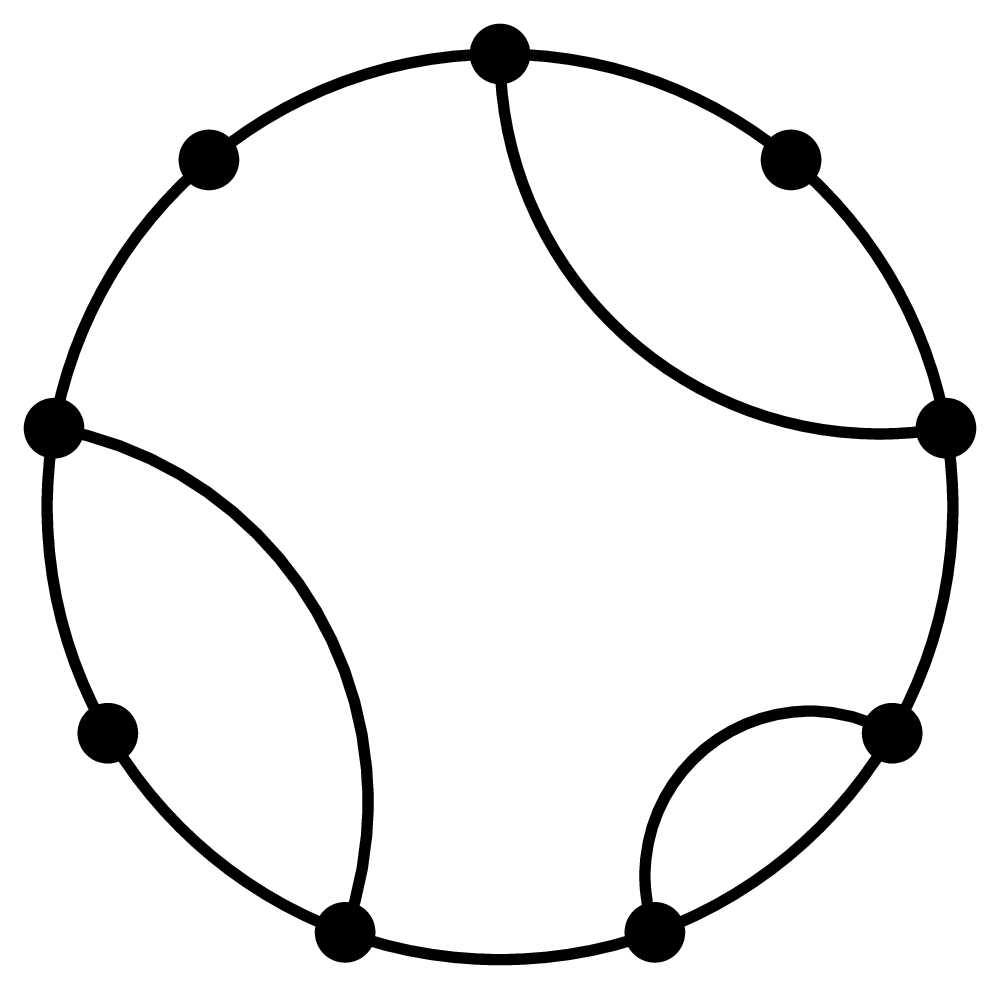}
\Bigg\}
\nonumber\\
\end{eqnarray}

\subsubsection{Open-strand growth model \textbf{G'}}
\label{sss-G'}
Let us now define a slightly different model \textbf{G'} (open
model).

At time $t=0$ we start from an open line with no point. At time
$t=1$ we deposit a point on the line. At time $t$, we assume that we
have already deposited $t$ points on the line and constructed a
maximal planar arch system between these points. Namely there are
$n_a(t)$ arches and $n_f(t)=t-2n_a(t)$ free points such that any
link between two of these free points necessarily intersect one of
the $n_a(t)$ existing arches. At time $t+1$ we then deposit a
$(t+1)^{\mathrm{th}}$ point, with equiprobability $1/(t+1)$ on the
$t+1$ intervals separated by  the $t$ already deposited points. If
it is possible to draw a planar arch between this last point and one
of the free points  we add this arch, otherwise the new point stays
free.

Note that model \textbf{G'} can also be viewed as model \textbf{G} with an additional inactive point, marking the cut. 

\begin{figure}[h]
\begin{center}
\includegraphics[width=0.95\textwidth]{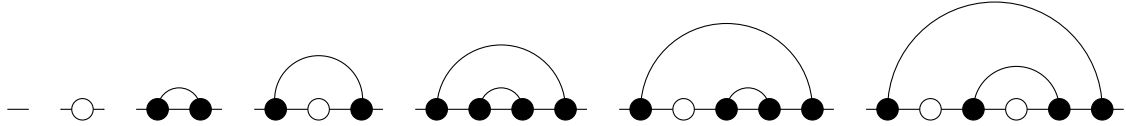}
\caption{Model \textbf{G'}: successive deposition of points on an
open line. Unlinked points are marked in white, linked points in black.} \label{k2}
\end{center}
\end{figure}

\subsubsection{Relation between model \textbf{G} and model \textbf{G'}}
\label{sss-G'=G}
\begin{figure}[h]
\begin{center}
\includegraphics[width=0.95\textwidth]{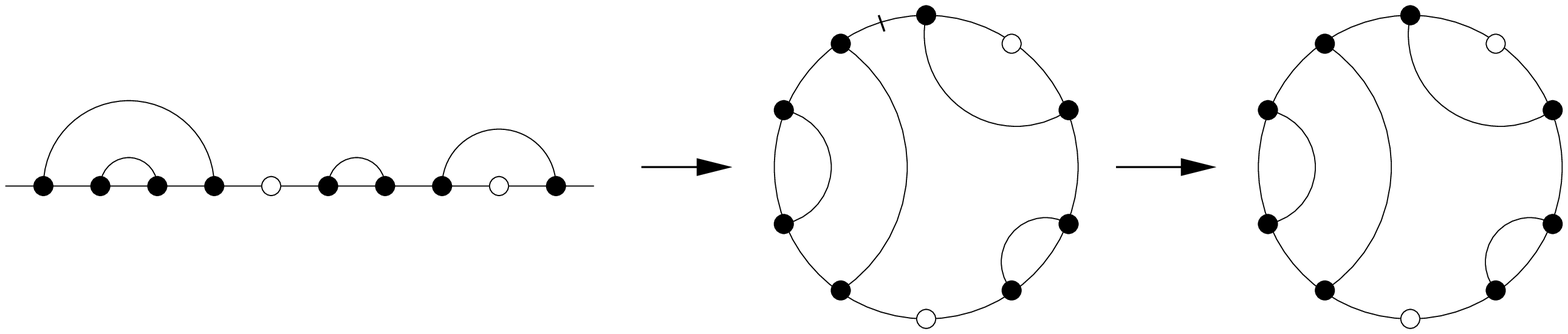}
\caption{From a configuration $\mathcal{C}$ of \textbf{G'} to a
configuration $\overline{\mathcal{C}}$ of \textbf{G}}
\label{arches-to-circles}
\end{center}
\end{figure}

It is clear that a configuration $\overline{\mathcal{C}}$ of
model \textbf{G} can be obtained from a configuration $\mathcal{C}$
of model \textbf{G'} by closing the line and that all configurations
$\mathcal{C}$ that are equivalent by a discrete rotation give the
same $\overline{\mathcal{C}}$ (see figure \ref{arches-to-circles}). In
other words, the configurations $\overline{\mathcal{C}}$ of model
\textbf{G} are the $\mathbb{Z}_n$-orbits of the configuration space
of \textbf{G'} under the action of discrete rotations.

\subsection{Equivalence between the growth model \textbf{G'} and the deposition model \textbf{A}}
\label{ss-G'=A}
It is clear that the arch configurations $\mathcal{C}$ of model
\textbf{G'} are the same as the arch configurations of model \textbf{A}.
It is less obvious that the probability for each configuration in both models are the same.
\medskip

\noindent\textbf{Theorem:} The probability $P(\mathcal{C})$ of any
configuration  (i.e.\ class of diagrams) $\mathcal{C}$ in models
\textbf{G'} and \textbf{A} are the same.
\begin{equation}
\label{k3} P_{A}(\mathcal{C})=P_{G'}(\mathcal{C}) \ .
\end{equation}

To prove the theorem we start from the recursion equation (\ref{eqn:decomp}) for
the configuration probabilities in the arch-deposition model
\textbf{A}, that we obtained in section~\ref{archmodel.def}, and rewrite here for completeness 
\begin{equation}
\label{k4} P_A(\mathcal{C})=\sum_{\mathrm{arch}\,a\in \mathcal{C}}
{2\over n(n-1)}\, P_A(\mathcal{C}_1)\,P_A(\mathcal{C}_2)
\end{equation}
This recursion relation, together with the initial condition $P=1$
for the $n=0$ and the $n=1$ configurations (no point and  a single
free point), is sufficient to obtain all probabilities.

\begin{figure}[h]
\begin{center}
\includegraphics[width=0.9\textwidth]{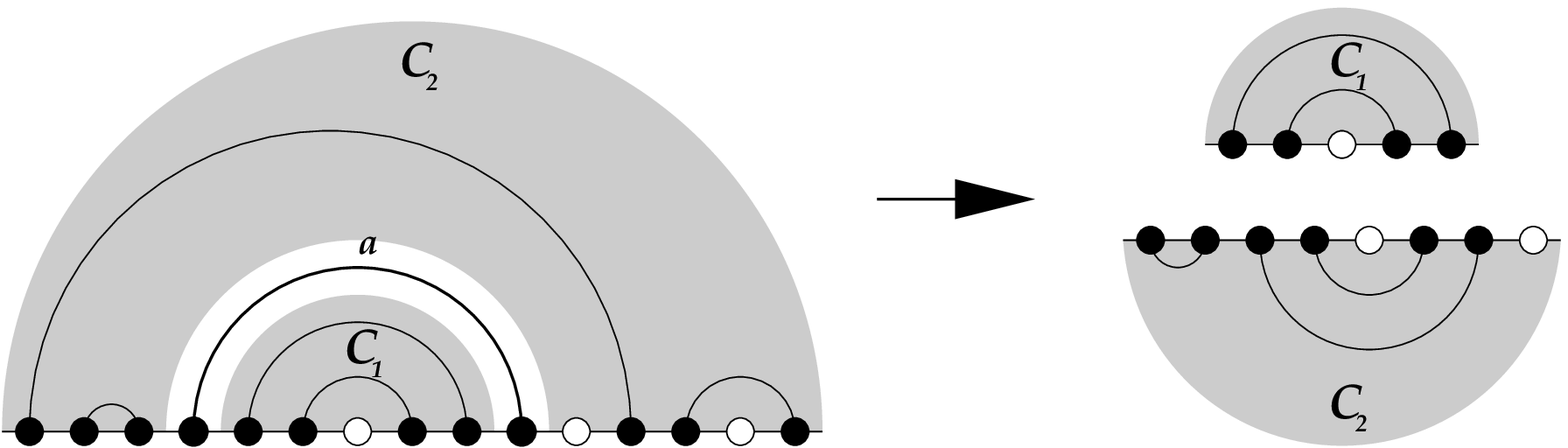}
\caption{Probability recursion (\ref{k4}) as decomposition of a configuration $\mathcal{C}$ in model
\textbf{A}} 
\label{decomposition-AA}
\end{center}
\end{figure}

We now prove that the probabilities in model \textbf{G'} obey the
same recursion relation. For this we first need to relate the
propabilities in model \textbf{G} to those in model \textbf{G'}.
\medskip

\noindent\textbf{Lemma:} Let $\mathcal{C}$ be a configuration with
$n$ points in model \textbf{G'} (successive deposition of points on
a line) and $\overline{\mathcal{C}}$ its equivalent configuration in
model \textbf{G} (successive point depositions on a circle). Let
$s(\overline{\mathcal{C}})$ be the symmetry factor of the
configuration $\overline{\mathcal{C}}$, i.e. the number of cyclic
rotations that leave
$\overline{\mathcal{C}}$ invariant. Then
\begin{equation}
\label{lemma1}
P_{\textbf{G'}}(\mathcal{C})={s(\overline{\mathcal{C}})\over
n}\,P_{\textbf{G}}(\overline{\mathcal{C}})
\end{equation}
\noindent\textbf{Proof of the lemma:} It is clear that in model
\textbf{G'} any deposition process of $n$ points on the line
is uniquely specified by the bijection $i\to x(i)$ where $x(i)$ is
the position at time $t=n$ of the point deposited at time $i$. $x$
is a bijection on $\{1,n\}$, i.e. a permutation. Any process is
equiprobable, therefore the probability for any $x$ is $p(x)=1/n!$.
It is equivalent to successively create the arches as soon as this
is possible, or to create all the arches at time $n$, with the
constraint that any point $x(i)$ can only be connected to the points
$x(j)$ with $j<i$. To any permutation $x$ is associated a unique
arch system $\mathcal{C}$ and the probability for ${\cal C}$ is
\begin{equation}
\label{k6} P_{\mathbf{G'}}(\mathcal{C})={1\over
n!}\,{\mathrm{number\, of}\,x\to \mathcal{C}} ={1\over
n!}\,\mathrm{card}\{x:\,x\to \mathcal{C}\} \ .
\end{equation}
Two configurations $\mathcal{C}$ and $\mathcal{D}$ of model
\textbf{G'} are equivalent in model \textbf{G} if they are
equivalent by some $\mathbb{Z}_n$ rotation $r$.
\begin{equation}
\label{k7} \mathcal{C}\equiv \mathcal{D}\ \ \iff\ \ \overline{
\mathcal{C}}=\overline{\mathcal{D}} \ .
\end{equation}
This means that if $x$ is a permutation for $\mathcal{C}$, $y=r\circ
x$ is a permutation for $\mathcal{D}$. Hence there are as many
permutations for $\mathcal{C}$ as for $\mathcal{D}$.
\begin{equation}
\label{k8} \mathcal{C}\equiv \mathcal{D}\ \ \Rightarrow\ \
P_{\mathbf{G'}}(\mathcal{C})=P_{\mathbf{G'}}(\mathcal{D}) \ .
\end{equation}
We now count the number of $\mathcal{C}$ which are equivalent by
rotation and give $ \overline{ \mathcal{C}}$. This is obviously
\begin{equation}
\label{k9} \mathrm{number\,of}\ \mathcal{C}\to\overline{
\mathcal{C}}={n\over s(\overline{\mathcal{C}})} \ .
\end{equation}
Now we go back to the model $G$. Any point deposition process on the
circle is also in bijection with a permutation, but now with one
point fixed, for instance $x(1)=1$. Therefore
\begin{equation}
\label{k10} P_\mathbf{G}(C)=\sum_{\mathcal{C}\to\overline{
\mathcal{C}}}P_{\mathbf{G'}}(\mathcal{C})={n\over
s(\overline{\mathcal{C}})}P_{\mathbf{G'}}(\mathcal{C}) \ .
\end{equation}

\begin{figure}[h]
\begin{center}
\includegraphics[width=0.7\textwidth]{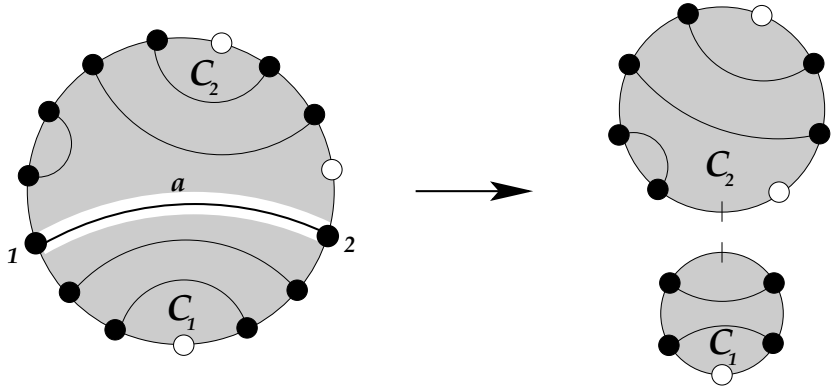}
\caption{The decomposition of a configuration of model \textbf{G}
used in the proof. To be compared with figure \ref{decomposition-AA}.}
\label{decomposition-G}
\end{center}
\end{figure}
We now go back to the proof of the theorem. In model \textbf{G} any
configuration ${\overline{\mathcal{C}}}$ (with $n$ points) can be
constructed by first depositing a couple of points $(1,2)$, which
form a first arch $a_1$ and then by depositing $n_1$ points to the
right of $a$ and $n_2$ points to the left, with of course
$n_1+n_2=n-2$. Let us denote $\mathcal{C}_1$ and $\mathcal{C}_2$ the
arch configurations to the right and to the left of $a_1$ in
${\overline{\mathcal{C}}}$. These configurations are arch
configurations of model \textbf{G'}, not of model \textbf{G}, since
the arch $a_1$ cuts the circle into two segments. Once the first two
points are deposited, amongst the $(n-1)!$ possible ways to deposit
successively the last $n-2$ points, each either to the left or to
the right of $a_1$,  there are $(n-2)!$ possible ways to deposit
$n_1$ points to the right and $n_2$ points to the left,
independently of $(n_1,n_2)$. In other words, the distribution for
$(n_1,n_2)$ is uniform.
\begin{equation}
\label{k11} \mathrm{prob}(n_1,n_2)={1\over n-1}\,,\qquad
n_1+n_2=n-2 \ .
\end{equation}
This can be shown easily by using the recursion relation
\begin{equation}
\label{k12}
\fl\mathrm{prob}(n_1,n_2)=\mathrm{prob}(n_1-1,n_2)\,{n_1\over
n_1+n_2+1}+\mathrm{prob}(n_1,n_2-1)\,{n_2\over n_1+n_2+1}\ ,
\end{equation}
with initial condition $\mathrm{prob}(0,0)=1$.
Once this is done, the conditional probabilities to obtain
$\mathcal{C}_1$ and $\mathcal{C}_2$  are independent, and given by
$P_{\mathbf{G'}}(\mathcal{C}_1)$ and
$P_{\mathbf{G'}}(\mathcal{C}_2)$.

The total probability to obtain a configuration
${\overline{\mathcal{C}}}$ in model \textbf{G} is therefore given by
a sum over all (first) arches $a_1$ in ${\overline{\mathcal{C}}}$.
Each term of the sum is the probability that $a_1$ is the first
deposited arch, and that one obtains $\mathbf{C}_1$ and
$\mathbf{C}_2$ in process \textbf{G'}. There is a counting factor
$2/s(\overline{\mathcal{C}})$ associated to each initial arch $a_1$,
where the factor of $2$ accounts for the two possible choices for
the first point $1$ on $a_1$, and the symmetry factor
$1/s(\overline{\mathcal{C}})$ is there to avoid multiple counting
when several arches are equivalent. Therefore we have finally
\begin{equation}
\label{k14}
P_{G}({\overline{\mathcal{C}}})=\sum_{\mathrm{arches\,}a\in{\overline{\mathcal{C}}}}\
{2\over s(\overline{\mathcal{C}})} \,{1\over n-1}\,
P_{\mathbf{G'}}(\mathcal{C}_1)\, P_{\mathbf{G'}}(\mathcal{C}_2) \ .
\end{equation}
Using Lemma (\ref{lemma1}) the symmetry factor disappears and we
obtain for the probability in model \textbf{G'} the recurrence
equation
\begin{equation}
\label{k15} P_{\mathbf{G'}}(\mathcal{C})=\sum_{\mathrm{arch}\,a\in
\mathcal{C}} {2\over n(n-1)}\,
P_{\mathbf{G'}}(\mathcal{C}_1)\,P_{\mathbf{G'}}(\mathcal{C}_2) \ .
\end{equation}
This is exactly the same recurrence relation as for model
\textbf{A}. The initial condition are the same for $n=0$ and
$n=1$, which proves the theorem.

\subsection{Equivalent tree growth processes}
\label{ss-treegrowth}
\subsubsection{Duality with trees}
\label{sss-dualtree}
\begin{figure}[h]
\begin{center}
\includegraphics[width=9cm]{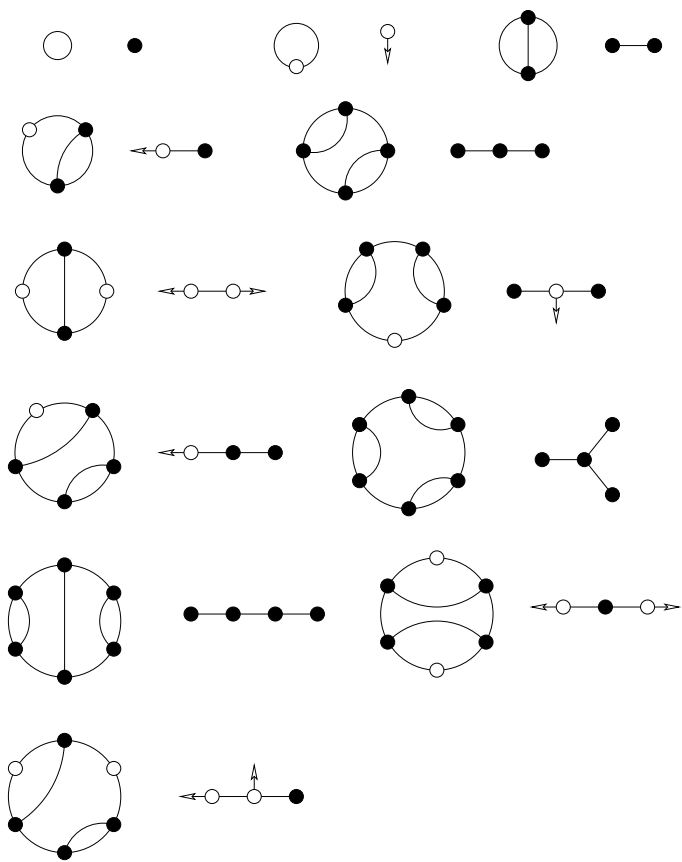}
\caption{First planar arches configurations and their dual decorated
trees configurations (here $n=0$ to $n=6$)} 
\label{circles-to-trees}
\end{center}
\end{figure}

There is a well-known dual description of planar arch systems in
terms of planar trees. Represent faces by vertices, and arches
by links between two vertices. In our model, we have also free
points deposited on the external circle but not yet linked to
another point by an arch. Every face of the planar arch system has
at most one such free point. We represent such a face by a white
vertex  $\circ$ with a white arrow pointing towards the free point.
Every face with no free vertex is represented by a black vertex
$\bullet$. We thus obtain a dual description in term of decorated
planar trees with at most one arrow per vertex (see figure \ref{circles-to-trees}).
Within this dual description, the model \textbf{G} is a planar tree
growth model \textbf{T} defined as follows.

\subsubsection{Tree growth processes \mbox{\bf T}}
\label{sss-modelT}

\begin{figure}[htpb]
\begin{center}
\includegraphics[width=0.5\textwidth]{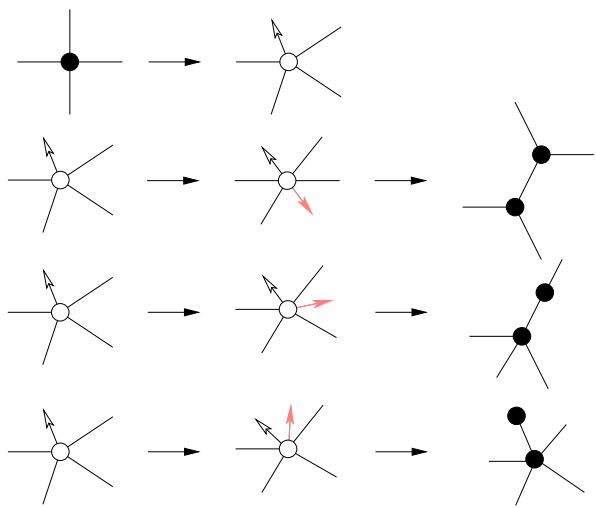}
\caption{Elementary growth steps for the decorated tree model: at
each step we add an arrow to some vertex; if there is already an
arrow the vertex splits in two.
} \label{k17}
\end{center}
\end{figure}

At $t=0$ we start from the tree with a single black vertex and no link. We define the tree-growth process as follows: As illustrated in figure \ref{k17}, at each time step we
\begin{description}
  \item[-] either add an arrow to any black vertex, so that it
  becomes a white vertex (for a black vertex with $k$ links, i.e.
  a $k$-vertex,  there are $k$ different ways to add an arrow);
  \item[-] or add a second arrow to a white vertex (for a white
  vertex with $k$ legs there are $k+1$ different possibilities);
  and then transform this white vertex onto two black vertex with
  a new link ortogonal to the two arrows.
\end{description}
See figure \ref{k17} for an illustration.
This internal budding process is a specific feature of our growth
model. 

\subsubsection{Another tree growth process \mbox{\bf T'}}
\label{sss-modelT'}
\begin{figure}[h]
\begin{center}
\includegraphics[width=0.7\textwidth]{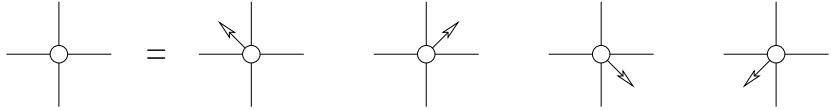}
\caption{The position of a arrow around a $\circ$ vertex is
uniformly distributed.} \label{arrow-no-arrow}
\end{center}
\end{figure}

A similar growth process is obtained if we forget about the arrow position for $\circ$-vertices.  One considers trees with
black vertices $\bullet$, and white vertices $\circ$ if there is an
arrow (see figure\ \ref{arrow-no-arrow}). Indeed it is easy to see
that at each step the position of the arrow around a white vertex is
equiprobable, i.e.\ there is a probability $1/k$ for an arrow to be
at a given position on a type $\circ$ $k$-vertex.
With this property, we consider undecorated trees made out
of $\bullet$- and $\circ$-vertices. We start
from a single $\bullet$ vertex at time $t=0$. At each time step
we
\begin{description}
  \item[-] either transform a black $k$-vertex into a white vertex
  with probability weight $w_{\bullet\to \circ}=k$ (where $k$ is the coordination
  number of the black vertex);
  \item[-] or transform a white $k$-vertex into a pair of black
  vertices, one $k_1$-vertex and one $k_2$-vertex, with $k_1+k_2=k+2$
  (and $k_1$ and $k_2>0$), with a uniform weight $w_{\circ\rightarrow \bullet\,\bullet}=2/k$
  for each occurrence (since for each ordered pair $(k_1,k_2)$ there
  are $k/2$ possible ways to split $\circ\to \bullet\,\bullet$ and there are $k+1$
  such ordered pairs).
\end{description}
\begin{figure}[h]
\centering
\includegraphics[width=0.5\textwidth]{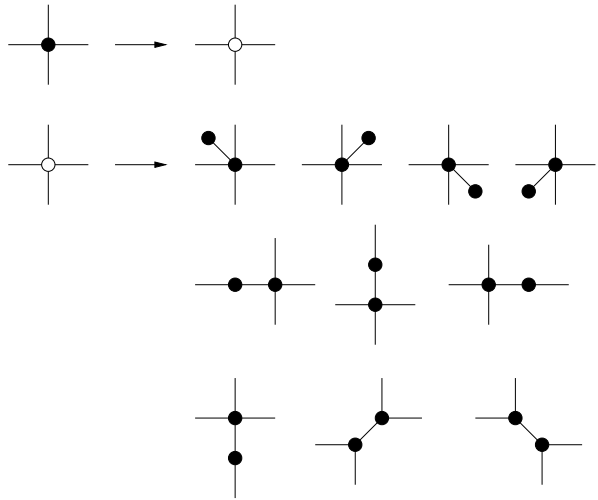}
\caption{Growth processes for a $k=4$ vertex in the undecorated
vertex model. For $k=4$ the process $\bullet\to\circ$ has
probability weight $w(k)=k=4$, each process
$\circ\to\bullet\,\bullet$ has $w(k)=2/k=1/2$.} \label{k18}
\end{figure}
\begin{equation}
\label{k19} w_{\bullet\to\circ}(k)=k
\qquad\mbox{and}\qquad w_{\circ\to\bullet\,\bullet}(k)={2\over
k} \ .
\end{equation}
Let us denote the number of $\bullet$- and $\circ$-vertices at time $t$ by $n_\bullet(t)$ and $n_\circ(t)$ respectively. For any tree created through this growth process up to time $t$, it holds the Euler relation
\begin{equation}\label{nb+nw}
2 n_\bullet(t)+3n_\circ(t)=t+2\ ,
\end{equation}
This relation proven by induction. For $t=0$, $n_{\bullet}=0$
and $n_{\circ}=1$. During a time step $t\to t+1$ we either have $(n_{\bullet},n_{\circ}) \to (n_{\bullet}-1,n_{\circ}+1)$, or
$(n_{\bullet},n_{\circ})\to (n_{\bullet}+2,n_{\circ}-1)$. In both cases $2 n_{\bullet}+3n_{\circ}$
increases by 1, as does $t+1$.

The transition probability $p_{C\to
C'}$ to go from $C\to C'$ is  defined from the probability weights
$w_{C\to C'}$ by
\begin{equation}
\label{k20} p_{C\to C'}={w_{C\to C'}\over \sum_{C''}w_{C\to C''}} \
.
\end{equation}
Thus the probability $P(C,t)$ at time $t$ to be in a configuration
$C$ is obtained recursively by
\begin{equation}
\label{k22} P(C,t)=\sum_{C'}p_{C'\to C}\,
P(C',t-1)=\sum_{C'}{w_{C'\to C}\over\sum\limits_{C''}w_{C'\to
C''}}\,P(C',t-1) \ .
\end{equation}

\begin{figure}[h]
\begin{center}
\includegraphics[width=0.8\textwidth]{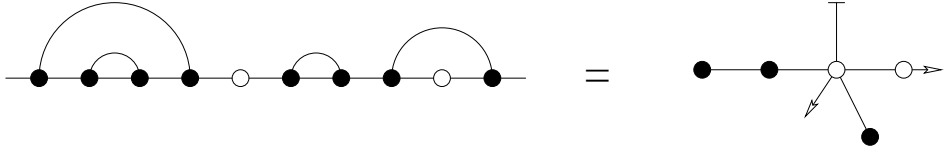}
\caption{Open planar arch systems are dual to rooted trees. The root
is indicated by the barred line pointing to the top.}
\label{g-prime-to-rooted-tree}
\end{center}
\end{figure}

Note also that the dual of an arch configuration $\mathcal{C}$ in
model \textbf{G'} is a rooted tree $\mathcal{T}_\mathrm{rooted}$
(decorated with white arrows, or with black and white vertices as
explained above). Thus model \textbf{G'} is dual to a growth model
for rooted trees (see figure \ref{g-prime-to-rooted-tree}).

\subsection{Mean-field calculation of local observables}
\label{ss-MFcalc}

In this section we present a mean-field theory for the point deposition model $\mathbf{G}$. It allows to compute local observables defined on the tree structures such as average vertex densities, coordination numbers etc. in the large-size or long-time limit. More precisely, the mean-field approximation amounts to neglect fluctuations in the limit $t\to \infty$ which are of order $1/t$. This allows to transform the problem into a stationary process.

\subsubsection{Vertex densities}
The simplest observables are local observables, such as the average
number of vertices of a given type. Let us denote by
$n_\bullet(k,\mathcal{T})$ and $n_\circ(k,\mathcal{T})$ the total
number of $k$-vertices of type $\bullet$ and $\circ$ in a tree
configuration $\mathcal{T}$ obtained at time $t$, starting from
$\bullet$ at time $t=0$. We have shown above in (\ref{nb+nw})
that at any time $t$ and for any tree configuration $\mathcal{T}$
\begin{equation}
\label{k28} \sum_{k=1}^\infty\left(
2n_\bullet(k,\mathcal{T})+3n_\circ(k,\mathcal{T})\right)=t+2 \ .
\end{equation}
Starting from $\mathcal{T}$ at time $t$, the total number of
weighted moves $t\to t+1$ (i.e.\ of ways to add a new point on the
dual configuration) is
\begin{equation}
\label{sumrule4n} \Sigma(t)=\sum_{k=1}^\infty\,\left(
k n_\bullet(k,\mathcal{T})+(k+1)\,n_\circ(k,\mathcal{T})\right)=t\ .
\end{equation}
For simplicity we denote by $b_k(t)={\langle n_\bullet(k,\mathcal{T})\rangle} _t$ and $w_k(t)={\langle n_\circ(k,\mathcal{T})\rangle}_t$ the average number of black and white $k$-vertices at time $t$. We write down a master equation for their time evolution during a step $t\to t+1$. To this end, we have to evaluate the transition probabilities for transformations of black and white vertices.

During the time step $t\to t+1$ the probability for a given black
$k$-vertex  to become white is
\begin{equation}
\label{k30}
p(\bullet\to\circ) = {k\over \Sigma(t)}={k\over t} \ .
\end{equation}
Similarly, the probability for a given white $k$-vertex to
split into a pair of black $k_1$- and $k_2$-vertices 
(with $k_1+k_2=k+2$) is
\begin{equation}
\label{k31}
p(\circ_{k}\to\bullet_{k_1}\bullet_{k_2}) = {k+1\over
\Sigma(t)}={k+1\over t} \ .
\end{equation}
Hence the master equations  for the vertex numbers are given by
\begin{eqnarray}
\label{k32}
b_k(t+1)&=b_k(t)+{1\over t}\left( -k\,b_k(t)+2\sum_{q\ge k-1}\,w_q(t)\right)\\
w_k(t+1)&=w_k(t)+{1\over t}\Bigl(k\,b_k(t)-(k+1)\,w_k(t)  \Bigr)\ .
\end{eqnarray}
In the large time limit $t\to\infty$ we expect the vertex numbers $b_k(t)$ and $w_k(t)$ to be extensive, i.e. proportional to $t$. Therefore, we define (assuming that the
limit exists) the density of black and white vertices as
\begin{equation}
\label{k33} \beta_k=\lim_{t\to\infty}{b_k(t)\over t} \qquad\mbox{and}\qquad
\omega_k=\lim_{t\to\infty}{w_k(t)\over t} \ .
\end{equation}
Consequently, from the master equations we find two coupled recurrence equations
\begin{equation}
\label{k34} \beta_k={2\over k+1}\sum_{q\le k-1}\omega_q \quad,\qquad
\omega_k={k\over k+2}\,\beta_k
\end{equation}
whereas relation (\ref{sumrule4n}) implies
\begin{equation}
\label{k35} \sum_k k\,\beta_k+(k+1)\,\omega_k =1\ .
\end{equation}
The solution of these equations for the densities is (see Appendix
\ref{MeanFieldApp} for the derivation)
\begin{equation}
\label{k36} \beta_k={1\over e^{2}}{2^k\over (k+1)!},\qquad
\omega_k={1\over e^2}{k\ 2^k\over (k+2)!} \ .
\end{equation}

\subsubsection{Results for vertices and related local observables}
The explicit expressions for the vertex densities allow to determine some interesting quantities.
For a system with $t$ bases, the average number of black
and white vertices are
\begin{eqnarray}
\label{k38}  n_\bullet(t)=t\times\,\sum_{k>0}\beta_k=t\times\,{e^2-3\over 2\,e^2}
\qquad\mbox{and}\\ n_\circ(t)=\,t\times\sum_{k>0}\omega_k = t\times\,{1\over e^2} \ .
\end{eqnarray}
Therefore, the average number of vertices is given by
\begin{equation}
  n_\bullet(t)+n_\circ(t) = t \times\left(1-e^{-2}\right)
\end{equation}
We are already familiar with the expression on the left-hand side: because of the duality between trees and arch diagrams, we have just calculated twice the average number of arches in the large strand limit. However, this is nothing but the number of bases (here $t$) times the single base probability (\ref{eqn:h1asymp}).
In fact, this observation is consistent with the value for the fraction of white vertices
\begin{equation}
\label{k37} \omega=\frac{n_\circ}{t}=\sum_{k>0}\omega_k=e^{-2}=0.135335
\end{equation}
because of the relation $\lim_{n\to\infty}h(1,n)=1-\omega$.

In order to learn more about the average tree structure, we compute the average coordination numbers in mean-field theory:
\begin{equation}
\label{k41} \fl\langle{k_\bullet}\rangle =
{\sum\limits_{k>0}k\,\beta_k\over
\sum\limits_{k>0}\beta_k}={e^2+1\over e^2-3}=1.91136,\qquad
\langle{k_\circ}\rangle = {\sum\limits_{k>0}k\,\omega_k\over
\sum\limits_{k>0}\omega_k}={e^2-3\over 2}=2.19453 \ .
\end{equation}
On average vertices have two legs.
{The probability for a branching, i.e.\ the probability to have a vertex with at least 3 points is 
\begin{equation}
\frac{\sum_{k=3}^{\infty} \omega_{k}+\beta_{k}}{\sum_{k=1}^{\infty} \omega_{k}+\beta_{k}} = 
\frac{3 \rme^2-17}{3
   \left(\rme^2-1\right)} = 0.269584\ .
\end{equation}
More specifically, the probabilities $p({k})$ to have a branching with a black or white $k$-vertex are: $p({1})=0.41738$, $p({2})= 0.313035$, $p({3})=0.166952$, $p({4})=0.0695634$, $p({5})=0.0238503$, \ldots. 
We thus conclude that  branchings (i.e.\ vertices with at least three legs) are not rare.}

\subsubsection{Substructures and exterior arch statistics}
\begin{figure}[htpb]
\centering
  \includegraphics[width=0.8\textwidth]{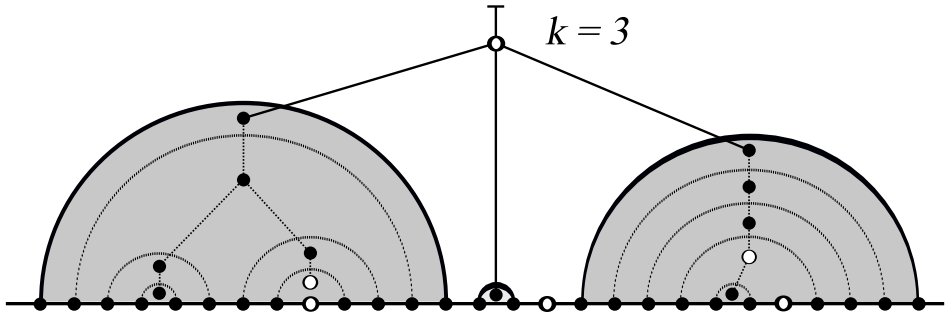}
  \caption{Example of a configuration with $k=3$ substructures and white-coloured root.}
\label{fig:substructure}
\end{figure}

The arch-tree duality allows us to use the tree growth model to analyse the number of substructures of arch diagrams. A substructure is defined as a maximal (or exterior) arch which has no further arch above itself (see figure \ref{fig:substructure}).

\noindent
We characterise an arch diagram by $(k,\sigma)$ where $k$ denotes the number of substructures (number of maximal arches) and $\sigma = \bullet$ or $ \circ$ if the root vertex is black or white.  
We are interested in the large-time probability distribution $p(k,\sigma)$ that  the ``state'' of the arch system is $(k,\sigma)$. 
Consequently, the probability that the arch diagram has $k$ substructures is given by $p(k) = p(k,\bullet)+p(k,\circ)$. 
A $(k,\bullet)$ state is dual to a tree with a black $k$-vertex with a marked leg. The same holds for $(k,\circ)$ states. Taking into account the combinatorial factor of $k$
 for marking a {\em black} $k$-vertex (or equivalently cutting open a circle at a $k$-vertex), and a similar factor of $k+1$ for a white circle (remembering that the additional white point, see figure \ref{fig:substructure}, allows for one more option to cut the circle open) the probabilities $p(k,\bullet)$ and $p(k,\circ)$ are 
proportional to the fraction of black or white vertices respectively
\begin{eqnarray}
p(k,\bullet)&={k\beta_k\over\sum_k (k\beta_k+(k+1) \omega_k)} = k\beta_k,
\\
p(k,\circ)&={(k+1)\omega_k\over\sum_k (k\beta_k+(k+1)\omega_k)} = (k+1)\omega_k\ ,
\end{eqnarray}
where we have used the Euler relation $\sum_k (k\beta_k+(k+1)\omega_k) =1$ to simplify the results. 
Using (\ref{k36}) we obtain
\begin{equation}
  p(k,\bullet) =   \frac{2^{k} }{ e^{2}(k+1) (k-1)!}\qquad\mbox{and}\qquad
  p(k,\circ) = \frac{2^{k} }{e^2  (k+2)(k-1)!}.
\end{equation}
With the probability distribution for the number $k$ of substructures 
\begin{equation}
p(k) = p(k,\bullet)+p(k,\circ) = \frac{2^{k}k(2k+3) }{e^2 (k+2)!}\ , 
\end{equation}
we are able to evaluate its moments $\langle k^m\rangle =\sum_k k^m p(k)$ in order to characterise the arch diagrams. 
The average number of substructures is $\langle k \rangle=(5 e^{2}+1)/(2 e^{2})\approx 2.56767$ to be compared with the result $\langle k\rangle_{\mbox{Catalan}}=3$ found for Catalan structures \cite{Francesco1997}. For its variance we find $\langle k^2\rangle -\langle k\rangle^2 = (9e^4-16e^2-1)/(4e^4)\approx 1.70408$ which is smaller than the corresponding value $\langle k^2\rangle_{\mbox{Catalan}}-\langle k\rangle_{\mbox{Catalan}}^2=4$. We therefore conclude that hierarchically constructed structures fluctuate less than generic, equiprobable Catalan structures.}

\subsection{Dynamical correlations}
\label{ss-Dyncor}
We can also compute \emph{dynamical quantities} in the tree growth model
\textbf{G} (and \textbf{G'}). The dynamics of this model is interesting in its
own, but note that its dynamics is  different
from the arch deposition dynamics of model \textbf{A}.
Let us give a few examples.

\subsubsection{Probability of non-immediate pairing}
Consider process \textbf{G}. Having deposited a point at time $t$, one might ask for the probability that it does not get paired
immediately with a free point already present. This equals
the probability that we add an arrow to a black vertex, not
to a white one. At large $t$ it is 
\begin{equation}
\label{k42} \sum_{k\ge 1} k\,\beta_k={1+e^2\over 2\,e^2}=0.567668\ .
\end{equation}

\subsubsection{Time-dependent pairing propabilities}
What is in model {\bf G} the probability $\Psi(i,j)$ that the point
deposited at time $t_1=i$ is paired with the point deposited at time
$t_2=j>i$, as a function of $i$ and $j$? This amounts to the following
event. At time $t=t_1=i$ a point is deposited on the circle so that no arch
is formed, i.e. a certain black $k$-node is converted to a white
$k$-node. The probability of this event reads $b_k(i)\,k/i$, see (\ref{k30}).
This particular node then remains white up to time $t=t_2=j$ where it is
converted to a pair of black nodes. In the timestep $t\to t+1 \le j$
the probability of keeping the white $k$-node unchanged is
$1-(k+1)/t$, the probability of splitting it is $(k+1)/t$, see (\ref{k31}).
Thus, if we start from a $k$-node at time $t=1$, the probability is
\begin{equation}
\Psi_k(i,j) = \frac{b_k(i)\,k}{i}\left[\prod_{t=i+1}^{j-1}
  \left(1-\frac{k+1}{t}\right)\right]\frac{k+1}{j}
\end{equation}
and the total probability is
\begin{equation}
\Psi(i,j)=\sum_k
  \frac{b_k(i)\,k}{i}\left[\prod_{t=i+1}^{j-1}
  \left(1-\frac{k+1}{t}\right)\right]\frac{k+1}{j}
\ .
\end{equation}
It is interesting to consider the large-time, i.e.\ large-size limit $t\to \infty$ with $i, j\ \to\ \infty$, ${i/j}=\mathcal{O}(1)$.
Indeed, using $b_k(t)\simeq t\,\beta_k$ and
\begin{equation}
\label{k44}
 \prod_{t=i+1}^{j-1}\left(1-\frac{k+1}{t}\right)\ \approx\ \left({i\over j}\right)^{k+1}
\ ,
\end{equation}
the time-dependent pairing probability takes a
simple scaling form 
\begin{equation}
  \Psi(i,j) = \left[\sum_{k=1}^{\infty}
  \frac{1}{j}\left(\frac{i}{j}\right)^{k+1}\right] \frac{2^k}{e^2\,(k-1)!}
  =
  \frac{2}{e^2}\frac{i^2}{j^3} e^{2i/j}
  =: \frac1j \psi(i/j)
 \ .
\end{equation}
Note also that in the large-time, i.e.\ large-size limit $t\to\infty$ a point
deposited at a finite time $i$ gets paired with probability one.
Indeed
\begin{equation}
\label{k45} \int_{i}^\infty\Psi(i,j)\,dj\ =\ {1+e^2\over 2\,e^2}\ =\
0.567668\ ,
\end{equation}
is the probability (\ref{k42}) of non-immediate pairing.

%% file: conclusion.tex
\section{Conclusions and outlook}

To summarise, inspired from the subject of RNA folding, we have introduced and  studied a growth model of planar arch structures (which can be viewed as an arch deposition process). The construction of arch structures is similar to processes generated by greedy algorithms.
The arch-growth model turns out to be amenable to analytical calculations. We have calculated the generating functions for the local height, and their moments. This allowed us to obtain the scaling exponent $\zeta$ for the height, the exponent $\rho$ for the pairing probability, the corresponding scaling functions in the limit of long strands $n\to \infty$, as well as finite-size corrections. We also proved the absence of multicriticality. These results were then confirmed by numerical simulations for systems of sizes up to $n=6500$. 

In a second step, we have defined an equivalent tree-growth model. This model involves growth by vertex splitting as well as by vertex attachment.
This growth process allows to generate RNA configurations with arbitrarily large strands (number of bases).
This allows us to obtain quantities as e.g.\ the probability, that a point gets paired, analytically. 

This work leaves open many interesting questions:
\begin{enumerate}
  \item[-] Some properties (e.g.\ distances on the tree, fractal dimension) are easy to study in the arch-deposition formulation, while some other properties (e.g. substructure statistics) are easier in the tree-growth formulation. It would be interesting to have a better understanding of this fact.
  \item[-] The equivalence between the arch-deposition process and the tree-growth process is very specific to models \textbf{A} and \textbf{G}. We have not been able to find a tree-growth process which is equivalent to the compact arch-deposition model $\bar{\mathbf{A}}$, although this model $\bar{\mathbf{A}}$ is in the same universality class as the non-compact arch-deposition model \textbf{A}.
  \item[-] Is there a tree-growth process which gives the statistics of planar arches in the high temperature phase where all arch structures have the same probability (i.e. the statistics of the so-called ``generic trees" or mean-field branched polymers)?
  \item[-] Arch structures and trees appear in many problems in physics, mathematical physics, combinatorics, computer sciences, etc., in particular in integrable systems (Razumov-Stroganof conjecture, loops models), random permutations, random matrix models, and interface growth.  Are the kind of models introduced in this article related to these problems?
%  \item[-] Finally, to go back to our initial biophysical(ly-inspired) motivation, is it possible to extend these kind of models  beyond simple greedy-like dynamics and to obtain statistics closer to that of random RNA? \red{*** I would not dare writing this: , or even real RNA? ***}
\end{enumerate}

{Finally, since our scaling exponent $\zeta=(\sqrt{17}-3)/2$ deviates from the 
value found for random RNA $\zeta\approx 0.66$, we conclude that the low-temperature phase of random RNA is governed by rules which are more complicated than the greedy algorithm. It would be interesting to find a refined scheme that yields statistics closer to random RNA in order to comprehend the nature of the glassy phase of random RNA.}

%\subsection*{Points to be discussed}
%Discuss the facts that
%\begin{enumerate}
%  \item Global properties (distances on the tree, fractal dimension) are
%  difficult to compute in the growth model formulation
%  \item Local properties much easier
%  \item No equivalence arch deposition = point deposition for compact
%  arch model ${\bar{\mathbf{A}}}$, although the recurrence equations are
%  very similar and the scaling limit is the same (same universality class). Why?
%  \item Is there a growth model equivalent to generic trees (Catalan statistics)?
%  \item Are there other interesting growth models with different and interesting
%  scaling properties?
%  \item Can we build some interesting connections with random permutations,
%  matrix models, integrable models, Temperley-Lieb algebras and all that?
%\end{enumerate}

\section*{Acknowledgements}
{
This work is supported by the EU ENRAGE network
(MRTN-CT-2004-005616) and the Agence Nationale de la Recherche
(ANR-05-BLAN-0029-01 \& ANR-05-BLAN-0099-01). 
The authors thank the KITP (NSF
PHY99-07949), where this work was started, for its hospitality. We are very grateful to M. M\"uller for stimulating discussions, and providing us with a copy of his PhD thesis. We also thank R.\ Bundschuh, P.\ Di Francesco, T.\ Jonsson, L.\ Tang and A.\ Rosso for useful discussions.}

%% file: appendix.tex
\appendix
\section{Mean field equation for the vertex densities}
\label{MeanFieldApp}
This appendix contains a detailed presentation of the computation of the vertex densities within the framework of mean field theory. We start from the Euler relation
\begin{equation}
\label{euler}
  \sum_{k=1}^{\infty}\left(k\;\beta_k+(k+1)\omega_k\right)=1
\end{equation}
and 
\begin{equation}
  \beta_k = \frac{2}{k+1}\sum_{q=k-1}^{\infty}\omega_q
  \, ,\qquad
  \omega_k = \frac{k\beta_k}{k+2}
\end{equation}
Taking these two last relations together, we find
\begin{eqnarray}
\fl\frac{(k+1)\beta_k}{2} = \omega_{k-1} + \frac{(k+2)\beta_{k+1}}{2}
 = \frac{(k-1)\beta_{k-1}}{k+1}+\frac{(k+2)\beta_{k+1}}{2}\\
 \Rightarrow (k+2)^2\beta_{k+1} =(k+2)(k+3)\beta_{k+1} + 2k\beta_k
\end{eqnarray}
This leads to the following differential equation for the generating
function $\mathcal{B}(z)=\sum_{k=1}^\infty \beta_k z^k$:
\begin{equation}
  \fl\left(z\frac{\mbox{d}}{\mbox{d}z}+1\right)^2\left(\mathcal{B}(z) -
  \beta_1z\right) = \frac{\mbox{d}}{\mbox{d}z}\left(z\frac{\mbox{d}}{\mbox{d}z}+1\right)\left(\mathcal{B}(z) -
  \beta_1z-\beta_2z^2\right)+2z^2\mathcal{B}'(z)
\end{equation}
Using the fact that $\beta_2 = 2\beta_1/3$, one finds
\begin{equation}
  z(z-1)\mathcal{B}''(z) - (2z^2-3z+2)\mathcal{B}'(z) +\mathcal{B}(z)
  =  -2\beta_1
\end{equation}
whose solution is
\begin{equation}
  \mathcal{B}(z) = -2\beta_1 + C_1 \;\frac{e^{2z}}{z} + C_2
  \;\frac{1-2z}{z}
\ .
\end{equation}
The simple pole at $z=0$ is removed via setting $C=C_1=-C_2$.
Furthermore, consistency requires $\mathcal{B}(z=0)=0$ what yields
$C=\beta_1/2$. Thus, the generating function is determined up to a
factor:
\begin{equation}
  \mathcal{B}(z) = \beta_1\;\frac{e^{2z}-1-2z}{2z}=
  \beta_1\sum_{k=1}^{\infty}\frac{2^k}{(k+1)!} z^k
\quad \Rightarrow\quad
\beta_k = \frac{\beta_1\,2^k}{(k+1)!}
\end{equation}
The overall factor $\beta_1$ is obtained by insertion of these expression into
\ref{euler}. The result
is
\begin{equation}
  \beta_1\sum_{k=1}\frac{k(2k+3)}{(k+2)!}\;2^k = \beta_1e^2 = 1
\end{equation}
so that $\beta_1=e^{-2}$. Therefore, we obtain the vertex densities
\begin{equation}
  \beta_k = \frac{2^k}{e^2\,(k+1)!} \quad,\qquad \omega_k=
  \frac{2^k \, k}{e^2\, (k+2)!}
\ .
\end{equation}

\section{The compact arch deposition model $\mathbf{\bar A}$}
\label{a-cAmod}
In the compact arch deposition model one deals with strands with an even number of bases
$\ell=2n$, and the arches are always between an even and an odd base
$a=(\mathrm{even},\mathrm{odd})$ or $(\mathrm{odd},\mathrm{even})$.
At the end of the deposition process, there are no free base and there are always $n=\ell/2$ arches.
The recursion relation (\ref{eqn:decomp}) for the probabilities $P_{\bar{\mathbf{A}}}(\mathcal{C})$ for the configurations $\mathcal{C}$  becomes in this model
\begin{equation}
\label{k5}
P_{\bar{\mathbf{A}}}(\mathcal{C})=\sum_{\mathrm{arch}\,a\in
\mathcal{C}} {1\over n^2}\,
P_{\bar{\mathbf{A}}}(\mathcal{C}_1)\,P_{\bar{\mathbf{A}}}(\mathcal{C}_2)
\ .
\end{equation}
The recursion relation for the generating function of the height 
\begin{equation}
F_{\mathbf{\bar A}}(u,v)=\sum_{n=0}^\infty \sum_{i=0}^{\ell=2n} u^i v^{\ell-i}\ \langle h(i,\ell)\rangle_{\mathbf{\bar A}}
\end{equation}
is easily derived and reads
\begin{eqnarray}
\fl    {1\over 4}\left(u{\partial\over\partial u}+v{\partial\over\partial v}\right)^2  F_{\mathbf{\bar A}}(u,v)= uv\left({1+uv\over (1-u^2)(1-v^2)}\right)^2
    \nonumber\\
 \fl   +\left[{u^2\over 1-u^2}\left(u{\partial\over\partial u}+1\right)
    + {v^2\over 1-v^2}\left(v{\partial\over\partial v}+1\right)+ 2uv{1+uv\over (1-u^2)(1-v^2)}\right]F_{\mathbf{\bar A}}(u,v) 
\end{eqnarray}
(to be compared with the equation (\ref{LPDE4F}) obtained for the non-compact model $\mathbf{A}$).

The scaling limit $\ell\to \infty$ is still given by the singularity at $u,v\to 1$.
In this limit the dominant (most singular) terms are
\begin{equation}
   \fl {1\over 4}\left({\partial\over\partial u}{+}{\partial\over\partial v}\right)^2  F_{\mathbf{\bar A}}(u,v)=   \left[ {1\over 2(1-u)}{\partial\over\partial u}
    {+} {1\over 2(1-v)}{\partial\over\partial v}{+}{1\over (1-u)(1-v)}\right]F_{\mathbf{\bar A}}(u,v) 
\end{equation}
This is the same equation than equation (\ref{LPDE4Fs}) for the scaling limit of the generating function $F(u,v)$ for the non-compact model $\mathbf{A}$.
Therefore the scaling limit for the non-compact model $\mathbf{A}$ and the compact model  $\mathbf{\bar A}$ are the same. The same result holds for the higher moments correlation functions and the $N$-points correlators.

Finally, let us mention that, although the deposition models $\mathbf{A}$ and $\mathbf{\bar A}$ are very similar, we have not been able to construct a growth model (i.e. a point deposition model) which could be equivalent to the compact arch deposition model $\mathbf{\bar A}$.

\section{Multicorrelators}
\label{a-mcorr}\begin{figure}[h]
\begin{center}
\includegraphics[width=0.25\textwidth]{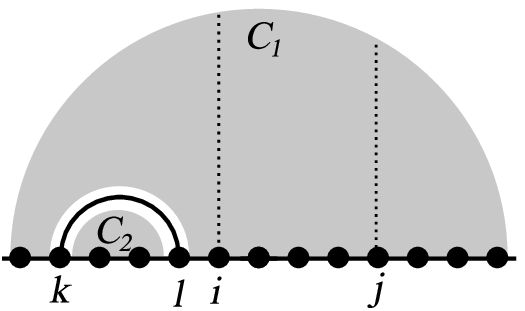}
\quad
\includegraphics[width=0.25\textwidth]{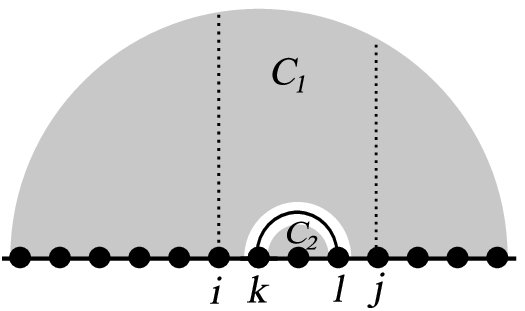}
\quad
\includegraphics[width=0.25\textwidth]{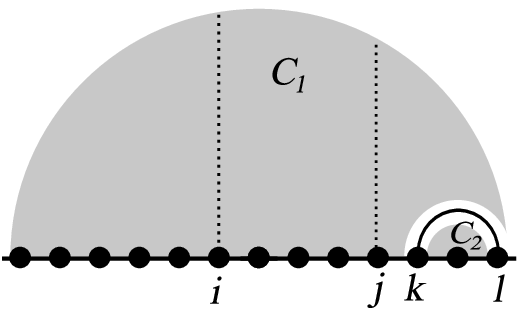}\\
\includegraphics[width=0.25\textwidth]{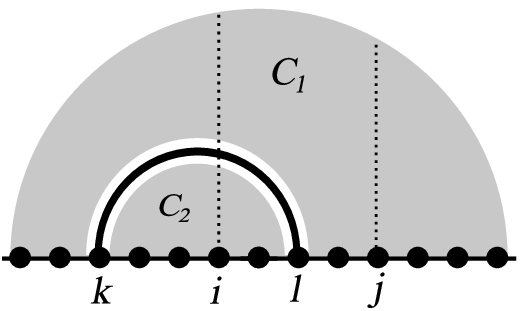}
\quad
\includegraphics[width=0.25\textwidth]{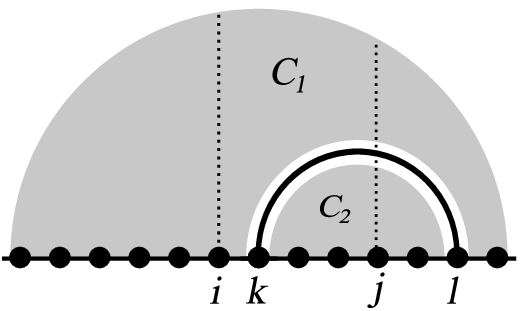}
\quad
\includegraphics[width=0.25\textwidth]{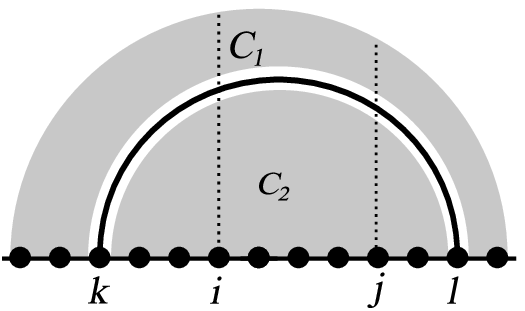}
\caption{The six different ways to deposit the first arch $(k,l)$ w.r.t. the two points $i$ and $j$ correspond to the six different terms in the r.h.s. of (\ref{PDE4G2})}
\label{FigA2pt}
\end{center}
\end{figure}

We can extend the recurrence equations (\ref{rec4h}) and (\ref{lf49}) to compute correlation functions for heights at several points of the strand. 
Let us consider the $2$-point  correlators. They are the expectation values, at two points $i$ and $j$, of
\begin{equation}
\langle h_{\scriptscriptstyle{\mathcal{C}}}(i_1,n)^{k_1} 
 h_{\scriptscriptstyle{\mathcal{C}}}(i_2,n)^{k_2} \rangle\ .
\end{equation}
A generating function for these correlators is
\begin{equation}
G_2(u,v,w;z_1,z_2)=\sum_{0\le i<j\le n}\, u^{i}\, v^{j-i}\, w^{n-j}
\langle
\mathrm{e}^{z_1 h_{\scriptscriptstyle{\mathcal{C}}}(i,n)}
\mathrm{e}^{z_2 h_{\scriptscriptstyle{\mathcal{C}}}(j,n)}
\rangle
\end{equation}
The recurrence equation is obtained by considering all the possible positions for the first arch $(k,l)$ with respect to the two points $i$ and $j$ (see figure \ref{FigA2pt})
Denoting by $\mathcal{L}$ the {\em length operator}
\begin{equation}
\mathcal{L}:=\left(u{\partial\over\partial u}+v{\partial\over\partial v}+w{\partial\over\partial w}\right) 
\end{equation}
and $G(u,v;z)$ the 1-point function studied in Sect.~\ref{ss-scalhk}, we obtain the {linear PDE} for $G_2$
\begin{eqnarray}
\label{PDE4G2}
\fl  {\frac 1 2}\mathcal{L}(\mathcal{L}{-}1)G_2(u,v,w;z_1,z_2){=} 
  \left[  
  {u^2\over 1-u}\left(u{\partial\over\partial u}{+}1\right)
  {+}{v^2\over 1-v}\left(v{\partial\over\partial v}{+}1\right)
  {+}{w^2\over 1-w}\left(w{\partial\over\partial w}{+}1\right)
  \right.
   \nonumber\\
  \fl\left.{
  \vphantom{{u^2\over 1-u}\left(u{\partial\over\partial u}+1\right)}
  {+}uv\,\mathrm{e}^{z_1}\,G(u,v;z_1)
  {+}vw\,\mathrm{e}^{z_2}\,G(v,w;z_2)
  {+}uw\,\mathrm{e}^{z_1+z_2}\,G(u,w;z_1+z_2)
 } \right]
  G_2(u,v,w;z_1,z_2)
  \end{eqnarray}
Each term in the r.h.s. of (\ref{PDE4G2}) corresponds to one of the positions in figure \ref{FigA2pt} (in the same order).
 The boundary conditions are given by the cases $u=0$, $v=0$ or $w=0$ where the 2-point function $G_2$ reduces to a 1-point function $G=G_1$ or a 0-point function $G_0$.
 
Similarly, we may consider the 3-point function 
$$G_3(u,v,w,x;z_1,z_2,z_3)=\sum_{0\le i\le j \le k\le n}\langle u^i v^{j-i} w^{k-l} x^{n-k} \mathrm{e}^{z_1h(i,n)+z_2h(j,n)+z_3h(k,n)}\rangle$$
$G_3$ satisfies a linear PDE with coefficients involving the 1-point and 2-point functions $G$ and $G_2$.